\documentclass[reprint,
preprintnumbers,
nofootinbib,
amsmath,amssymb,
aps,
prd]{revtex4-2}

\usepackage{graphicx}
\usepackage{dcolumn}
\usepackage{bm}
\usepackage[hidelinks]{hyperref}
\usepackage[all]{hypcap}
\usepackage{xcolor}
\usepackage{enumitem}

\usepackage{tikz}
\usetikzlibrary{decorations.markings,arrows.meta}

\tikzset{->-/.style={decoration={
  markings,
  mark=at position .68 with {\arrow{Latex}}},postaction={decorate}}}

\tikzset{-->--/.style={decoration={
  markings,
  mark=at position .59 with {\arrow{Latex}}},postaction={decorate}}}

\def\plaq{\tikz[baseline=-0.6ex]{
\fill (0,0.75ex) circle (1.5pt) coordinate (A);
\fill (4ex,0.75ex) circle (1.5pt) coordinate (B);
\fill (4ex,4.75ex) coordinate (C);
\fill (0,4.75ex) coordinate (D);
\draw[black,->-] (B)--(C);
\draw[black,->-] (C)--(D);
\draw[black,->-] (D)--(A);

\fill (0,-0.75ex) circle (1.5pt) coordinate (A);
\fill (4ex,-0.75ex) circle (1.5pt) coordinate (B);
\fill (4ex,-4.75ex) coordinate (C);
\fill (0,-4.75ex) coordinate (D);
\draw[black,->-] (B)--(C);
\draw[black,->-] (C)--(D);
\draw[black,->-] (D)--(A);}
}

\def\upperonebytworect{\tikz[baseline=-0.6ex]{
\fill (0,-4ex) circle (1.5pt) coordinate (A);
\fill (4ex,-4ex) circle (1.5pt) coordinate (B);
\fill (4ex,4ex) coordinate (C);
\fill (0,4ex) coordinate (D);
\draw[black,-->--] (B)--(C);
\draw[black,->-] (C)--(D);
\draw[black,-->--] (D)--(A);}
}

\def\loweronebytworect{\tikz[baseline=-0.6ex]{
\fill (0,4ex) circle (1.5pt) coordinate (A);
\fill (4ex,4ex) circle (1.5pt) coordinate (B);
\fill (4ex,-4ex) coordinate (C);
\fill (0,-4ex) coordinate (D);
\draw[black,-->--] (B)--(C);
\draw[black,->-] (C)--(D);
\draw[black,-->--] (D)--(A);}
}

\def\forwardtwobyonerects{\tikz[baseline=-0.6ex]{
\fill (0,0.75ex) circle (1.5pt) coordinate (A);
\fill (4ex,0.75ex) circle (1.5pt) coordinate (B);
\fill (8ex,0.75ex) coordinate (C);
\fill (8ex,4.75ex) coordinate (D);
\fill (0,4.75ex) coordinate (E);
\draw[black,->-] (B)--(C);
\draw[black,->-] (C)--(D);
\draw[black,-->--] (D)--(E);
\draw[black,->-] (E)--(A);

\fill (0,-0.75ex) circle (1.5pt) coordinate (A);
\fill (4ex,-0.75ex) circle (1.5pt) coordinate (B);
\fill (8ex,-0.75ex) coordinate (C);
\fill (8ex,-4.75ex) coordinate (D);
\fill (0,-4.75ex) coordinate (E);
\draw[black,->-] (B)--(C);
\draw[black,->-] (C)--(D);
\draw[black,-->--] (D)--(E);
\draw[black,->-] (E)--(A);}
}

\def\backwardtwobyonerects{\tikz[baseline=-0.6ex]{
\fill (0,0.75ex) circle (1.5pt) coordinate (A);
\fill (4ex,0.75ex) circle (1.5pt) coordinate (B);
\fill (4ex,4.75ex) coordinate (C);
\fill (-4ex,4.75ex) coordinate (D);
\fill (-4ex,0.75ex) coordinate (E);
\draw[black,->-] (B)--(C);
\draw[black,-->--] (C)--(D);
\draw[black,->-] (D)--(E);
\draw[black,->-] (E)--(A);

\fill (0,-0.75ex) circle (1.5pt) coordinate (A);
\fill (4ex,-0.75ex) circle (1.5pt) coordinate (B);
\fill (4ex,-4.75ex) coordinate (C);
\fill (-4ex,-4.75ex) coordinate (D);
\fill (-4ex,-0.75ex) coordinate (E);
\draw[black,->-] (B)--(C);
\draw[black,-->--] (C)--(D);
\draw[black,->-] (D)--(E);
\draw[black,->-] (E)--(A);}
}

\def\mbynclover{\tikz[baseline=-0.6ex]{
\fill (0.5ex,0.5ex) circle (1.5pt) coordinate (A);
\fill (8.5ex,0.5ex) coordinate (B);
\fill (8.5ex,4.5ex) coordinate (C);
\fill (0.5ex,4.5ex) coordinate (D);
\draw[black,-->--] (A)--(B);
\draw[black,->-] (B)--(C);
\draw[black,-->--] (C)--(D);
\draw[black,->-] (D)--(A);

\fill (-0.5ex,0.5ex) circle (1.5pt) coordinate (A);
\fill (-0.5ex,4.5ex) coordinate (B);
\fill (-8.5ex,4.5ex) coordinate (C);
\fill (-8.5ex,0.5ex) coordinate (D);
\draw[black,->-] (A)--(B);
\draw[black,-->--] (B)--(C);
\draw[black,->-] (C)--(D);
\draw[black,-->--] (D)--(A);

\fill (0.5ex,-0.5ex) circle (1.5pt) coordinate (A);
\fill (0.5ex,-4.5ex) coordinate (B);
\fill (8.5ex,-4.5ex) coordinate (C);
\fill (8.5ex,-0.5ex) coordinate (D);
\draw[black,->-] (A)--(B);
\draw[black,-->--] (B)--(C);
\draw[black,->-] (C)--(D);
\draw[black,-->--] (D)--(A);

\fill (-0.5ex,-0.5ex) circle (1.5pt) coordinate (A);
\fill (-8.5ex,-0.5ex) coordinate (B);
\fill (-8.5ex,-4.5ex) coordinate (C);
\fill (-0.5ex,-4.5ex) coordinate (D);
\draw[black,-->--] (A)--(B);
\draw[black,->-] (B)--(C);
\draw[black,-->--] (C)--(D);
\draw[black,->-] (D)--(A);}
}

\def\nbymclover{\tikz[baseline=-0.6ex]{
\fill (0.5ex,0.5ex) circle (1.5pt) coordinate (A);
\fill (4.5ex,0.5ex) coordinate (B);
\fill (4.5ex,8.5ex) coordinate (C);
\fill (0.5ex,8.5ex) coordinate (D);
\draw[black,->-] (A)--(B);
\draw[black,-->--] (B)--(C);
\draw[black,->-] (C)--(D);
\draw[black,-->--] (D)--(A);

\fill (-0.5ex,0.5ex) circle (1.5pt) coordinate (A);
\fill (-0.5ex,8.5ex) coordinate (B);
\fill (-4.5ex,8.5ex) coordinate (C);
\fill (-4.5ex,0.5ex) coordinate (D);
\draw[black,-->--] (A)--(B);
\draw[black,->-] (B)--(C);
\draw[black,-->--] (C)--(D);
\draw[black,->-] (D)--(A);

\fill (0.5ex,-0.5ex) circle (1.5pt) coordinate (A);
\fill (0.5ex,-8.5ex) coordinate (B);
\fill (4.5ex,-8.5ex) coordinate (C);
\fill (4.5ex,-0.5ex) coordinate (D);
\draw[black,-->--] (A)--(B);
\draw[black,->-] (B)--(C);
\draw[black,-->--] (C)--(D);
\draw[black,->-] (D)--(A);

\fill (-0.5ex,-0.5ex) circle (1.5pt) coordinate (A);
\fill (-4.5ex,-0.5ex) coordinate (B);
\fill (-4.5ex,-8.5ex) coordinate (C);
\fill (-0.5ex,-8.5ex) coordinate (D);
\draw[black,->-] (A)--(B);
\draw[black,-->--] (B)--(C);
\draw[black,->-] (C)--(D);
\draw[black,-->--] (D)--(A);}
}

\hypersetup{
	colorlinks,
	linkcolor={red!50!black},
	citecolor={blue!50!black},
	urlcolor={blue!80!black}
}

\DeclareMathOperator{\tr}{Tr}
\renewcommand{\Re}{\operatorname{Re}}
\renewcommand{\Im}{\operatorname{Im}}

\begin{document}

\preprint{ADP-23-30/T1239}

\title{Numerical evidence for fractional topological objects in SU(3) gauge theory}

\author{Jackson A. Mickley}
\author{Waseem Kamleh}
\author{Derek B. Leinweber}
\affiliation{Centre for the Subatomic Structure of Matter, Department of Physics, The University of Adelaide, South Australia 5005, Australia}

\begin{abstract}
The continued development of models that propose the existence of fractional topological objects in the Yang-Mills vacuum has called for a quantitative method to study the topological structure of SU($N$) gauge theory. We present an original numerical algorithm that can identify distinct topological objects in the nontrivial ground-state fields and approximate the net charge contained within them. This analysis is performed for SU(3) colour at a range of temperatures crossing the deconfinement phase transition, allowing for an assessment of how the topological structure evolves with temperature. We find a promising consistency with the instanton-dyon model for the structure of the QCD vacuum at finite temperature. Several other quantities, such as object density and radial size, are also analysed to elicit a further understanding of the fundamental structure of ground-state gluon fields.
\end{abstract}

\maketitle

\section{Introduction} \label{sec:intro}
The nonperturbative nature of quantum chromodynamics (QCD) precludes the analytic study of many of its most important phenomena, such as quark confinement. In SU($N$) gauge theory, an area-law behaviour of large Wilson loops,
\begin{equation}
    \langle W(C)\rangle \sim \exp\left(-\sigma A(C)\right) \,,
\end{equation}
is often taken as an indicator of confinement in the context of static heavy quarks \cite{WilsonLoopAreaLaw}. This picture is complicated by the presence of light quarks, which results in string breaking at large separations \cite{StringBreaking}. One can instead analyse the Schwinger function of the gluon propagator, where a transition to negative values at large Euclidean times implies the spectral density is not positive definite \cite{GluonPropConfinementI}. It follows that there is no K\"{a}llen-Lehmann representation of the gluon propagator, a manifestation that the corresponding physical states are confined.

This is found in theories with or without dynamical quarks \cite{GluonPropConfinementII, GluonPropConfinementIII}, suggesting it is the behaviour of the gluon fields that underpins confinement, though no complete theoretical mechanism is currently known. Pure SU($N$) Yang-Mills theory is known to experience a phase transition at a critical temperature $T_c$ above which confinement breaks down. This motivates exploring the evolution of the gauge fields through the phase transition to elicit fundamental properties that can be attributed to confinement.

Nonperturbative aspects of QCD are primarily studied through lattice QCD, wherein the theory is formulated on a discrete lattice in Euclidean spacetime. Modelling SU($N$) gauge theory in Euclidean spacetime brought about the discovery of the instanton \cite{YangMillsSolution}, a classical topological configuration which corresponds to the (anti-)self-dual local minima of the Yang-Mills action functional,
\begin{equation}
    S = \frac{1}{2} \int d^4x \, \tr\left(F_{\mu\nu} F_{\mu\nu} \right) \,.
\end{equation}
The instanton solution formed the basis of the instanton liquid model \cite{InstantonLiquidModelI}, which sought to model the QCD vacuum in terms of an ensemble of interacting semiclassical instantons, with fluctuations around the classical solution. The model was able to explain chiral symmetry breaking, though did not account for confinement \cite{InstantonLiquidModelI, InstantonLiquidModelII, InstantonLiquidModelIII, ILMConfinementI, ILMConfinementII, ILMConfinementIII}.

Models subsequently emerged that proposed the existence of fractional topological configurations. Analytic self-dual solutions to the Yang-Mills equations with fractional charge $\sim 1/N$ have been known to exist on the twisted torus $\mathbb{T}^4$ since the early 1980s \cite{TwistedSelfdual}. Cooling methods with twisted boundary conditions have isolated such ``fractional instantons" \cite{TwistedSelfdualCoolingI, TwistedSelfdualCoolingII}, and the solutions have been studied numerically in SU(2) \cite{TwistedSelfdualStudyI, TwistedSelfdualStudyII} and general SU($N$) \cite{TwistedSelfdualStudyIII, TwistedSelfdualStudyIV} gauge theory. Unlike the regular instanton, fractional instantons possess $\mathbb{Z}_N$ flux and have thus been put forth as a possible microscopic mechanism for confinement \cite{TwistedSelfdualCoolingI, TwistedSelfdualCoolingII, FractionalReview}. The original twisted $\mathbb{T}^4$ solution has since been extended to a vastly broader class of configurations \cite{FractionalGeneralisationI, FractionalGeneralisationII}.

However, the breakthrough came with the discovery of calorons \cite{InstantonDyonsI, InstantonDyonsII, InstantonDyonsIII, InstantonDyonsIV}, a finite-temperature generalisation of the instanton on $\mathbb{R}^3\times \mathbb{T}^1$. The caloron profile can be viewed as composed of $N$ monopole constituents, known as dyons, each possessing fractional charge depending on the Polyakov loop at spatial infinity. This naturally led to modelling the finite-temperature Yang-Mills vacuum in terms of semiclassical instanton-dyons \cite{ConfiningDyons, DyonConfinementDeconfinement, DyonLiquidModel, SU2DyonEnsemble}, which has successfully reproduced the second-order phase transition of SU(2) \cite{SU2DyonConfinementI, SU2DyonConfinementII, SU2DyonConfinementIII} and first-order phase transition of SU(3) \cite{SU3DyonConfinement} in numerical simulations.

By additionally varying the periods of the torus in each dimension, classical fractional solutions have been constructed on $\mathbb{R}^n \times \mathbb{T}^{4-n}$ for different $n$. For instance, whilst the caloron solution has $n=3$, Refs.~\cite{FractionalVortexI, FractionalVortexII, FractionalVortexIII} considered ``doubly-periodic" vortex-like solutions on $\mathbb{R}^2 \times \mathbb{T}^2$ ($n=2$) and Refs.~\cite{TwistedSelfdualStudyI, TwistedSelfdualStudyII, FractionalFibonacci} explored the ``Hamiltonian geometry" $\mathbb{R}\times \mathbb{T}^3$ ($n=1$). Although specific periodicities and twists are necessary to \textit{isolate} these fractional solutions, they are certainly not required for their existence \cite{FractionalReview}. More recently, there have also been constructions of fractional topological objects with charges $\sim 1/N$ in the confining phase through quantum fluctuations of an effective action \cite{FractionalSUN}.

Working in SU(3) pure gauge theory, we present a numerical algorithm that can identify distinct topological objects within an arbitrary distribution and approximate the net topological charge contained within each such object. This analysis is performed at a range of temperatures either side of $T_c$, providing a direct evaluation of the underlying structure in the gluon fields and how this evolves with temperature. The conclusions are subsequently compared against the instanton-dyon model to test the validity of its main predictions.

This paper is structured as follows. In Sec.~\ref{sec:charges}, our algorithm is presented in detail and tested. Section \ref{sec:smoothing} covers the smoothing applied to the gauge fields. The continuum limit is explored in Sec.~\ref{sec:scaling}, and our main results are subsequently presented in Sec.~\ref{sec:results}. Thereafter, the significance of the results is discussed in Sec.~\ref{sec:discussion}, along with an investigation into several other statistics available through our methods. Finally, we conclude our main findings in Sec.~\ref{sec:conclusion}.

\section{Calculating Charges} \label{sec:charges}
The core of our analysis involves a novel method to approximate the net charge of distinct topological objects within an arbitrary topological charge distribution. This is fundamentally a very complicated task, as we desire a technique that avoids any references to a specific topological configuration. A previous approach to studying topological structure, presented throughout Refs.~\cite{ThresholdAnalysisI, ThresholdAnalysisII, ThresholdAnalysisIII, ThresholdAnalysisIV, ThresholdAnalysisV, ThresholdAnalysisVI, ThresholdAnalysisVII}, proceeds by implementing a threshold that divides the distribution up into disconnected clusters. Though interesting properties can be studied through this method, it is unsuitable for our primary goal of providing an unbiased estimate of the charge within each distinct object. This is due to two primary reasons. First, points below the threshold that are not assigned to a cluster are disregarded, even though these may be an important contribution to the charge of some objects. Second, distinct objects can be connected by a path above the threshold and identified as part of the same cluster, which is a clear hindrance to our objective.
	
To overcome these, we have devised a more fundamental strategy presented below. Instead, we utilise an iterative procedure where objects are first identified by peaks in the topological charge density and then allowed to grow outwards one step at a time. Subject to a few assumptions, we can be confident the set of points ultimately assigned to an object is a reasonable representation of its distribution. The algorithm is described in detail below.

\subsection{Algorithm} \label{subsec:algorithm}
\begin{figure*}
    \centering
    \includegraphics[width=0.35\linewidth]{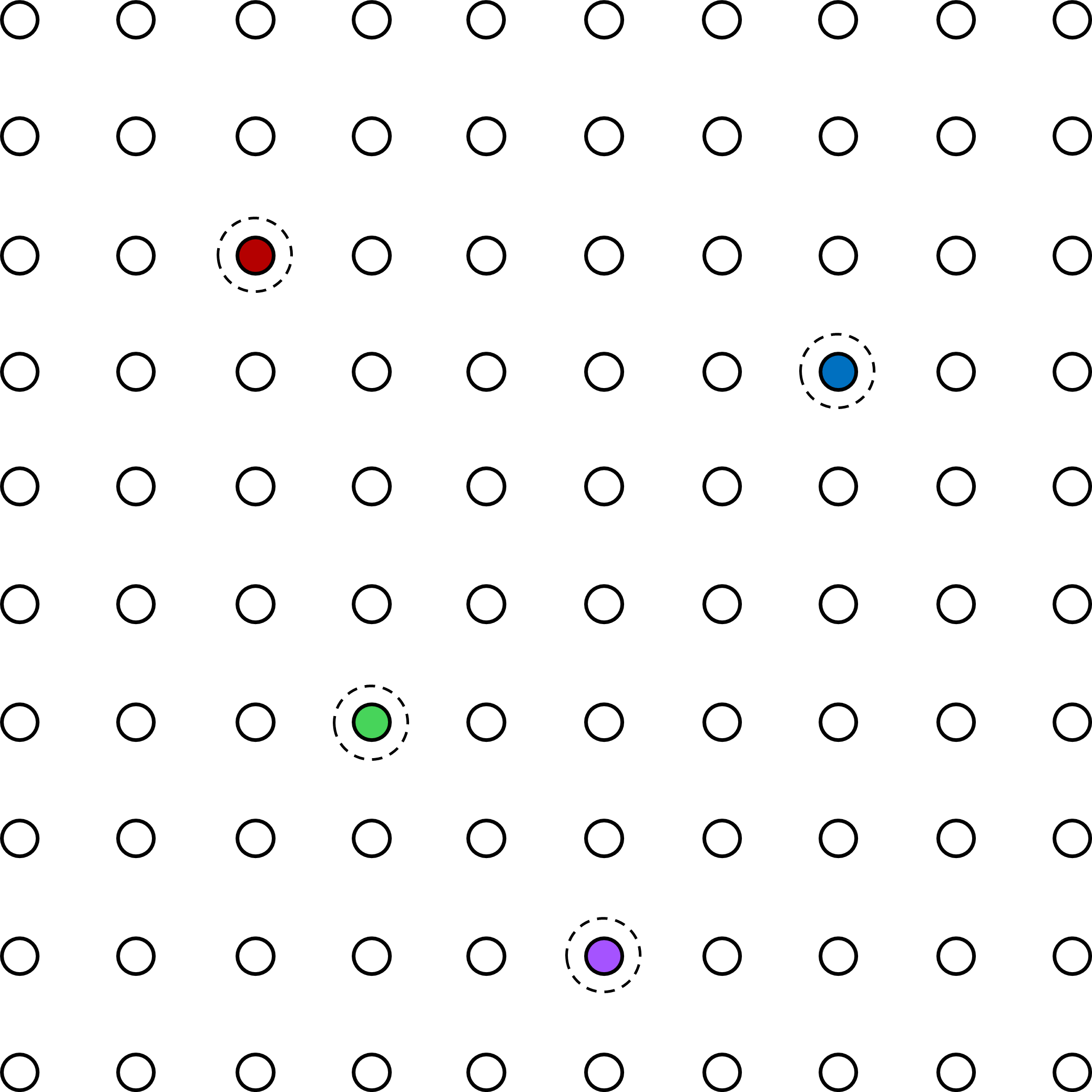}
    \hspace{5em}
    \includegraphics[width=0.35\linewidth]{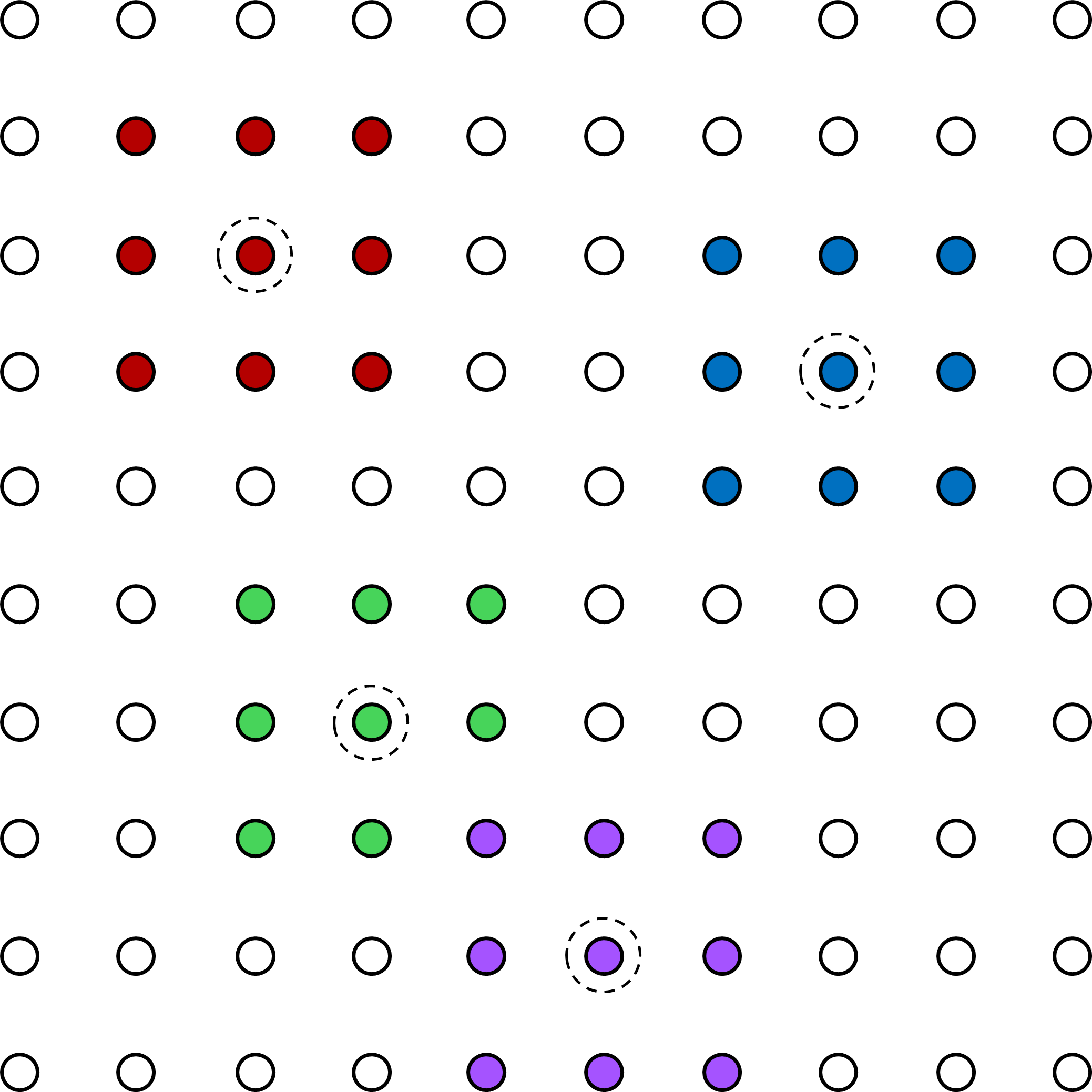}
    \vspace{3em}
    
    \includegraphics[width=0.35\linewidth]{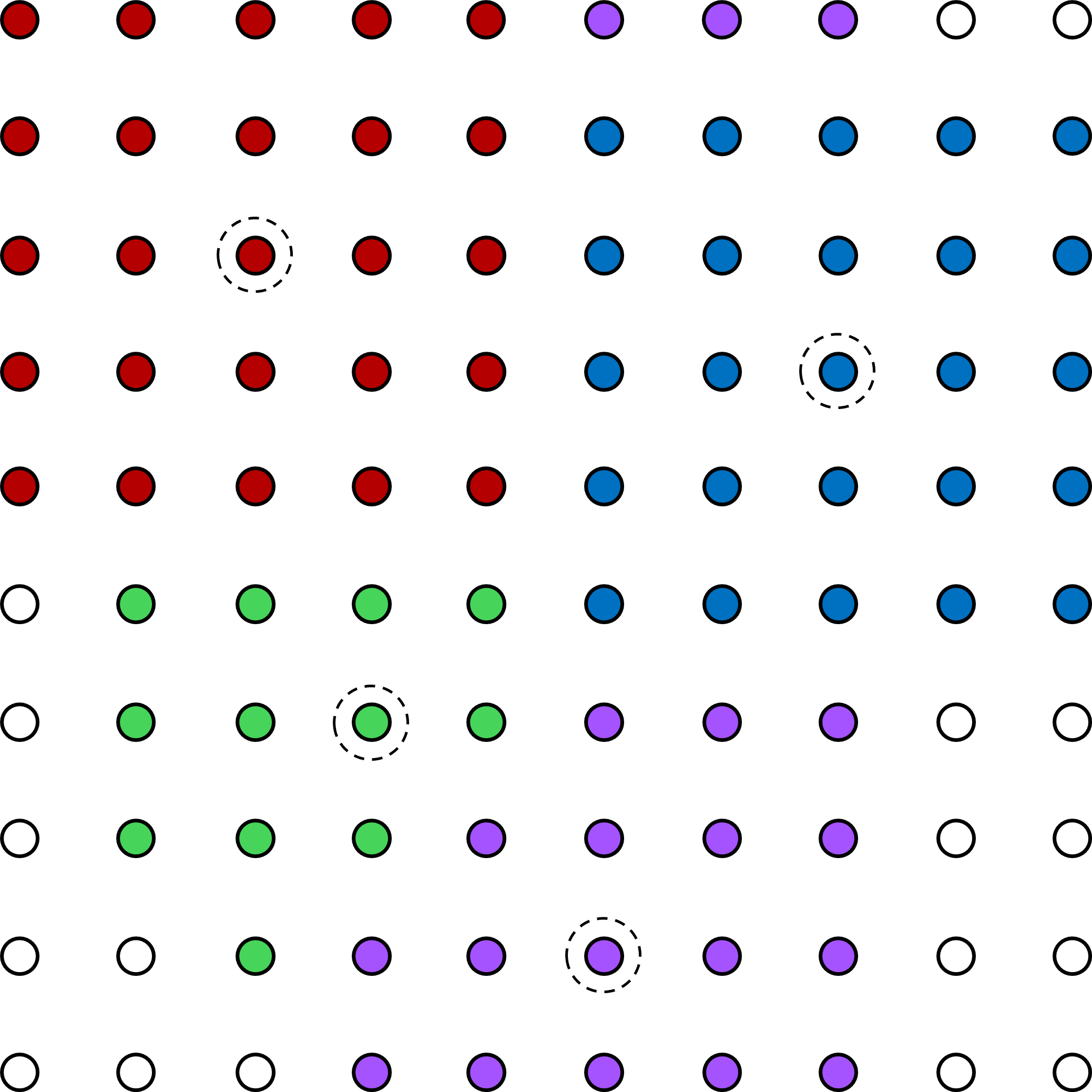}
    \hspace{5em}
    \includegraphics[width=0.35\linewidth]{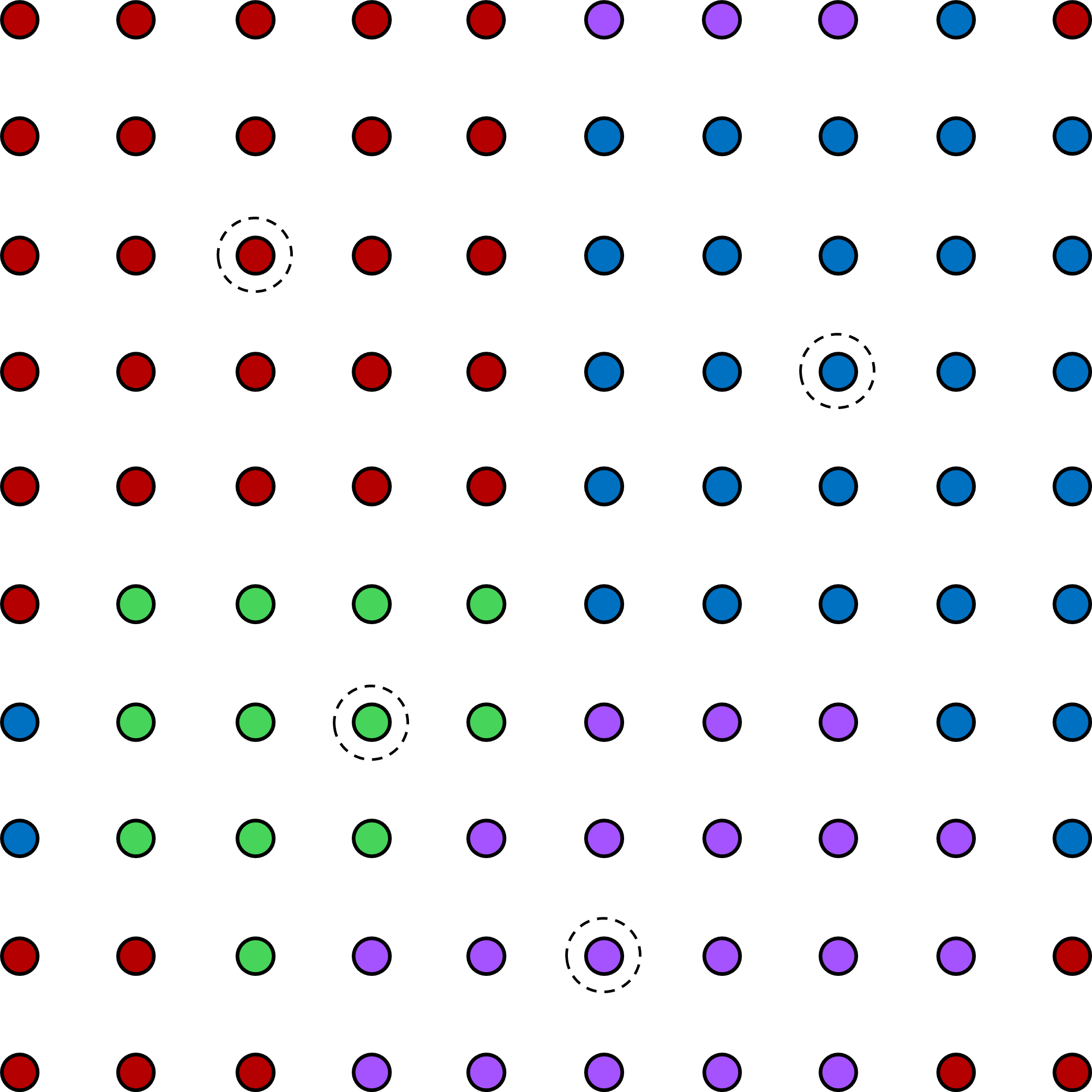}
    \vspace{3em}
    
    \includegraphics[width=0.35\linewidth]{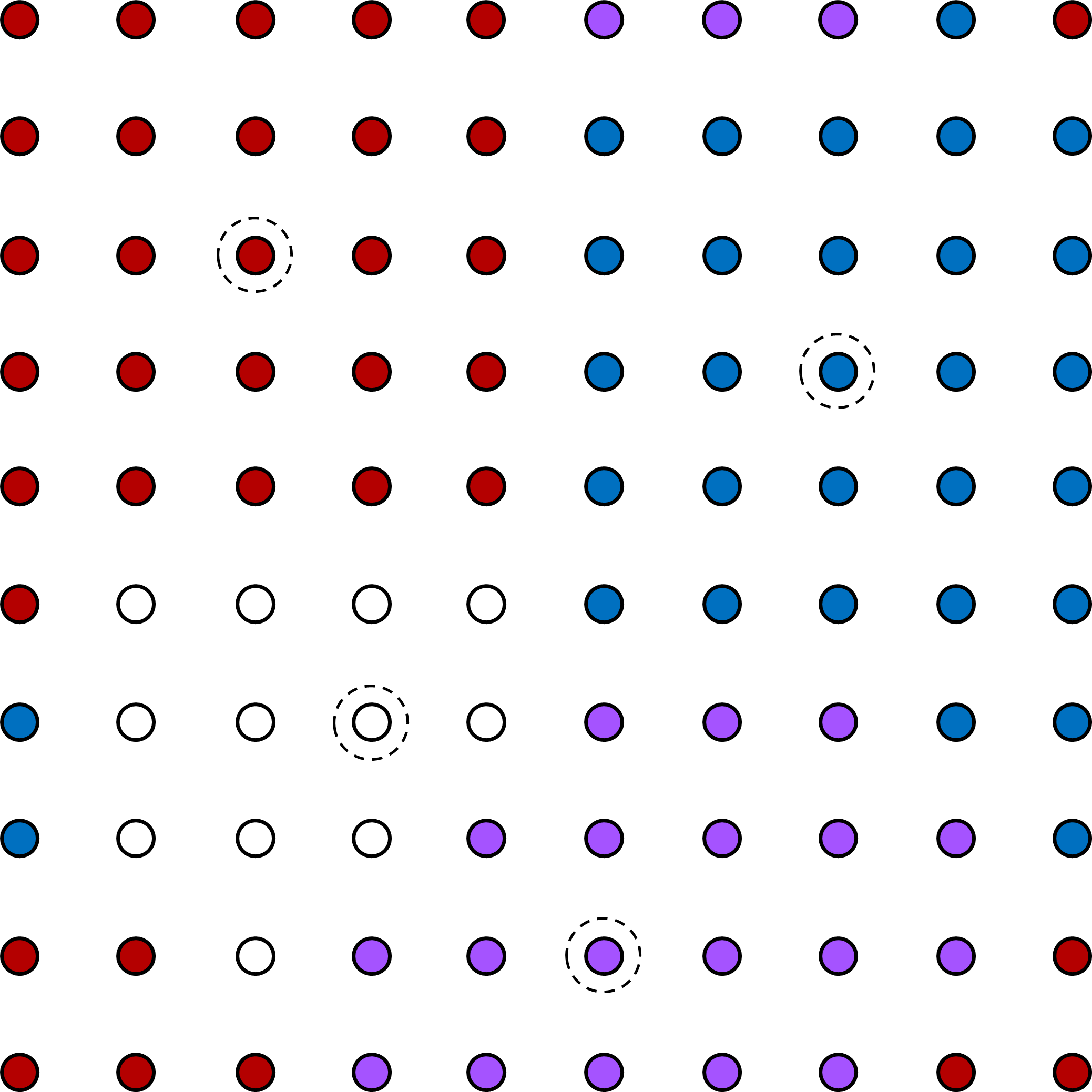}
    \hspace{5em}
    \includegraphics[width=0.35\linewidth]{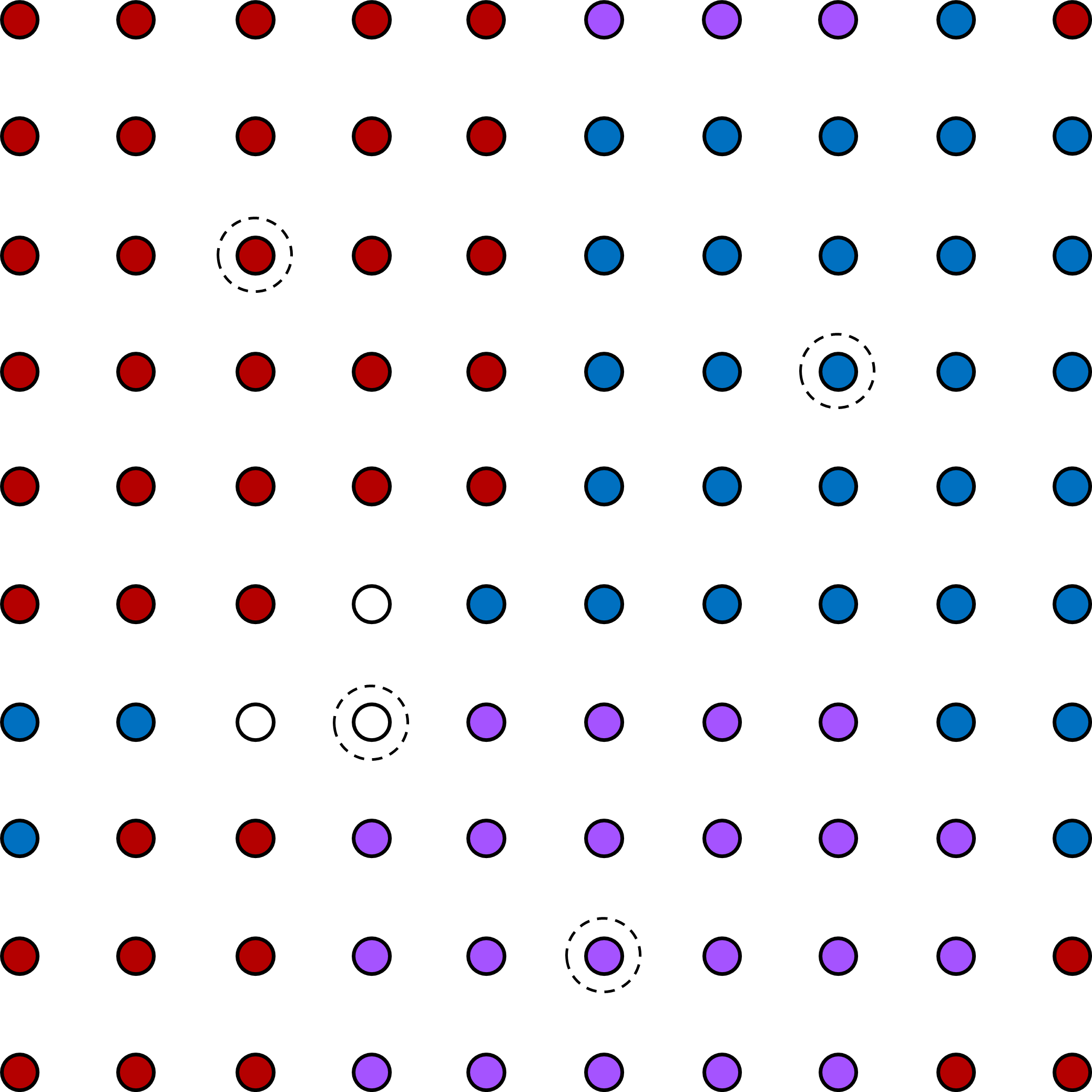}
    \caption{\label{fig:algorithmgraphic} A graphic demonstrating our algorithm, proceeding from left to right then top to bottom. The order in which objects grow is red $\rightarrow$ blue $\rightarrow$ purple $\rightarrow$ green (in greyscale, from darkest to lightest shade of grey). First, the peaks are identified and neighbouring hypercubes assigned (\textbf{top row}). The objects are then allowed to grow until all possible points have been assigned (\textbf{middle row}). Finally, we discard the peaks that fail to assign all points within the $3^4$ hypercube, and the remaining objects grow until no more points satisfy the required criteria (\textbf{bottom row}). In this case, these are the green (lightest grey) points.}
\end{figure*}

We start by presenting the full algorithm and discuss the motivation for each step in the following subsections. The algorithm is to be applied to a UV-smoothed configuration, as covered in detail in Sec.~\ref{sec:smoothing}. For a topological charge distribution $q(x)$:
\begin{enumerate}
    \item Identify the peaks of objects through local maxima [for $q(x)>0$] and minima [for $q(x)<0$] within a $3^4$ hypercube. Label each one with an identifying ``object number".
    \item Proceeding in order of smallest to largest peak, take all points currently assigned to the corresponding object number, $\{x\}$, and assign the same object number to neighbouring points, $\{x'\}$, in the $3^4$ hypercube centred at $x$ which:
    \begin{itemize}[leftmargin=*]
        \item have not yet been assigned an object number,
        \item have the same sign density ($q(x')\, q(x) > 0$), and
        \item have a smaller absolute value $|q(x')| \leq |q(x)|$.
    \end{itemize}
    \item Repeat Step 2 until no more valid points can be assigned as part of an object (i.e. no remaining lattice sites satisfy the above criteria).
    \item Filter out the peaks that fail to assign all points within the surrounding hypercube. Reinstate Step 3 for the surviving peaks to reallocate the newly available points.
    \item Calculate the net charge of each object by summing $q(x)$ over each set of points having the same object number.
\end{enumerate}
This procedure is demonstrated visually in Fig.~\ref{fig:algorithmgraphic}.

\subsubsection{Peak identification} \label{subsec:peakidentification}
Treating each object as a localised region of dense topological charge density, $q(x)$, it follows that each will have some distribution which decays away from a peak value. This structure has been seen before in visualisations \cite{SU3SmoothingCalibration, QCDVis}, and allows one to identify the approximate centres of distinct objects through local extrema. To determine whether a point qualifies as a local maximum or minimum, we consider the $3^4$ hypercube centred at that point, which is to say one point in every direction (including diagonals).

\subsubsection{Allocating points} \label{subsec:growing}
The substance of our algorithm consists of iteratively allocating points to one of the objects. In a specified order (see below), all points $\{x\}$ currently assigned a given object number assign the same object number to neighbouring points $\{x'\}$ within the $3^4$ hypercube surrounding $x$ that pass the below constraints.
\begin{itemize}
    \item $x'$ has not previously been assigned an object number. This simply ensures we do not overwrite information already associated with another object.
    \item $q(x')$ has the same sign as $q(x)$. This requirement treats positively and negatively charged topological excitations as entirely distinct.
    \item $x'$ has a lower absolute topological charge density value. In essence, this condition implements a natural boundary between objects, keeping them contained to a region of the lattice within which the topological charge is solely decreasing. A two-dimensional illustration of this idea is shown in Fig.~\ref{fig:algorithm},
\begin{figure}
    \centering
    \includegraphics[width=0.96\linewidth]{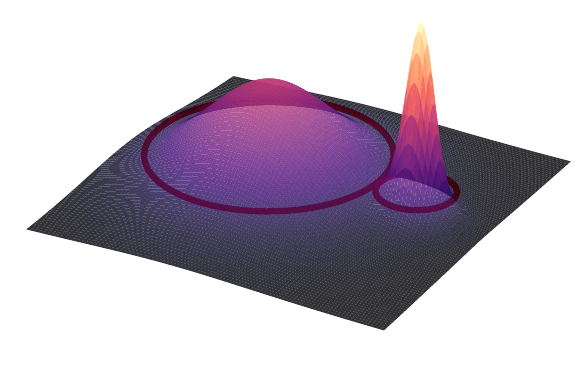}
    \caption{\label{fig:algorithm} A graphic illustrating the topological charge density as a function of two coordinates, highlighting a potential scenario our algorithm must deal with where two objects of differing sizes are located near each other. In order to stop the narrow object incorrectly obtaining points associated with the broader object, we enforce that each object can only grow downwards in topological charge value.}
\end{figure}
    a hypothetical scenario wherein a narrow object and a broader object are situated near each other. Each has a circle drawn near its base to show that one object is unable to grow within the circle associated with the other when it is restricted to grow downhill. In this way, we guarantee the sites assigned to each object reflect their distributions. This concept readily extends to four dimensions.
\end{itemize}

\subsubsection{Growing order} \label{subsec:groworder}
The order in which the objects assign their neighbouring points is observed to be important, encompassing ``in-between points" where it is ambiguous to which specific object they should be assigned. As mentioned in Step 2, our choice is to perform this process in order of ascending peak value. This is based on the below observations and assumptions:
\begin{itemize}
    \item Points farther away from lower-peaked objects still have a greater relative weight compared to sharply peaked objects.
    \item If two objects have \textit{similar} net charges (not necessarily identical), the lower-peaked one must have a broader distribution.
\end{itemize}
This order therefore introduces a bias towards smaller-peaked objects such that we conform to the above observations. In Sec.~\ref{subsec:testing}, we present test results for both our chosen ordering and the reverse order (decreasing peak value), emphasising the difference between the two extremes and demonstrating that our selection produces the more accurate results.

\subsubsection{Dislocation filtering} \label{subsec:filtering}
A nonzero lattice spacing $a$ gives rise to dislocations, fluctuations in the action and topological charge density on the scale of the lattice spacing. We stress that these are distinct from the ``genuine" topological features we are interested in, and thus desire a method to distinguish between the two to prevent our results from being skewed by the presence of lattice artefacts.

For this reason, we implement a cutoff such that any identified peak that fails to assign all points within a defined size is discarded. The objective is to filter out the dislocations with size $\sim\mathcal{O}(a)$. Accordingly, we investigate two different choices for the filter in terms of the lattice spacing:
\begin{enumerate}[label=\Alph*.]
    \item Nearest-neighbour filter: the peak must attain all points 1 unit away in each Cartesian direction (i.e. the peak must be resolved by the lattice spacing).
    \item Hypercube filter: the peak must attain \textit{all} neighbouring points, covering the full $3^4$ hypercube in addition to the points 1 unit away.
\end{enumerate}
Note that based on the conditions outlined in Step 2, a size cutoff enforces, as a minimum, that:
\begin{itemize}
    \item no two peaks overlap within the cutoff,
    \item the topological charge density $q(x)$ has the same sign at every point within the cutoff of each peak, and
    \item $|q(x)|$ exclusively decreases up to the cutoff away from each peak.
\end{itemize}
These constraints make intuitive sense on defining a topological object to have a minimum size. The remaining objects that survive the filter are then allowed to grow until no more points satisfy the criteria in Step 2.

A natural question that arises from introducing a scale-dependent filter is whether this dependence carries through to our final results. However, an appropriate cutoff will achieve the opposite, precisely because dislocations scale with the lattice spacing. Therefore, with no filter or an especially weak filter, our results would be sensitive to the size of the dislocations and would change if the lattice spacing is varied. Conversely, provided the filter is sufficiently strong to separate out a majority of the dislocations, the remaining topological features should be the same irrespective of the lattice spacing. Using this logic, we deduce that the simple nearest-neighbour filter is too weak, producing results that diminish with the lattice spacing. In contrast, the hypercube filter is found to ensure the desired scale independence. The details are provided in Sec.~\ref{sec:scaling}. Hence, we specifically mention in Step 4 to consider the hypercube for dislocation filtering, as an ``objective" choice of a correct filtering method.

\subsection{Classical Limit} \label{subsec:testing}
Having developed an algorithm, we next require to test it on a configuration with an expected outcome. To achieve this we employ gauge cooling \cite{CoolingI, CoolingII, CoolingIII}, which seeks to minimise the local action at each lattice site through a sequential update of the link variables $U_\mu(x)$. Zero-temperature gauge configurations under extended cooling are known to approach the classical limit consisting entirely of (anti-)instantons with integer topological charge \cite{CoolingII, CoolingIII}.

Following the procedure outlined in Appendix \ref{app:cooling}, 4000 sweeps of $\mathcal{O}(a^4)$-improved cooling is performed on five $32^3\times 64$ pure gauge configurations with $a=0.1$\,fm. The simulation details for these configurations are provided in Sec.~\ref{subsec:simdetails}. Their properties after cooling, including the action $S$, integrated topological charge $Q$ and number of identified objects are summarised in Table \ref{tab:cooledconfigs}.
\begin{table}[b]
\caption{\label{tab:cooledconfigs} $S/S_0$, $Q$ and the number of objects identified by our algorithm on five $32^3\times 64$ configurations, after 4000 sweeps of cooling. They are seen to be self-dual within an extremely small margin of error.}
\begin{ruledtabular}
\begin{tabular}{cD{.}{.}{2.4}D{.}{.}{2.4}D{.}{.}{2.0}}
Configuration number & \multicolumn{1}{c}{$S/S_0$} & \multicolumn{1}{c}{$Q$} & \multicolumn{1}{c}{No. objects} \\
\colrule
011 & 18.0004 & -18.0011 & 17 \\
015 & 8.0001 & 8.0002 & 7 \\
016 & 11.0004 & 11.0007 & 10 \\
017 & 17.0009 & 17.0007 & 16 \\
022 & 9.0015 & 9.0003 & 5 \\
\end{tabular}
\end{ruledtabular}
\end{table}
Based on the filtering detailed in Sec.~\ref{subsec:algorithm}, we find there are no dislocations present under extended cooling as defined by either filter. Therefore, the number of objects in this instance is precisely the number of local extrema.

The action is normalised by the single-instanton action $S_0 = 8\pi^2/g^2$ such that it can be directly compared to $Q$. Our observation is that the net topological charge consistently converges to within 1\% of an integer in $<75$ sweeps through the lattice and remains stable thereafter for the duration of the cooling, which is in agreement with previous work \cite{ImprovedFmunu}. The large number of cooling sweeps is to ensure the configurations satisfy self-duality, which is seen to be reached with the values of $S/S_0$ and $|Q|$ for each configuration agreeing to at least one part in one thousand. The number of extrema $N$ tends to be less than $S/S_0$ in each case, though the fact they are nevertheless similar indicates these features are instanton-like. Analytic self-dual solutions to the Yang-Mills equations are known to exist for arbitrary topological index \cite{Multi-instantonSolutionI, Multi-instantonSolutionII}, which we refer to as ``multi-instanton" objects. Additionally, the profile of overlapping instantons is known to result in ``hiding" instantons in extreme circumstances \cite{OverlappingInstantonsI, OverlappingInstantonsII}. Both of these could be the source of the slight discrepancy between $N$ and $S/S_0$.

\begin{figure}
    \centering
    \includegraphics[width=\linewidth]{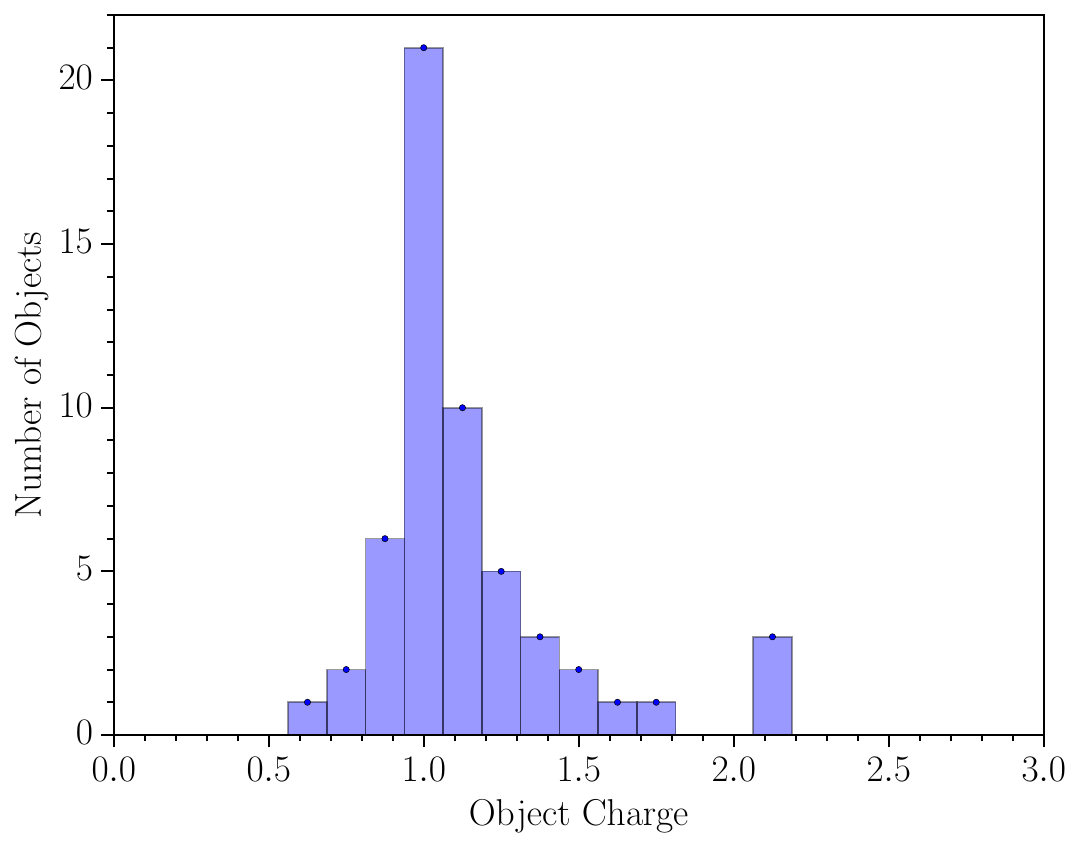}
    \includegraphics[width=\linewidth]{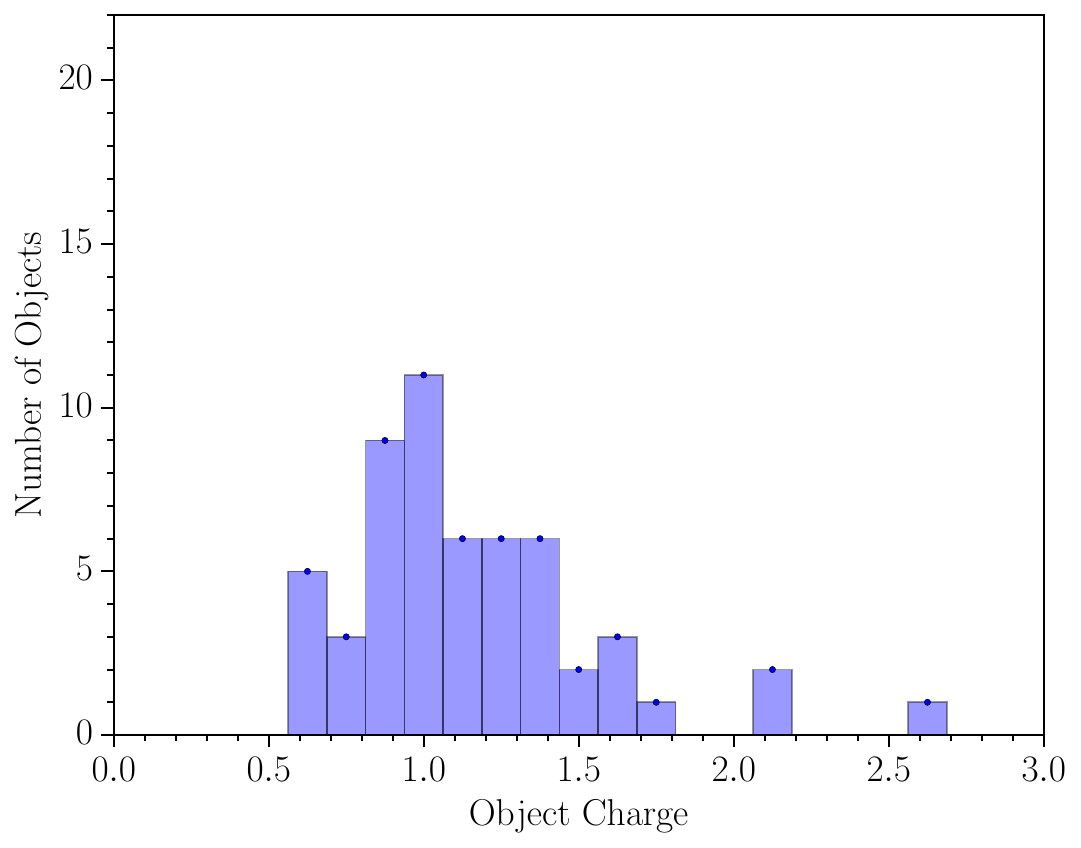}
    \caption{\label{fig:testresults} Histograms showing the results of our algorithm to five $32^3\times 64$ configurations after 4000 sweeps of $\mathcal{O}(a^4)$-improved cooling, for growing in order of ascending peak value (\textbf{top}) and the reverse order (\textbf{bottom}). Our choice of ascending peak order justified in Sec.~\ref{subsec:algorithm} is seen to produce substantially more accurate results, with sharper peaks in the histogram near integer values and a visibly smaller spread.}
\end{figure}

The collated results of applying the algorithm to the cooled configurations, for both ascending and descending peak order, are shown in Fig.~\ref{fig:testresults}.
In these histograms, the horizontal axis shows the absolute value of the charges we calculate for each identified object, whilst the vertical axis gives the number of identified objects for a given interval of values.

\begin{figure*}
    \centering
    \includegraphics[width=0.39\linewidth]{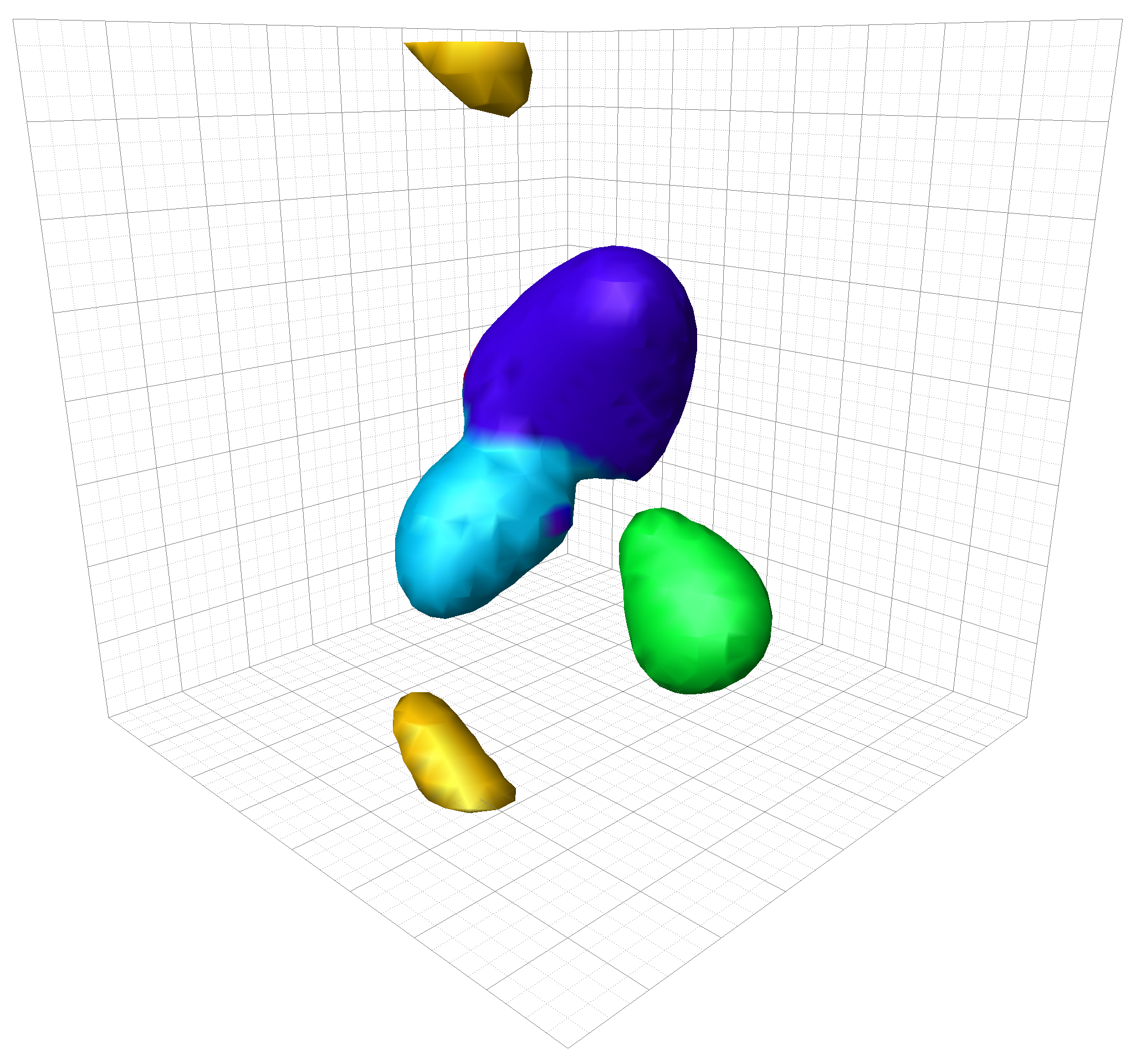}
    \hspace{3em}
    \includegraphics[width=0.39\linewidth]{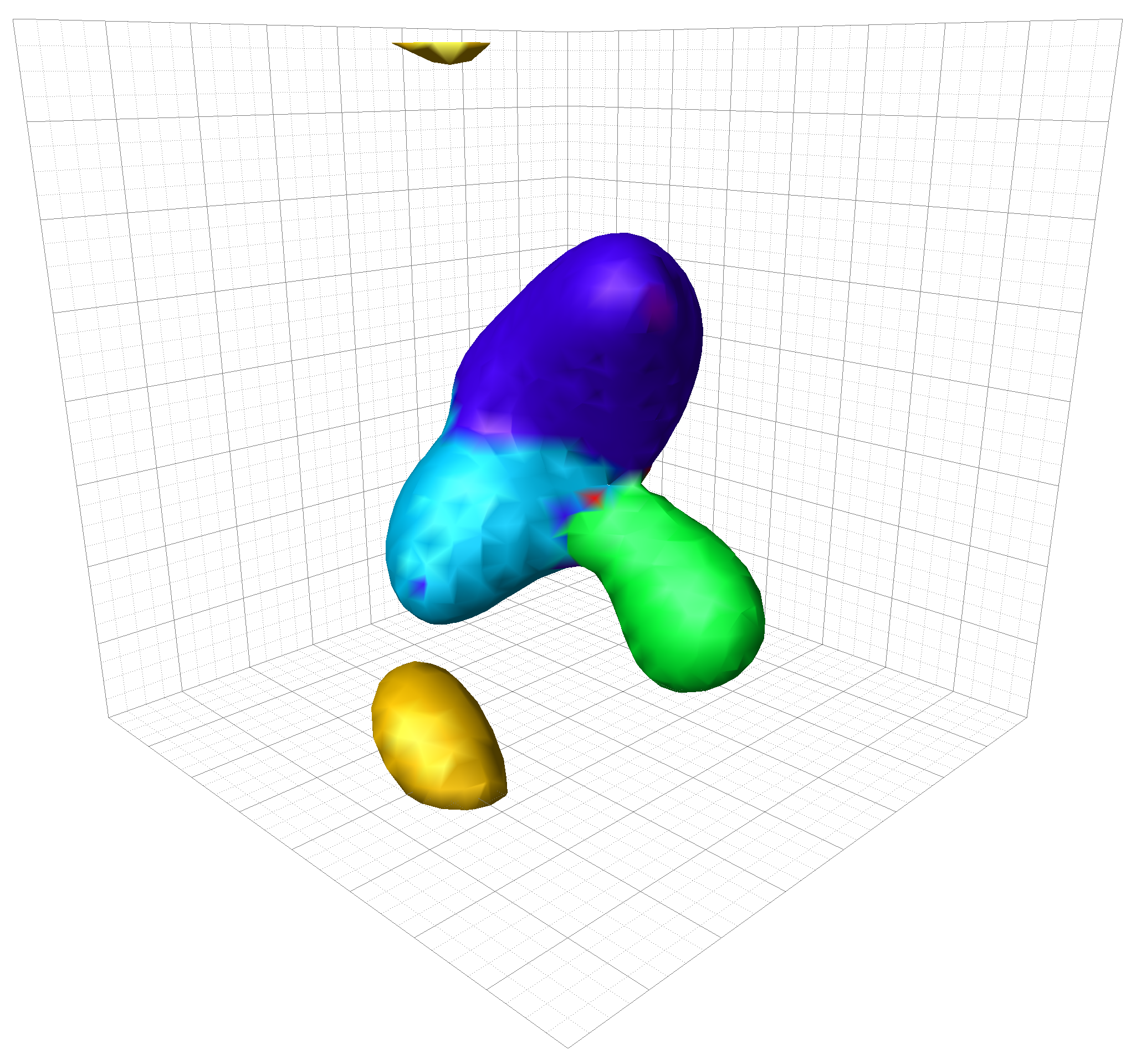}
    \includegraphics[width=0.39\linewidth]{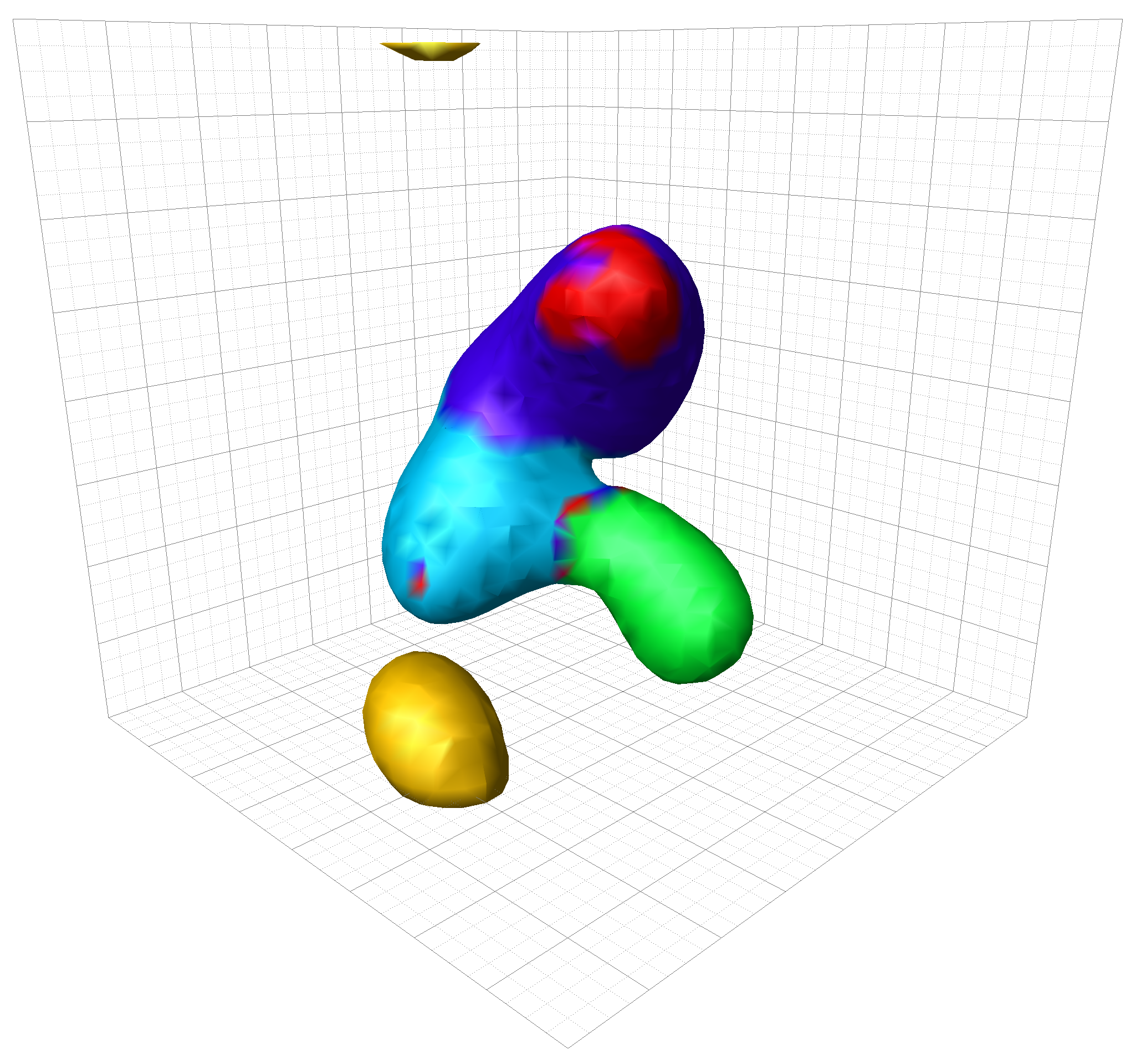}
    \hspace{3em}
    \includegraphics[width=0.39\linewidth]{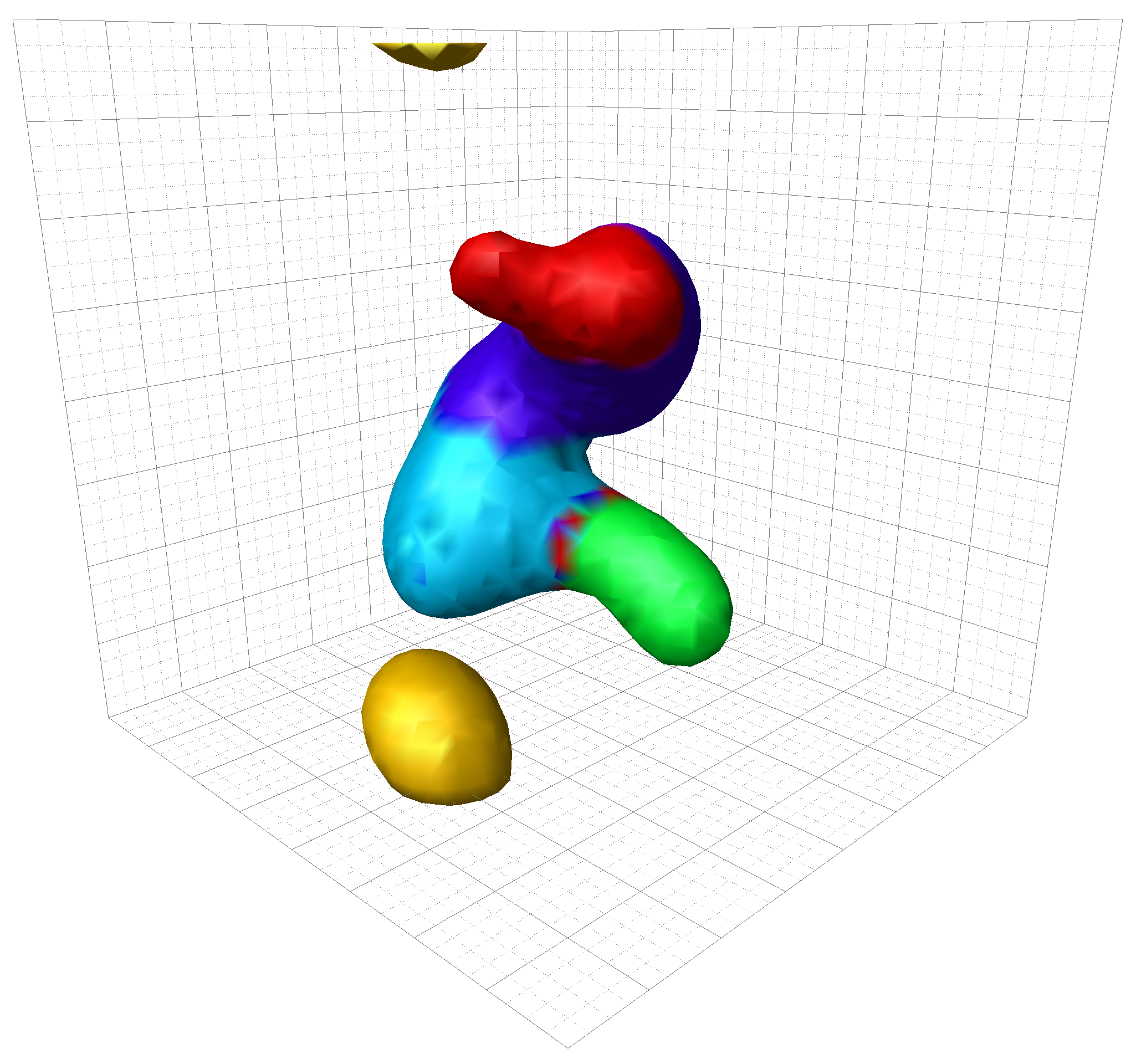}
    \includegraphics[width=0.39\linewidth]{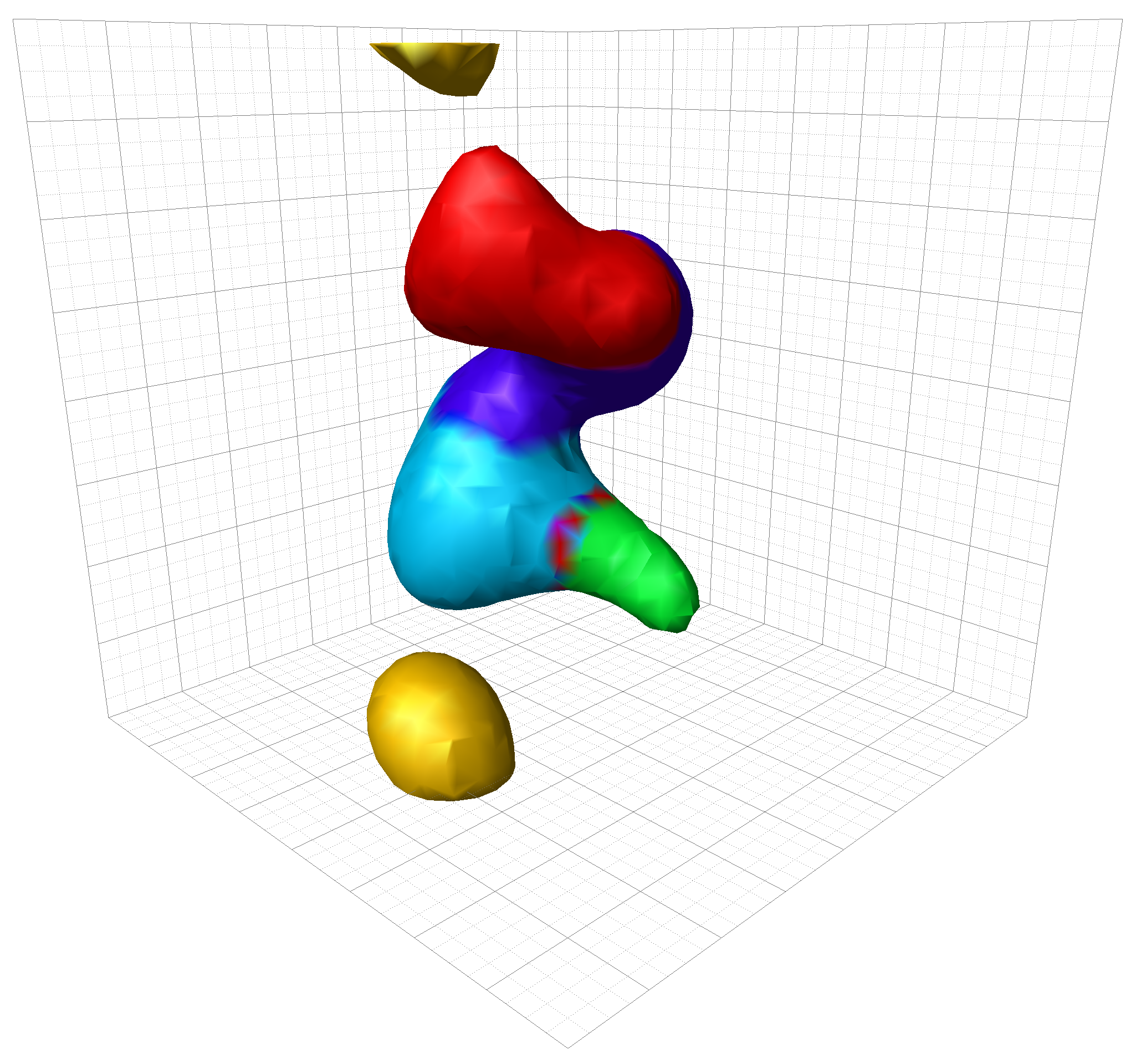}
    \hspace{3em}
    \includegraphics[width=0.39\linewidth]{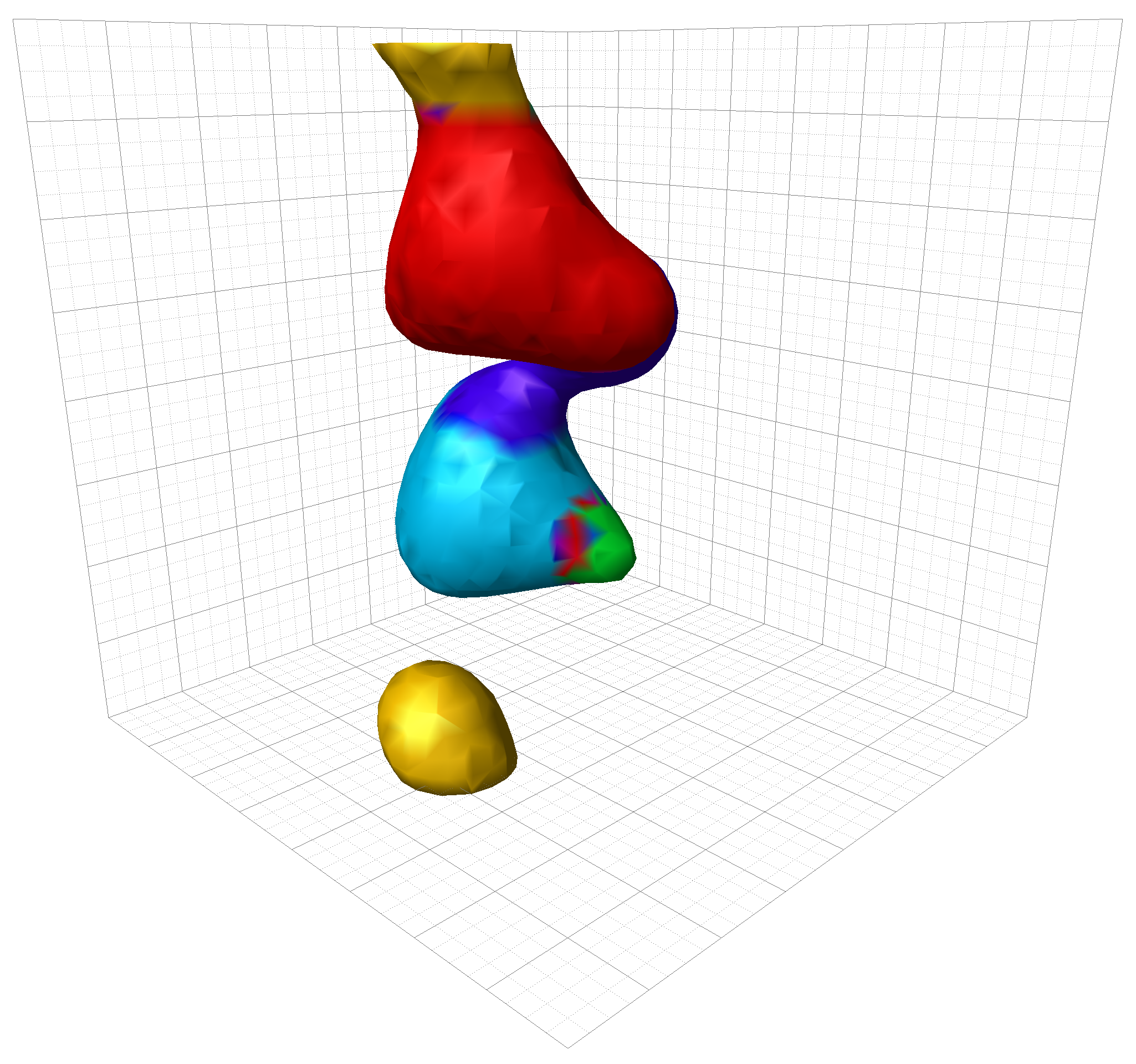}
    \caption{\label{fig:classicalvis} Visualisations of the topological charge density in the self-dual limit obtained by slicing along the temporal dimension and visualising the remaining three-dimensional spatial structure. These are six consecutive frames, displayed from left to right then top to bottom. The topological charge density is coloured (shaded) according to the object number in the algorithm, allowing insight into how it divides topological objects in four dimensions. We only visualise the topological charge density above some minimum threshold value to observe the behaviour of the algorithm on the most significant topological charge density; this also enables one to see into the three-dimensional space. The overlapping nature of the instantons, even in this ``classical limit", is apparent. The challenging four-dimensional nature of the problem is also revealed. For instance, the red object grows on top of the purple object as we advance in the temporal dimension, and subsequently merges with the yellow object. The algorithm is seen to implement an effective and reasonable boundary between each of the topological features.}
\end{figure*}

As mentioned, even this self-dual limit is not comprised of ideal well-separated instantons, and this has ramifications for accurately capturing the single-instanton properties. To highlight this complication, we visualise the topological charge density on a cooled self-dual configuration in Fig.~\ref{fig:classicalvis}.
The overlapping nature of individual topological features is clear, and by colouring the topological charge based on object number we can observe the effective boundary created by the algorithm between these objects. Consequently, we expect to find a distribution of charge values, representing fluctuations around the exact $|Q|=1$ solution that arise from overlapping distributions.

For growing in ascending peak order, the majority of charge values we calculate lie very near $|Q|=1$, with a small spread of values slightly farther away. The same pattern is seen for the reverse ordering although to a visibly lesser extent, corroborating our choice. On occasion, integer values of larger magnitude are observed, as several objects have a calculated charge near $|Q|=2$. As previously surmised, these could arise due to extreme cases of overlapping instantons which only have a single peak in their topological charge density, or signal the presence of multi-instanton objects after extended cooling.

These results provide strong evidence that we can take the mode of the distribution for a general configuration as a reliable indicator for the topological charge values which tend to comprise the gluon-field objects. Their distribution reflects both quantum fluctuations around those solutions and inherent uncertainties in assigning topological charge density to objects in four dimensions. The upshot is we have developed a means by which we can explore the nature of objects in ground-state fields and their evolution with increasing temperature.

\section{Smoothing} \label{sec:smoothing}
It is well known that lattice operators for the action and topological charge densities encounter renormalisation factors that differ significantly from 1 and are poorly controlled. For example, different $\mathcal{O}(a^4)$-improved lattice operators will produce significantly different topological charge densities, which is elaborated on in Sec.~\ref{subsec:topqcomparison}. To ensure reliable results, we accordingly seek to minimise these discretisation effects through the application of smoothing.

In addition to cooling, there are various smoothing algorithms in common use, including APE smearing \cite{APEI, APEII}, stout-link smearing \cite{Stout-linkSmearing}, gradient flow \cite{GradientFlowI, GradientFlowII} and their over-improved variants \cite{Over-improvedCooling, Over-improvedStout-linkSmearing}. Over-improvement proceeds by defining a one-parameter family of actions $S(\varepsilon)$, with $\varepsilon$ tuned to preserve instantons in the smoothing process with a size above some minimum dislocation threshold $\rho \geq \rho_0$. We specifically avoid such methods in our work so as to not bias our results towards a particular topological configuration. Instead, our dislocation filtering additionally serves to remove any topological features that shrink below the lattice spacing during the smoothing process. For this reason, the size cutoff remains important regardless of the level of smoothing.

At this stage we are no longer interested in the classical limit, and instead desire to minimise the amount of smoothing required to accurately resolve the topological objects of a typical vacuum configuration. Cooling is unsuitable for such a gradual process, and thus we implement gradient flow, described below.

\subsection{Gradient flow} \label{subsec:gradientflow}
The evolution of the link variables $U_\mu(x)$ under gradient flow is defined by the differential equation \cite{GradientFlowI}
\begin{align} \label{eq:gradientflow}
    \frac{d}{d\tau}U_\mu(x,\tau) &= iQ_\mu(x) \,U_\mu(x,\tau)\,, \\
    U_\mu(x,0) &= U_\mu(x)\,,
\end{align}
for dimensionless ``flow time" $\tau$, and where $Q_\mu(x)\in\mathfrak{su}(3)$ generates the infinitesimal field transformation $U\longrightarrow U + i\epsilon \,Q(U) \,U$. An explicit choice for $Q_\mu(x)$ is given in terms of the staple sum,
\begin{align} \label{eq:Qmu}
    \begin{split}
        Q_\mu(x) &= \frac{i}{2}\big(\Omega_\mu(x) - \Omega_\mu^\dagger(x) \big) \\
        &\hphantom{=} - \frac{i}{6}\tr\big(\Omega_\mu(x) - \Omega_\mu^\dagger(x)\big) \,,
    \end{split} \\
    \Omega_\mu(x) &= U_\mu(x) \,\sum_{\nu\neq\mu} \Sigma_{\mu\nu}(x) \,,
\end{align}
where $\Sigma_{\mu\nu}(x)$ is the staple product of links connecting $U_\mu(x)$ in the $\mu$-$\nu$ plane. $Q_\mu(x)$ is seen to be traceless Hermitian by construction.

In the interest of preserving locality during the smoothing process, we calculate a staple sum which includes the contributions of the plaquette, $1\times 2$ and $2\times 1$ rectangles. This is given pictorially by
\begin{equation} \label{eq:Oa2Staples}
    \begin{split}
        \Sigma_{\mu\nu} &= \frac{5}{3} \left(\;\, \plaq \,\;\right) - \frac{1}{12} \left(\;\,\upperonebytworect \; + \; \loweronebytworect \,\;\right) \\
        &\hphantom{=} - \frac{1}{12} \left(\;\,\forwardtwobyonerects \;\; + \;\; \backwardtwobyonerects \,\;\right) \,,
    \end{split}
\end{equation}
where the coefficients correspond to a standard Symanzik $\mathcal{O}(a^2)$-improved lattice action. Defining $P_{\mu\nu}^{(m\times n)}(x)\equiv \frac{1}{3}\Re\tr\,(m\times n$ Wilson loop at $x)$, this is
\begin{equation} \label{eq:symanzik}
\begin{split}
    S &= \beta \sum_{x,\,\mu>\nu} \bigg[\frac{5}{3}\big(1 - P_{\mu\nu}^{(1\times 1)}(x)\big) \\
    &\hphantom{=}- \frac{1}{12}\big(1-P_{\mu\nu}^{(2\times 1)}(x)\big) - \frac{1}{12}\big(1-P_{\mu\nu}^{(1\times 2)}(x)\big) \bigg] \,.
\end{split}
\end{equation}

Numerical integration of the Wilson flow is performed using an Euler integration scheme in which the link variables are updated successively in time steps of $\epsilon$ via
\begin{align} \label{eq:eulerintegration}
    U_\mu(x,\tau+\epsilon) = \exp\big(i\epsilon \,Q_\mu(x)\big) \,U_\mu(x,\tau)\,.
\end{align}
One can see that with the given choice of $Q_\mu(x)$, gradient flow corresponds to an annealed version of stout-link smearing with an extremely small isotropic smearing parameter. Indeed, for sufficiently small $\epsilon$ the finite transformation generated by Euler integration of the gradient flow is equivalent to stout-smeared links \cite{GradientFlowI, GradientFlowII, Stout-linkSmearing, GradientFlowCoolingComparison, GradientFlowSmoothingConnectionI, GradientFlowSmoothingConnectionII}. Previous work has shown that taking $\epsilon \lesssim 0.02$ is sufficiently small to accurately solve the differential equation and ensure the independence of $\epsilon$ \cite{GradientFlowCoolingComparison, GradientFlowSmoothingConnectionII}. That said, we desire to perform enough smoothing to guarantee discretisation errors are negligible, whilst at the same time retaining as many genuine topological features as possible. Based on this, we choose a comparatively small value of $\epsilon = 0.005$. This provides a greater degree of control over the level of smoothing, which is highly beneficial to our cause.

In addition to removing UV fluctuations, the gradient flow is understood to distort the distribution of topological objects such as through instanton and anti-instanton pair annihilation. One might be concerned about the potential effect this has on our results. To understand this, we draw on previous work comparing the effects of smoothing on the gluonic definition of the topological charge density to that obtained via the overlap Dirac operator with an ultraviolet cutoff, $\lambda_\mathrm{cut}$ \cite{Eigenmodes}. At low levels of stout-link smearing, the structure of the gluonic density is found to strongly coincide with the overlap definition for a specific $\lambda_\mathrm{cut}$. Given that the UV-filtered overlap topological charge density has no distortion effects, we can be assured, at the comparable amount of gradient flow performed in this work, that this issue is negligible.

An intuitive picture for this is realised by recalling that at leading order, the gradient flow corresponds to a simple convolution of the gauge field with a Gaussian of RMS radius $\sqrt{8\tau}$ \cite{GradientFlowII}. This obviously results in a smoothing effect at short flow times $\tau$. It takes extended cooling for instanton and anti-instanton pairs to walk across the lattice and begin to annihilate with each other, as revealed through visualisations \cite{QCDVis}.

\subsection{Comparison of improvement schemes} \label{subsec:topqcomparison}
The suppression of action and UV fluctuations induced by gradient flow can cause substantial changes to the topological charge density over a relatively small number of updates. This makes selecting the flow time at which to analyse the topological charge a nontrivial matter. Some inroads have formerly been made towards solving this problem by comparing different action lattice operators. Besides the ``standard" action, such as Eq.~(\ref{eq:symanzik}), one can define an alternate ``reconstructed action" via a lattice field strength tensor which can be substituted directly into the definition of the continuum action \cite{ImprovedFmunu},
\begin{equation} \label{eq:reconaction}
    S_R = \frac{1}{2} \sum_{x} \tr\big(F_{\mu\nu}(x) F_{\mu\nu}(x) \big) \,.
\end{equation}
These discretisations will experience different renormalisation effects, allowing the two actions to be compared when smoothing to gauge the size of the remaining discretisation artefacts. Nevertheless, it is still unclear exactly how similar the operators should be before one can be confident errors have been amply suppressed. Furthermore, it is not obvious whether comparing two discretisations of the \textit{action} translates directly to the topological charge density, which is our primary interest.

Still, motivated by this idea we instead utilise two different $\mathcal{O}(a^4)$-improved topological charge operators to assess the magnitude of the discretisation errors. The first of these is an ``Improved $F_{\mu\nu}$" scheme calculated by substituting an $\mathcal{O}(a^4)$-improved field strength tensor into the definition of the topological charge,
\begin{equation}
    q(x) = \frac{g^2}{32\pi^2} \varepsilon_{\mu\nu\rho\sigma} \tr\big(F_{\mu\nu}(x) F_{\rho\sigma}(x) \big) \,.
\end{equation}
The improvement of $F_{\mu\nu}$ proceeds as follows \cite{ImprovedFmunu}. First, the $m\times n$ clover term $\mathcal{C}_{\mu\nu}^{(m\times n)}(x)$ is defined as the sum of the $m\times n + n\times m$ Wilson loops in the $\mu$-$\nu$ plane touching the point $x$, depicted as
\begin{equation}
    \mathcal{C}_{\mu\nu}^{(m\times n)} = \;\mbynclover \;\,+\,\; \nbymclover \;\;.
\end{equation}
Each clover term gives an estimate of the field strength tensor as
\begin{equation} \label{eq:cloverfmunu}
    F_{\mu\nu}^{(m\times n)}(x) = \frac{1}{8} \Im\big(\mathcal{C}_{\mu\nu}^{(m\times n)}(x) \big) \,.
\end{equation}
The improved operator is then constructed by an appropriate linear combination of the above terms,
\begin{equation}
    ga^2F_{\mu\nu}(x) = \sum_{m,n} k^{(m\times n)} F_{\mu\nu}^{(m\times n)}(x) \,.
\end{equation}
To eliminate the $\mathcal{O}(a^2)$ and $\mathcal{O}(a^4)$ errors, it is sufficient to consider $(m,n)=(1,1)$, $(2,2)$, $(1,2)$, $(1,3)$, and $(3,3)$. The desired coefficients are
\begin{equation}
\begin{aligned}
    k^{(1\times 1)} &= \frac{19}{9} - 55\,k^{(3\times 3)} \,, \\
    k^{(2\times 2)} &= \frac{1}{36} - 16\,k^{(3\times 3)} \,, \\
    k^{(1\times 2)} &= -\frac{32}{45} + 64\,k^{(3\times 3)} \,, \\
    k^{(1\times 3)} &= \frac{1}{15} - 6\,k^{(3\times 3)} \,,
\end{aligned}
\end{equation}
where $k^{(3\times 3)}$, the coefficient of the $3\times 3$ clover term, is a free parameter.

Alternatively, one can proceed by defining a series of discretised topological charge operators $q^{(m\times n)}(x)$ for each clover term \cite{ImprovedActionTopQ},
\begin{equation}
    q^{(m\times n)} = \frac{1}{32\pi^2} \frac{1}{m^2 n^2} \varepsilon_{\mu\nu\rho\sigma} \tr\big( F_{\mu\nu}^{(m\times n)} F_{\rho\sigma}^{(m\times n)} \big) \,.
\end{equation}
Here, $F_{\mu\nu}^{(m\times n)}(x)$ is as defined in Eq.~(\ref{eq:cloverfmunu}), and the factor of $1/(m^2n^2)$ is included in the definition for convenience. These terms can subsequently be combined to produce a different improved topological charge operator,
\begin{equation}
    q(x) = \sum_{m,n} c^{(m\times n)} q^{(m\times n)}(x) \,.
\end{equation}
We refer to this as the ``Improved TopQ" scheme. Since $q(x)$ is nonlinear in $F_{\mu\nu}$, an $\mathcal{O}(a^4)$-improved operator via this method will have different renormalisation effects than the Improved $F_{\mu\nu}$ scheme. The same five clover terms can be used to eliminate the $\mathcal{O}(a^2)$ and $\mathcal{O}(a^4)$ corrections, with the coefficients in this case turning out as
\begin{equation}
\begin{aligned}
    c^{(1\times 1)} &= \frac{1}{9}\big(19 - 55\,c^{(3\times 3)}\big) \,, \\
    c^{(2\times 2)} &= \frac{1}{9}\big(1 - 64\,c^{(3\times 3)}\big) \,, \\
    c^{(1\times 2)} &= \frac{1}{45}\big(\!-\!64 + 640\,c^{(3\times 3)}\big) \,, \\
    c^{(1\times 3)} &= \frac{1}{5} - 2\,c^{(3\times 3)} \,.
\end{aligned}
\end{equation}
In fact, these coefficients are identical to those used in constructing an improved action from the same five planar $m\times n$ loops.

To compare the two improvement schemes, we consider three-loop and five-loop versions of both operators. The three-loop variants are unique, obtained by setting $k^{(3\times 3)}=1/90$ for Improved $F_{\mu\nu}$ and $c^{(3\times 3)}=1/10$ for Improved TopQ. We examine five-loop versions corresponding to $k^{(3\times 3)}=1/180$ \cite{ImprovedFmunu} and $c^{(3\times 3)}=1/20$ \cite{5-loopImprovedCooling, ImprovedActionTopQ}.

\begin{figure}
    \centering
    \includegraphics[width=\linewidth]{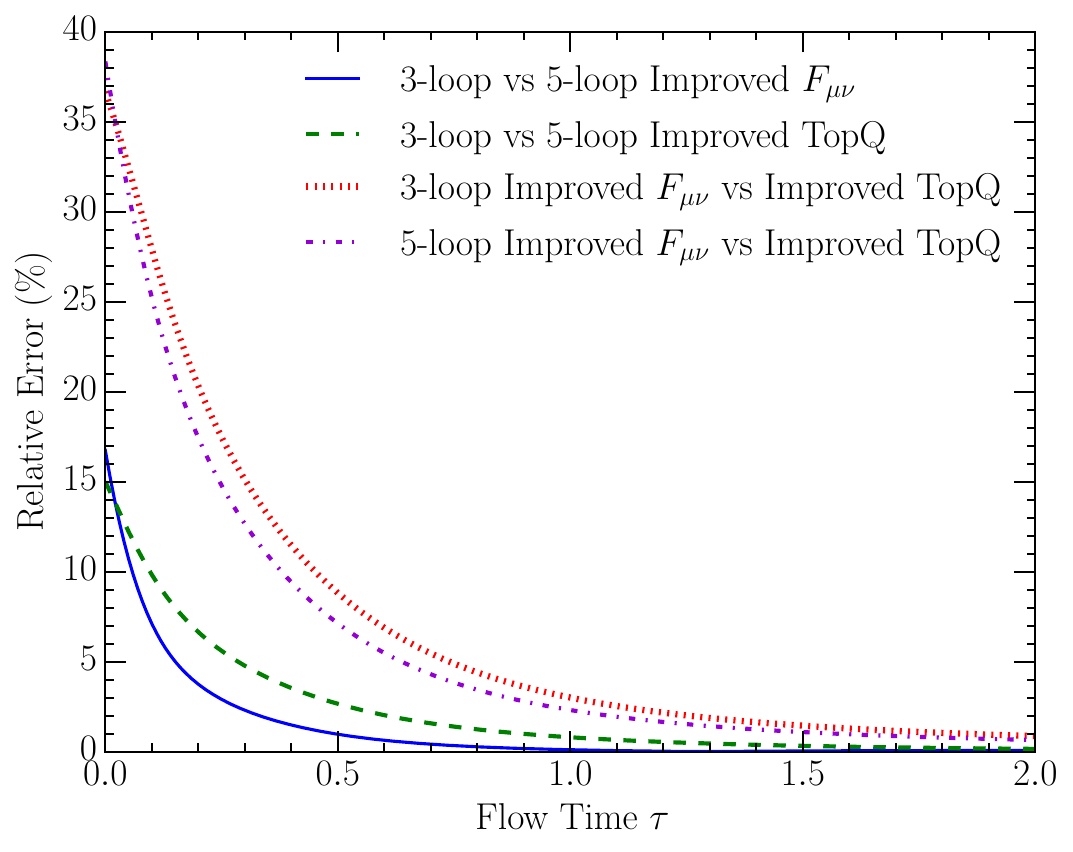}
    \caption{\label{fig:topqcomparison} The evolution of the relative error between the summed absolute topological charge density for several different $\mathcal{O}(a^4)$-improved lattice topological charge operators under gradient flow. The difference produced by the two improvement schemes is much greater than varying the number of loops within the same improvement scheme.}
\end{figure}

With these choices, we now present comparisons between various improved topological charge operators:
\begin{itemize}
    \item three-loop vs five-loop Improved $F_{\mu\nu}$,
    \item three-loop vs five-loop Improved TopQ,
    \item three-loop Improved $F_{\mu\nu}$ vs Improved TopQ, and
    \item five-loop Improved $F_{\mu\nu}$ vs Improved TopQ.
\end{itemize}
For each possibility we sum the absolute value of the topological charge $Q=\sum |q(x)|$, and compute the below ``pseudo-" relative error between the two forms in question:
\begin{equation}
    \text{RE} = \frac{|Q_1 - Q_2|}{\frac{1}{2}(Q_1 + Q_2)} \,.
\end{equation}
We normalise the difference by the average of the two values to provide a common base for comparison. Figure \ref{fig:topqcomparison}
shows the evolution of the relative error for each of the listed comparisons over a flow time $\tau=2$. From this, we conclude there is a greater disparity in the operators defined in the contrasting improvement schemes as opposed to using a different combination of loops within one of the improvement schemes. Hence, this allows us to proceed referring exclusively to the three-loop combination for the two improvement schemes as the most reliable indicator for when discretisation effects have been suppressed.

\subsection{Selecting the optimal smoothing level} \label{subsec:optimalsmoothing}
Even though we now have a technique for analysing the discretisation errors in the topological charge density, this is yet to single out the precise ideal smoothing level. A natural solution to this problem is provided by our methods. If we apply our algorithm in Sec.~\ref{subsec:algorithm} to the topological charge densities obtained through both improvement schemes, then provided renormalisation factors are significant the net charges obtained will in general be different. This emerges from three compounding effects: differences in the topological charge density value at lattice points we identify as part of topological objects, inherent distinctions in the number and locations of local extrema, and variations in how the lattice points are distributed between the objects.

\begin{figure}
    \centering
    \includegraphics[width=\linewidth]{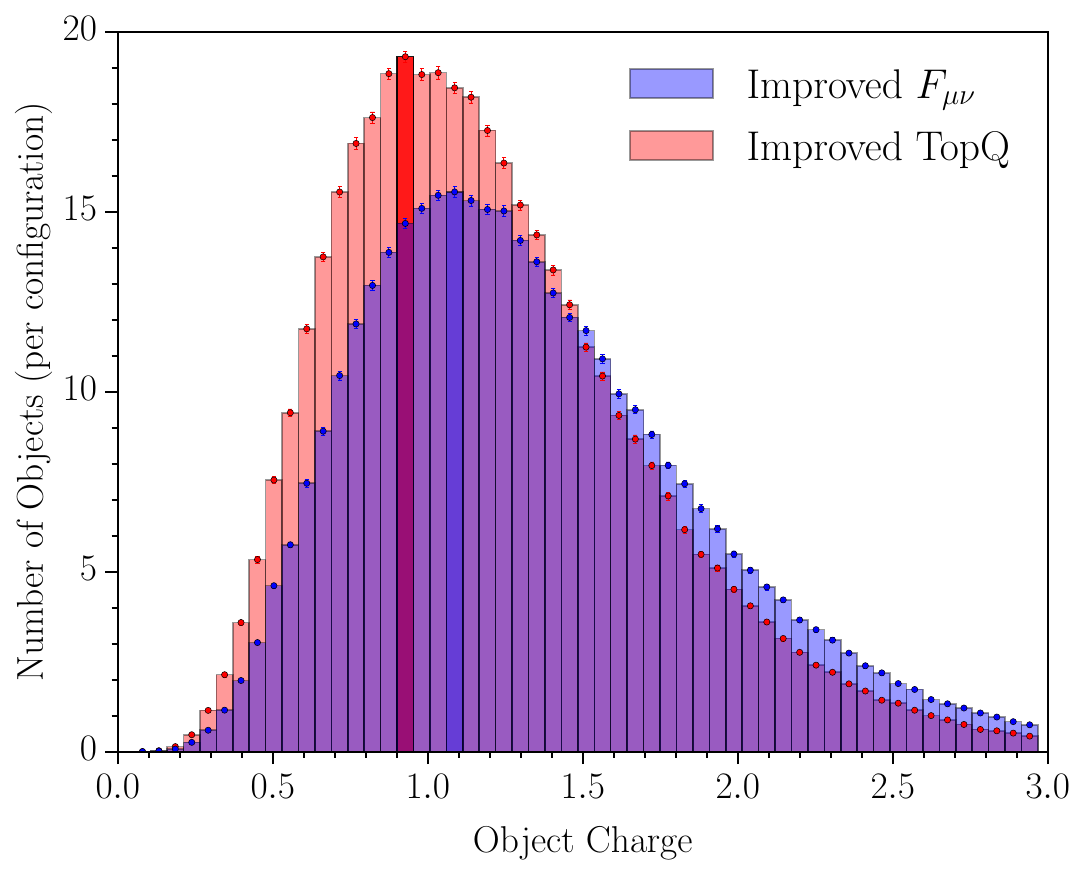}
    \caption{\label{fig:s32t64 n0050} Histograms of the net object charges obtained with the hypercube dislocation filter for both $\mathcal{O}(a^4)$-improved topological charge operators. This example is taken from our $32^3\times 64$ ensemble at $\tau=1$. The modes are marked by a darker colour and are visibly shifted from each other by $\approx 0.15$, which is certainly not an insignificant difference. This implies the level of smoothing is insufficient.}
\end{figure}

As a consequence, the histograms of the charge assigned to each object reveal divergent modes, inhibiting our ability to draw the same conclusion from both improved topological charge operators. An example of this is presented in Fig.~\ref{fig:s32t64 n0050}.
The distribution produced by the Improved TopQ scheme is visibly shifted to the left from the Improved $F_{\mu\nu}$ scheme, with the modes clearly incompatible at this smoothing level. Recall that the modes (and fluctuations thereabout) provide an indicator of the underlying topological structure. It follows that the conclusions we would infer on the net charge of distinct topological objects would differ from each other.

This motivates our criterion for the optimal smoothing level as the minimum number of updates required for the modes of the two histograms to agree. This is a necessary condition to ensure that discretisation errors have been adequately suppressed, with the two improvement schemes providing consistent conclusions. At a foundational level, this requirement is justified from both improvement definitions being valid ways to calculate the topological charge density on the lattice. Therefore, either should be able to be used to the same effect. At the same time, we do not desire to perform any more smoothing than what is necessary as it risks distorting or destroying genuine topological features. Throughout our results in Sec.~\ref{sec:results} we will continue to display both histograms to emphasise that this criterion has been satisfied and to illustrate any remaining systematic uncertainties.

Before proceeding, we establish that this criterion is allowed to depend on both the lattice spacing and the filter used. The former of these will be considered in greater detail in the next section. The latter is because the size and number of lattice sites associated with the topological objects differs greatly between the two dislocation cutoffs under investigation. This could induce a greater discrepancy between the improved operators when the hypercube filter is applied compared to the nearest-neighbour filter. Nevertheless, as discussed at the end of Sec.~\ref{subsec:filtering} and detailed in the following section, our best results correspond to the hypercube dislocation criterion.

\section{Continuum Limit} \label{sec:scaling}
Before proceeding to present our finite-temperature findings, it is important to establish that our results scale properly in the continuum limit; that is, they are independent of the lattice spacing $a$. To achieve this, we utilise two ensembles with equal physical volumes: a $32^3\times 64$ ensemble with $a\approx 0.10\,$fm, and a $48^3\times 96$ ensemble with $a\approx 0.067\,$fm. For simplicity, we refer to these as the ``coarse" and ``fine" ensembles respectively throughout this section.

We consider two different possibilities for taking the continuum limit. The first of these is a ``fixed lattice dislocation filter" method in which the dislocation filter is applied identically on both ensembles. This allows for \textit{physically} smaller topological objects to be considered as $a\to 0$. The second is a ``fixed scale" method, for which care is taken to fix the physical scale for resolving the topological objects as the continuum limit is approached.

\subsection{Fixed lattice dislocation filter} \label{subsec:fixedcutoff}
In this approach, we apply a consistent dislocation filter across the two ensembles. In Sec.~\ref{subsec:filtering}, the intention to investigate ``nearest-neighbour" and ``hypercube" filters was discussed. These are both expressed in terms of the lattice spacing, and thus have the prospect to admit physically smaller topological features on the finer lattice. This allows us to take advantage of the improved resolution provided by the finer lattice spacing to probe vacuum structure at a smaller scale. We are interested in determining whether the charge contained within each such objects nonetheless remains invariant, for instance because their topological charge profiles are sharper. In this case, it is crucial to allow for the possibility that less smoothing is required on the finer lattice (as per the criterion from Sec.~\ref{subsec:optimalsmoothing}).

\begin{figure}
	\centering
	\includegraphics[width=\linewidth]{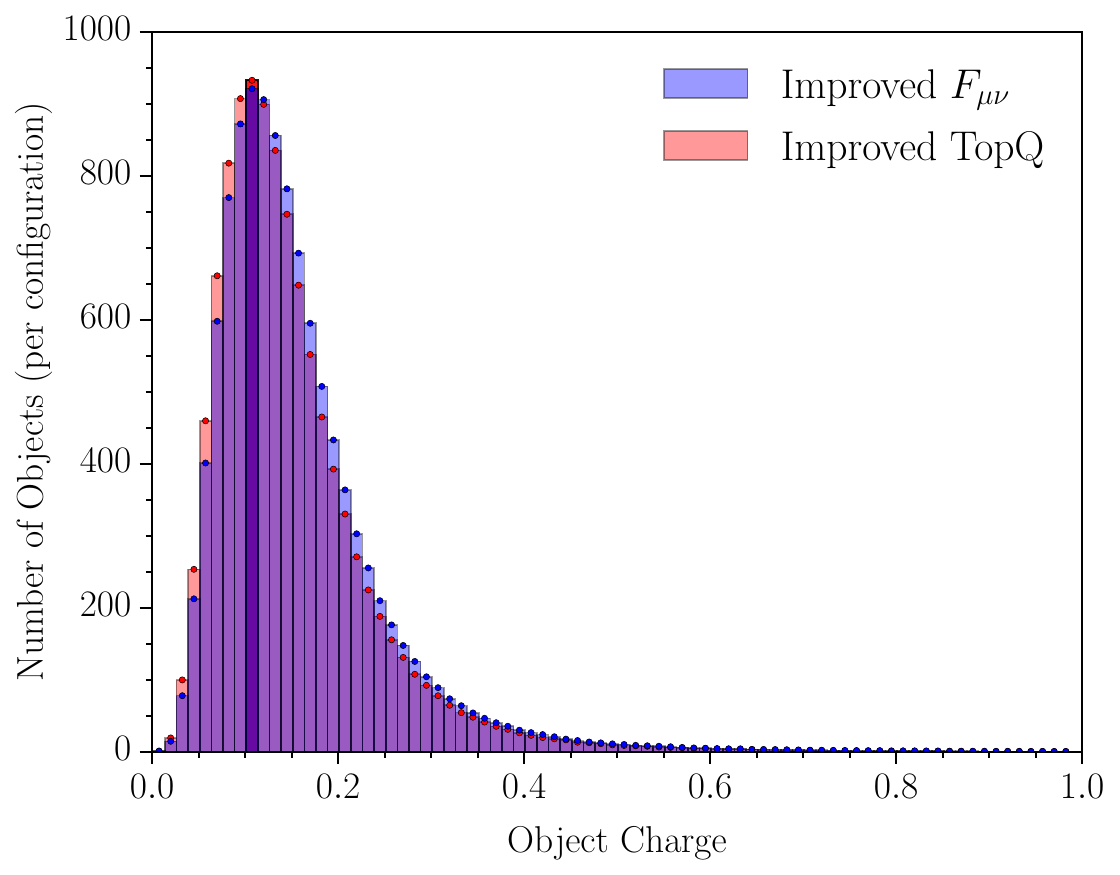}
	\includegraphics[width=\linewidth]{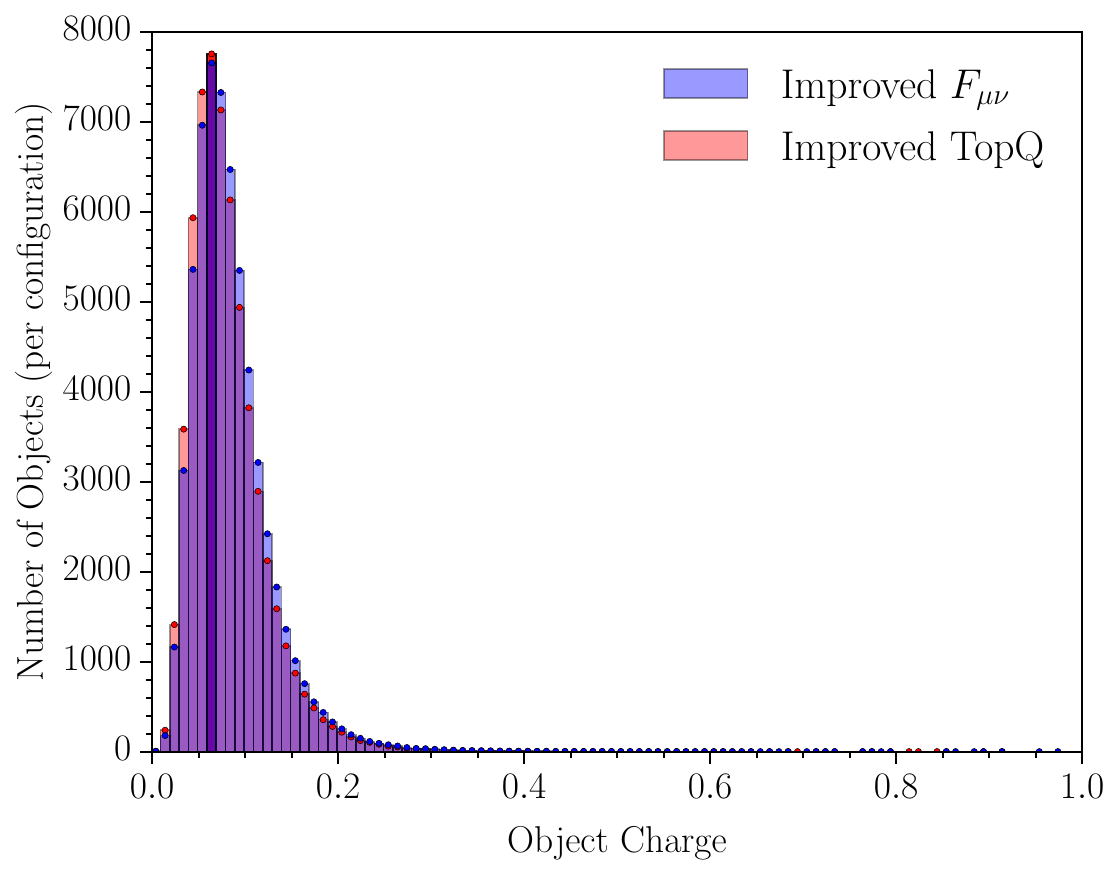}
	\caption{\label{fig:1unit} The results of our algorithm with the nearest-neighbour dislocation filter applied to the coarse (\textbf{top}) and fine (\textbf{bottom}) ensembles. The mode for the former is $0.11$, but this shifts to $0.064$ for the latter, suggesting that a nearest-neighbour cutoff is insufficient to ensure scale independence.}
\end{figure}

We start by comparing the results obtained with the nearest-neighbour filter between the two ensembles. The histograms are shown in Fig.~\ref{fig:1unit}.
We can see that both distributions have been sufficiently smoothed such that the two topological charge definitions produce consistent results. For the coarse ensemble, this is achieved after a flow time $\tau=0.625$, whilst for the fine ensemble it is slightly less at $\tau=0.525$. This is the expected outcome.

The modes occur at $|Q| = 0.111(6)$ and $|Q| = 0.064(5)$, which are inconsistent with each other. This embodies a considerable relative difference, with an $\approx 40\%$ decrease in value moving to the finer lattice. Given the precision with which the modes have been resolved, we can be assured this is a statistically significant discrepancy arising from the smaller lattice spacing. Indeed, one can observe the extent to which the histogram for the fine ensemble has fallen off from the mode by $|Q|\approx 0.11$. From this, we deduce that the topological structure revealed by the simple nearest-neighbour cutoff is scale dependent. This provides evidence that this filter is insufficient to minimise the effects of dislocations on the lattice, with the results being sensitive to their size $\sim\mathcal{O}(a)$.

\begin{figure}
	\centering
	\includegraphics[width=\linewidth]{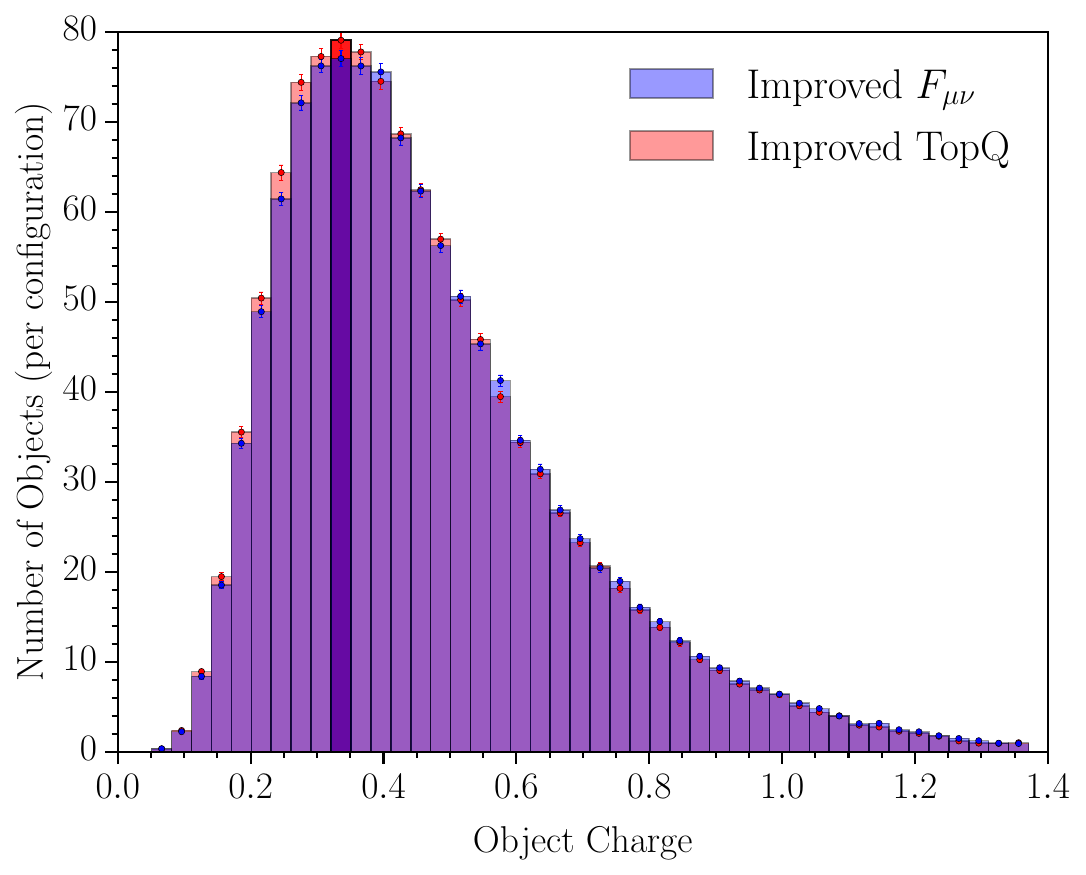}
	\includegraphics[width=\linewidth]{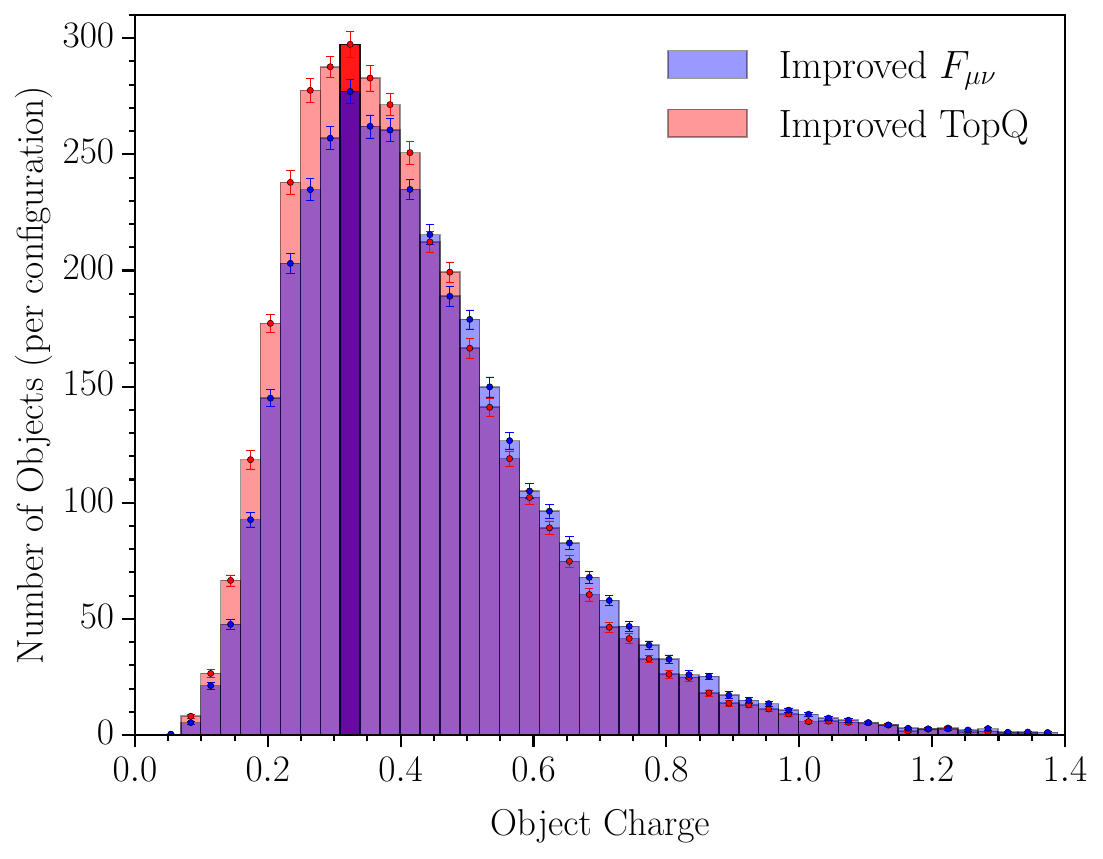}
	\caption{\label{fig:hypercube} The results of our algorithm with the hypercube dislocation filter applied to the coarse (\textbf{top}) and fine (\textbf{bottom}) ensembles. The mode is $0.336$ for the former and $0.324$ for the latter. These are consistent with each other, suggesting that a hypercube cutoff is sufficient to ensure proper scaling in the continuum limit.}
\end{figure}

Next, we repeat the above process by applying the hypercube filter to the definition of a topological object. The results are presented in Fig.~\ref{fig:hypercube}.
The hypercube filter is observed to be substantially stronger than the nearest-neighbour cutoff, preserving fewer than 10\% of all objects accepted by the nearest-neighbour filter. This is especially pronounced on the finer lattice. The size of the objects and necessary degree of smoothing accordingly increase, achieved with $\tau=1.45$ on the coarse ensemble and $\tau=1.25$ on the fine ensemble.

The corresponding modes are now located at $|Q| = 0.336(15)$ and $|Q| = 0.324(15)$. Although there is still a slight discrepancy in central value, the difference here is insignificant compared to the larger charge values calculated and broader bin width required to maintain a smooth distribution. The two modes are observed to overlap within the uncertainty provided by the bin width, meaning we can draw the same conclusions in both cases: the topological charge is predominantly comprised of individual objects with net charges near $|Q|\approx 0.33$, far from the instanton charge of $1$.

Based on this we are confident, up to statistical fluctuations, that the hypercube filter is sufficient to render our results insensitive to the size of dislocations and therefore independent of the lattice spacing. Hence, implementing a fixed hypercube cutoff and allowing for less gradient flow provides a valid procedure for taking the continuum limit in the usual manner where the scale of short-distance physics included in the calculation reduces with the lattice spacing. This is what we sought to achieve.

In Sec.~\ref{subsec:finiteT}, we will consequently present the finite-temperature analysis exclusively for the hypercube filter, as our preferred definition for what is considered a ``genuine" topological object.

\begin{figure}
	\centering
	\includegraphics[width=\linewidth]{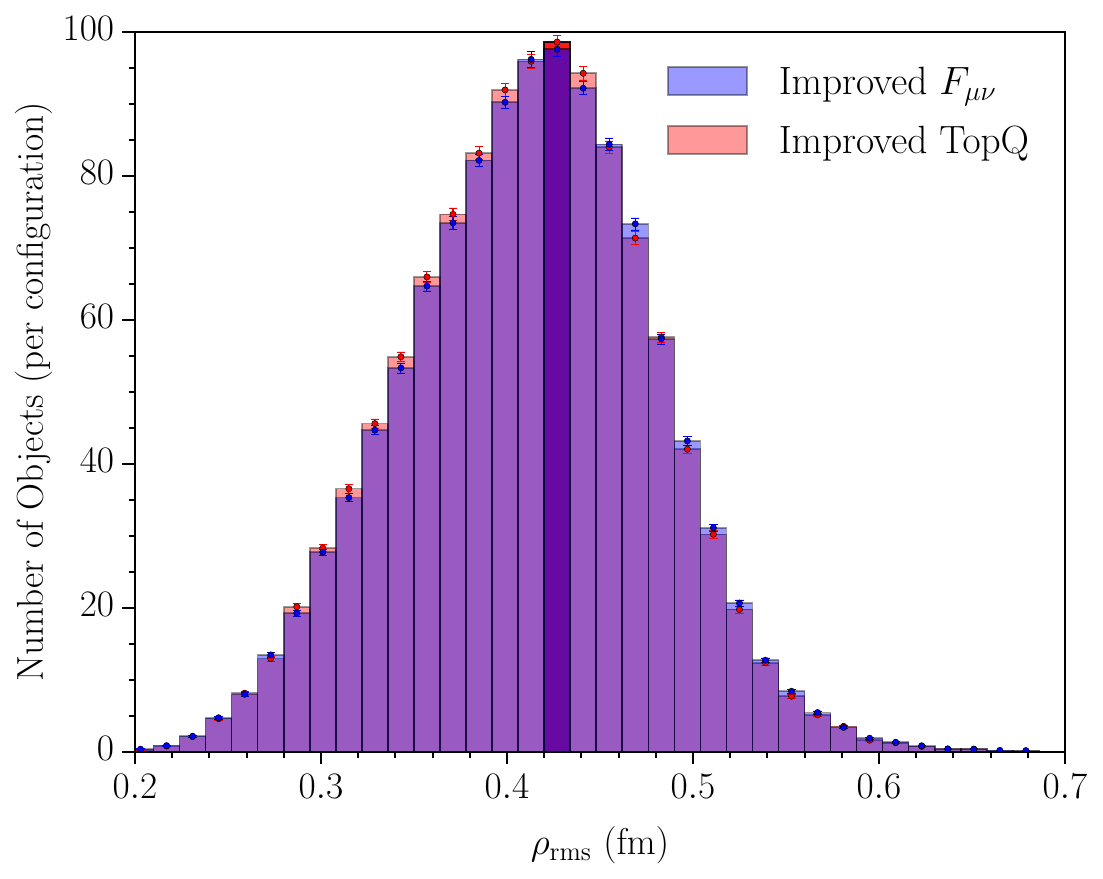}
	\includegraphics[width=\linewidth]{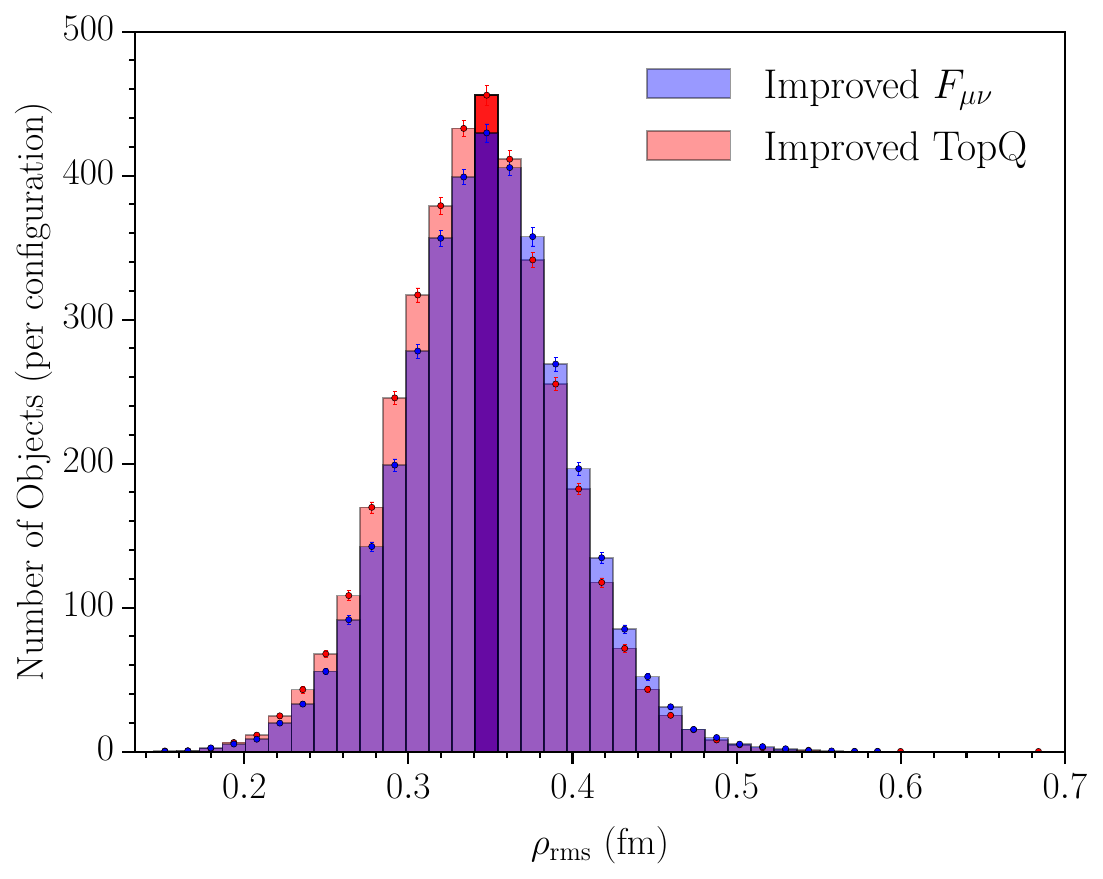}
	\caption{\label{fig:rmsradius_hypercube} Histograms showing the normalised RMS topological charge density radius results on the coarse (\textbf{top}) and fine (\textbf{bottom}) ensembles. The radial size of the objects tends to be smaller for the smaller lattice spacing.}
\end{figure}

Before proceeding, it is insightful to examine the radial size of the topological objects under the hypercube filter, having now established that their charges are scale independent. An RMS estimate of their radii is calculated by summing the squared distance between each point assigned to a given object and its centre $x_0$ (approximated by the local extremum), weighted by the ratio of the topological charge density to the net charge $Q$ of the object. These weights sum to unity over the entire object and discriminate between broad and sharply peaked features. Mathematically, this normalised topological charge density radius is
\begin{equation} \label{eq:rmsradius}
	\rho_{\mathrm{rms}} = \sqrt{\frac{1}{Q} \sum_{x\in\text{obj}} q(x) (x - x_0)^2 } \,.
\end{equation}
The calculations are performed at the same flow time as for the charges, and we find that the modes of the RMS radii distributions from our two different topological charge definitions also match. We contrast the $\rho_\mathrm{rms}$ results between the two ensembles in Fig.~\ref{fig:rmsradius_hypercube}.

This reveals a decrease in the typical radial size of objects with the lattice spacing. Although this may seem curious given the consistency between the net charge values, upon further investigation one finds that this is to be anticipated. The lattice spacing introduces a cutoff such that any topological features smaller than the lattice spacing fail to be resolved. As previously surmised, by utilising a filter that scales with the lattice spacing, one would therefore expect the resulting distribution to be comprised of topological objects with a smaller radial size. In a similar vein to the instanton solution, one could propose that the topological objects have a free size parameter which can be varied whilst keeping their net charge constant. The combination of Figs.~\ref{fig:hypercube} and \ref{fig:rmsradius_hypercube} strongly suggests this is the pattern underlying changes to the gauge field as the lattice spacing is decreased.

\subsection{Fixed scale method} \label{subsec:fixedscale}
Besides probing the topological structure at the improved resolution provided by a finer lattice, the other continuum limit we turn our attention towards is a fixed scale method. This entails maintaining a fixed physical scale for resolving the topological objects under consideration. There are two different aspects at play to ensure this occurs.

First, there is the matter of the smearing scale, $r_\mathrm{sm}/a = \sqrt{8\tau}$, induced by gradient flow. To set a fixed size, this smoothing radius $r_\mathrm{sm}$ must remain unchanged (in physical units) between the two lattice spacings. Using primed symbols for the finer lattice, this clearly requires
\begin{equation} \label{eq:flowtime}
	\sqrt{\frac{\tau'}{\tau}} = \frac{a}{a'} \implies \tau' \approx 2.25\, \tau \,,
\end{equation}
where we have substituted the values $a = 0.10\,$fm and $a' = 0.067$\,fm.

The second factor concerns the filter applied in the algorithm. To ensure an equal footing between the ensembles, it is vital to implement a dislocation filter of fixed physical size for each ensemble. Instead of the hypercube filter examined in the previous section, which scaled with the lattice spacing, we choose here a radial cutoff $r_\mathrm{cut}$ in physical units which is applied on both ensembles. Motivated by the success of the hypercube filter in eliminating dislocations, however, we choose the minimum physical radius needed to cover the hypercube on the coarse ensemble. This is a radius of two lattice units, dictating $r_\mathrm{cut} = 0.2\,$fm. On the fine ensemble, this is three lattice units.

\begin{figure}
	\centering
	\includegraphics[width=\linewidth]{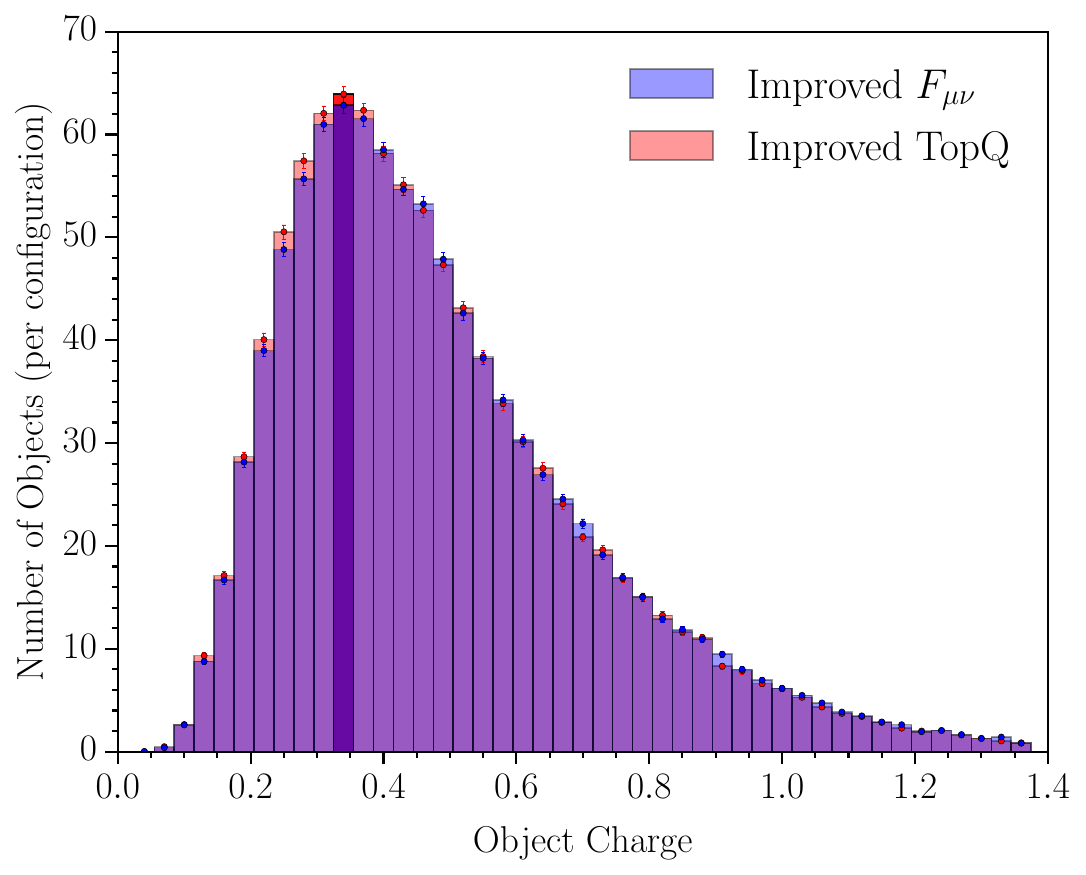}
	\includegraphics[width=\linewidth]{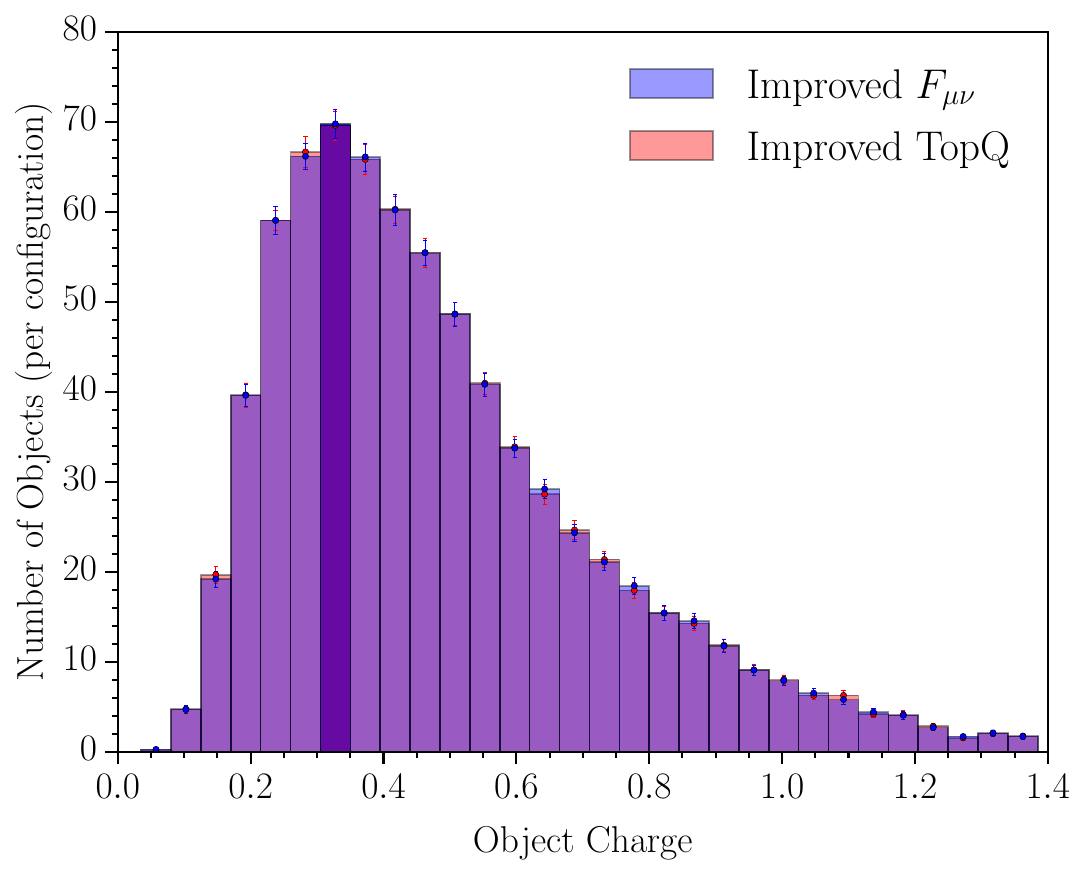}
	\caption{\label{fig:fixedscale} The results of our algorithm with a fixed scale, realised by a physical radial cutoff $r_\mathrm{cut} = 0.2\,$fm, applied to the coarse ensemble at a flow time $\tau = 1.65$ (\textbf{top}) and the fine ensemble at $\tau' = 3.71$ (\textbf{bottom}). The histogram modes again match, implying that this is an equally valid procedure for taking the continuum limit.}
\end{figure}

As for the flow time $\tau$, we note that a radial cutoff of two lattice units on the coarse ensemble is stronger than the hypercube filter previously applied. Thus, a mild increase in smoothing level is necessary to match the charge values. This is satisfied by $\tau = 1.65$, implying $\tau' = 3.71$ on the fine ensemble as per Eq.~(\ref{eq:flowtime}). With this setup, the charge histograms in the fixed scale method are shown in Fig.~\ref{fig:fixedscale}.

\begin{figure}
	\centering
	\includegraphics[width=\linewidth]{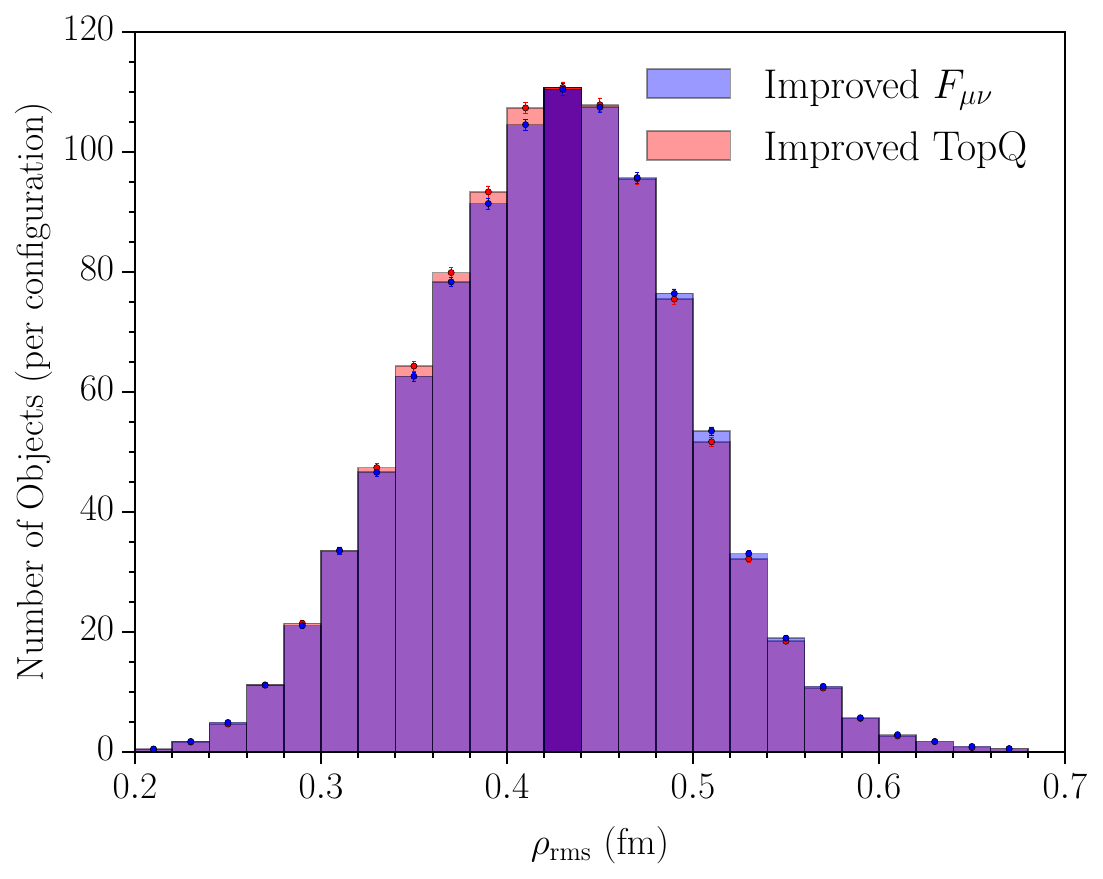}
	\includegraphics[width=\linewidth]{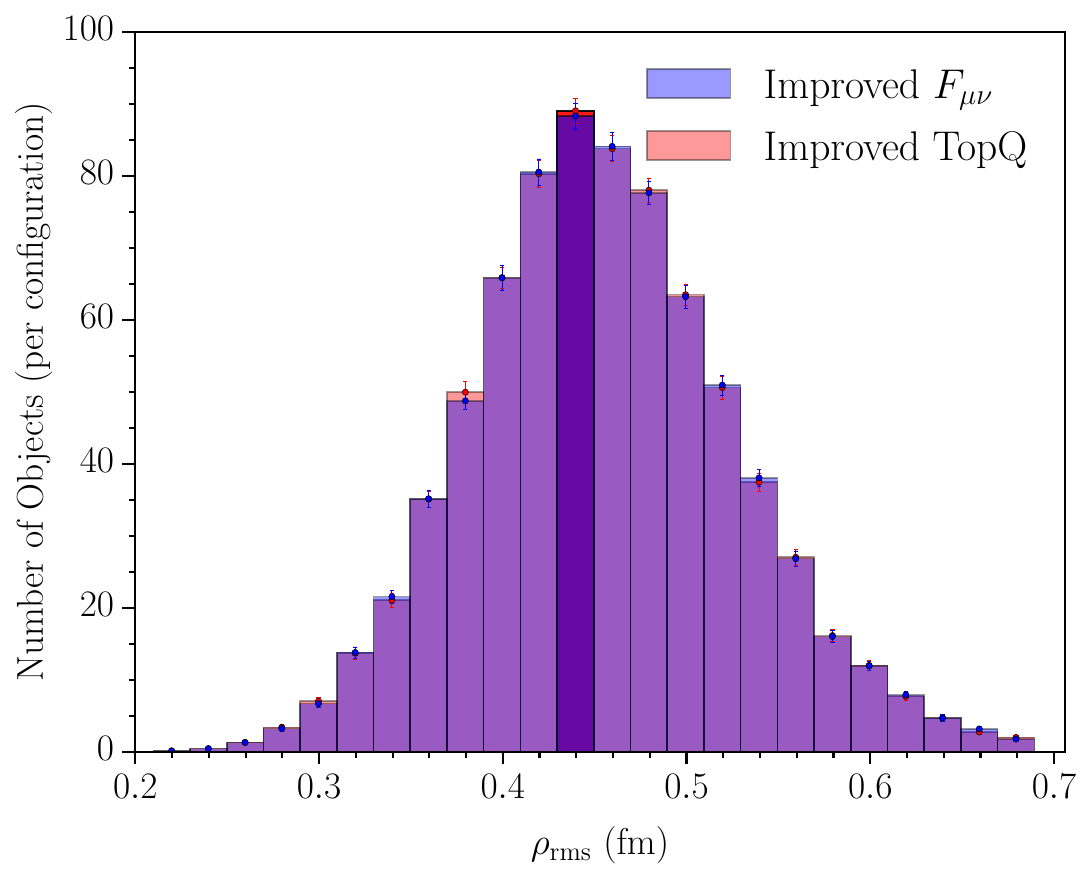}
	\caption{\label{fig:rmsradius_fixed} Histograms showing the normalised RMS topological charge density radius results for the fixed physical scale continuum limit on the coarse (\textbf{top}) and fine (\textbf{bottom}) ensembles. The typical radial sizes of the objects are indistinguishable between the lattice spacings.}
\end{figure}

Remarkably, we once again find consistent histogram modes, indicating we have successfully uncovered similar topological structures between the two lattice spacings. Therefore, setting a fixed physical scale provides an alternative continuum limit. As with the fixed cutoff limit, we also investigate the radial size of the topological objects within this framework to ascertain whether their physical sizes are indeed coincident, as one might expect. These $\rho_\mathrm{rms}$ histograms are given in Fig.~\ref{fig:rmsradius_fixed}.

The modes are observed to be consistent with each other within uncertainty, indicating we have successfully held the typical radial size for resolving topology constant whilst decreasing the lattice spacing. The centre of the modal bin for the fine ensemble is marginally to the right of that for the coarse ensemble, though one can easily imagine this might be due to statistical deficiencies or a slightly inaccurate setting of the smoothing scale.

To summarise, we have considered two possibilities for taking the continuum limit. In one, a consistent hypercubic dislocation filter was applied to analyse physically smaller topological features as $a\to 0$, whilst in the second method a fixed physical scale for resolving topology was utilised. Both provide topological charge distributions insensitive to the lattice spacing, as illustrated in Figs.~\ref{fig:hypercube} and \ref{fig:fixedscale}. Thus either can be used in the subsequent analysis.

Our preferred method is the fixed hypercube cutoff, due to its ability to probe vacuum structure at the improved resolution provided by a smaller lattice spacing. This approach is more in accord with traditional continuum limits where short-distance physics is allowed to enter the calculations as $a\to 0$. This makes such a limit more interesting, and it is notable that the object charge values remain well defined.

\section{Results} \label{sec:results}
Having established the behaviour of the continuum limit, we now present our findings on the evolution of the topological structure with temperature. This is performed by applying the hypercube filter for distinguishing topological objects over dislocations. We ensure the flow time is independent of temperature such that the results on each ensemble can be easily compared, set to $\tau = 1.45$ as per the $32^3\times 64$ ensemble in Sec.~\ref{subsec:fixedcutoff}.

\subsection{Simulation details} \label{subsec:simdetails}
To explore the evolution of the topological structure with temperature, we generate ensembles consisting of 100 configurations at three temperatures below the critical temperature $T_c$, and three temperatures above $T_c$. Each has a spatial volume of $32^3$ and fixed isotropic lattice spacing $a=0.1\,$fm, with the temporal extent of the lattice varied to change the temperature. The details are provided in Table \ref{tab:ensembledetails}, where we take $T_c = 270\,$MeV \cite{CriticalTemp}.
\begin{table*}
\caption{\label{tab:ensembledetails} The statistics for each of our $32^3\times N_t$ ensembles, including: the number of sites $N_t$ in the temporal dimension, the corresponding temperatures, histogram modes, Polyakov loop values $\langle P\rangle = \frac{1}{3}\langle\tr P\rangle$ and respective holonomy parameters. The histogram mode is determined by the centre of the corresponding bin, with uncertainties quoted as half the bin width. The trend displayed by the holonomy parameter matches with the modes of our histograms.}
\begin{ruledtabular}
\begin{tabular}{lD{.}{.}{3.2}D{.}{.}{1.4}D{.}{.}{1.7}D{.}{.}{1.8}D{.}{.}{1.8}}
$N_t$ & \multicolumn{1}{c}{$T$ (MeV)} & \multicolumn{1}{c}{$T/T_c$} & \multicolumn{1}{c}{Histogram mode} & \multicolumn{1}{c}{$\langle P\rangle$} & \multicolumn{1}{c}{$\nu$} \\
\colrule
64 & 30.8 & 0.114 & 0.336(15) & 0.0047(2) & 0.3321(1) \\
12 & 164.4 & 0.609 & 0.338(19) & 0.0208(11) & 0.3277(3) \\
8 & 246.6 & 0.913 & 0.312(19) & 0.0355(20) & 0.3237(5) \\
6 & 328.8 & 1.218 & 0.202(16) & 0.562(16) & 0.1944(40) \\
5 & 394.6 & 1.461 & 0.156(15) & 0.677(19) & 0.1639(54) \\
4 & 493.3 & 1.827 & 0.102(10) & 0.8634(6) & 0.1037(2)
\end{tabular}
\end{ruledtabular}
\end{table*}

The ensembles are generated using Hybrid Monte Carlo \cite{HMCI, HMCII} with an Iwasaki action \cite{IwasakiActionI, IwasakiActionII} given by
\begin{equation}
    S = \beta \sum_{x,\mu>\nu}\left[c_0\, P_{\mu\nu}^{(1\times 1)}(x) + c_1\, P_{\mu\nu}^{(1\times 2)}(x) \right] \,,
\end{equation}
with $c_1 = -0.331$ and $c_0 = 1-8c_1 = 3.648$. A coupling $\beta=2.58$ gives the desired lattice spacing of $a=0.1$\,fm. A unit trajectory length is used for the Hamiltonian dynamics evolution, with 50 accepted trajectories between sampling a configuration following thermalisation.

For each histogram, we analyse 100 gauge field configurations. We obtain 100 bootstrap resamples on the set of calculated charge values and extract the bin counts for each such resample. This allows an error to be placed on the histogram bins, which we display in our results. The precise location of the mode is ascertained by shifting the bins to maximise the height of the modal bin, and is singled out by a darker colour so the associated charge value is visually clear. We find that the position of the mode shows no variation in the bootstrap resamples such that its uncertainty is governed by the bin width.

\subsection{Finite Temperature} \label{subsec:finiteT}
\begin{figure*}
    \centering
    \includegraphics[width=0.478\linewidth]{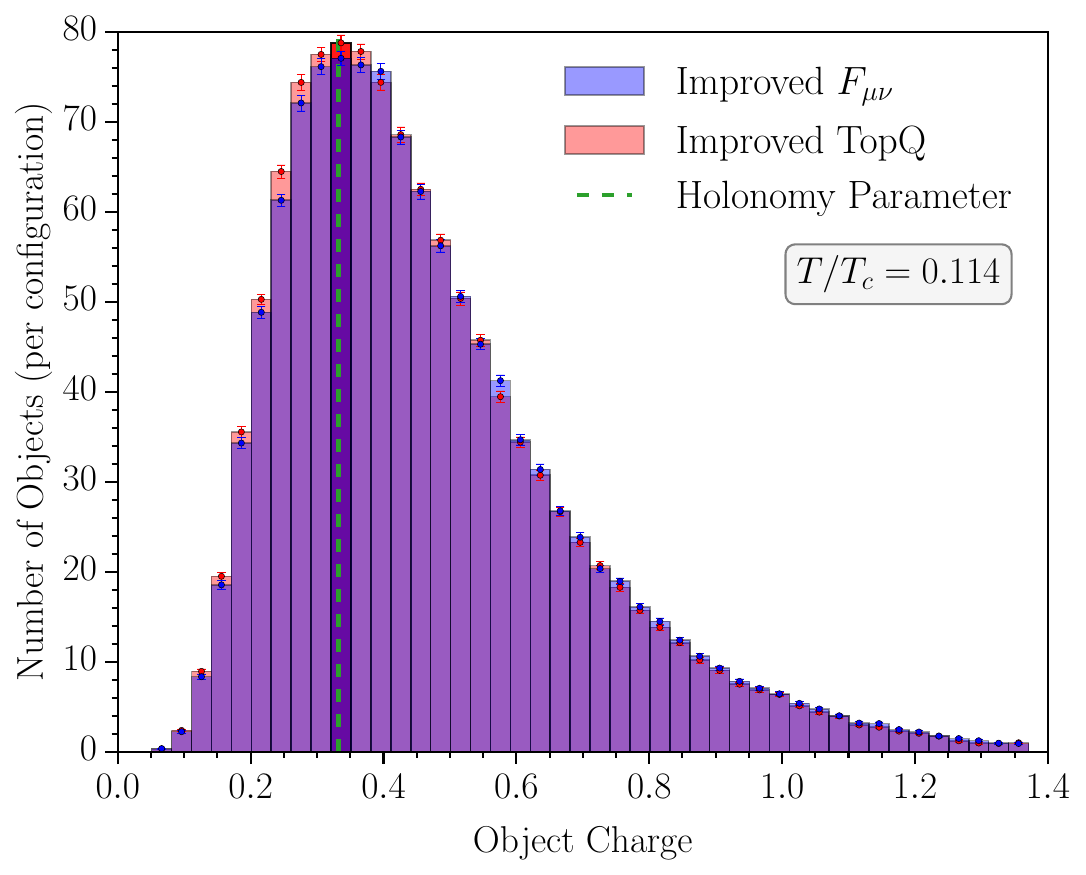}
    \includegraphics[width=0.478\linewidth]{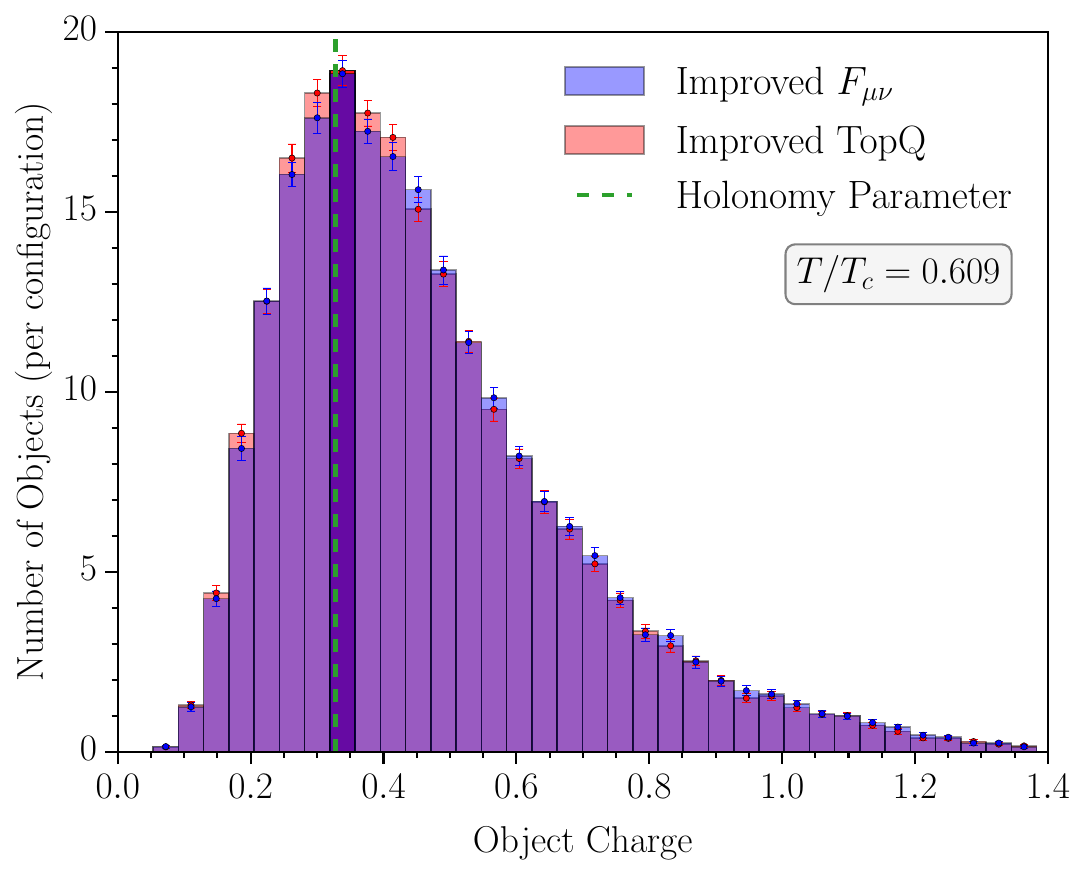}
    \includegraphics[width=0.478\linewidth]{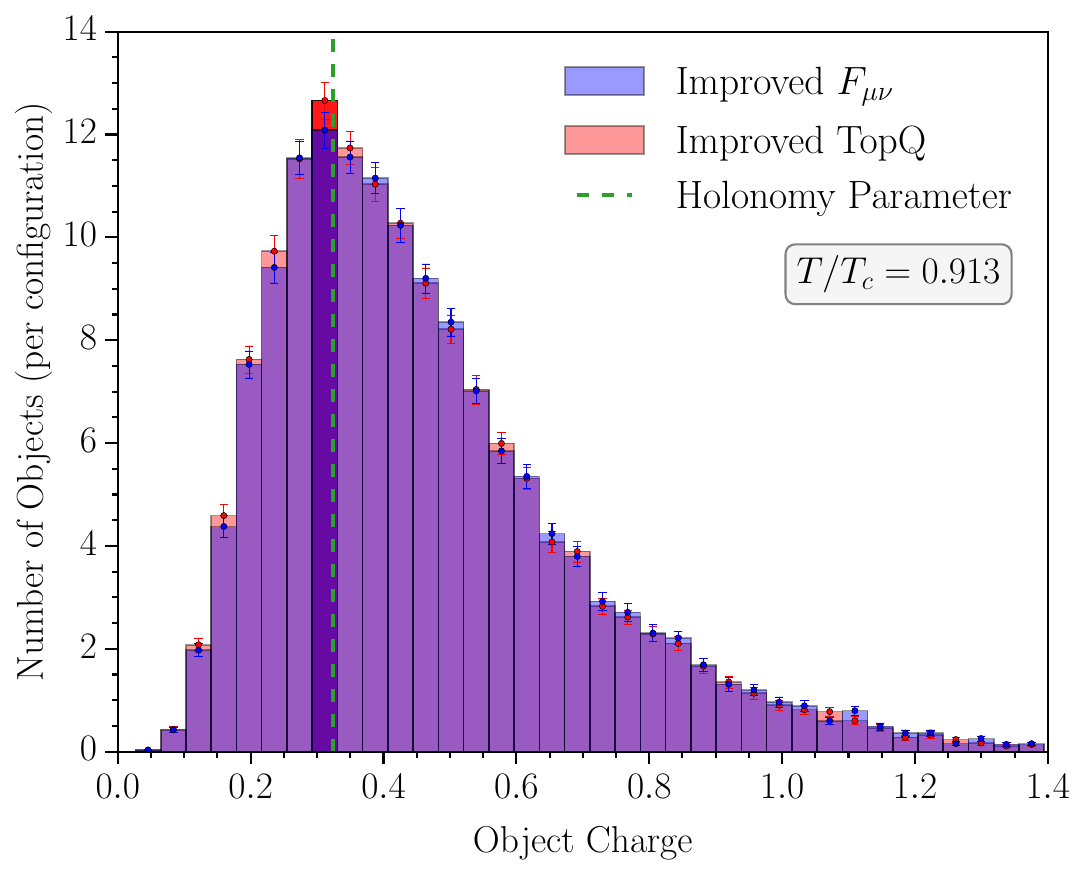}
    \includegraphics[width=0.478\linewidth]{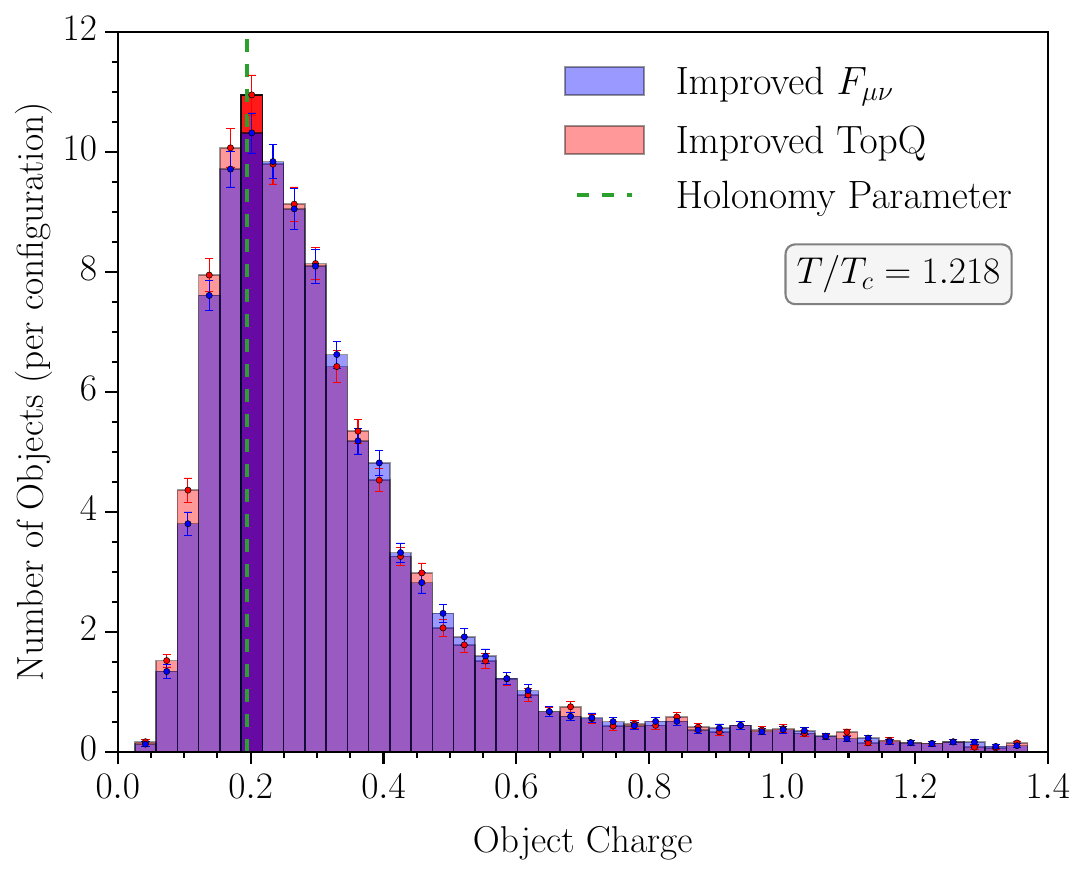}
    \includegraphics[width=0.478\linewidth]{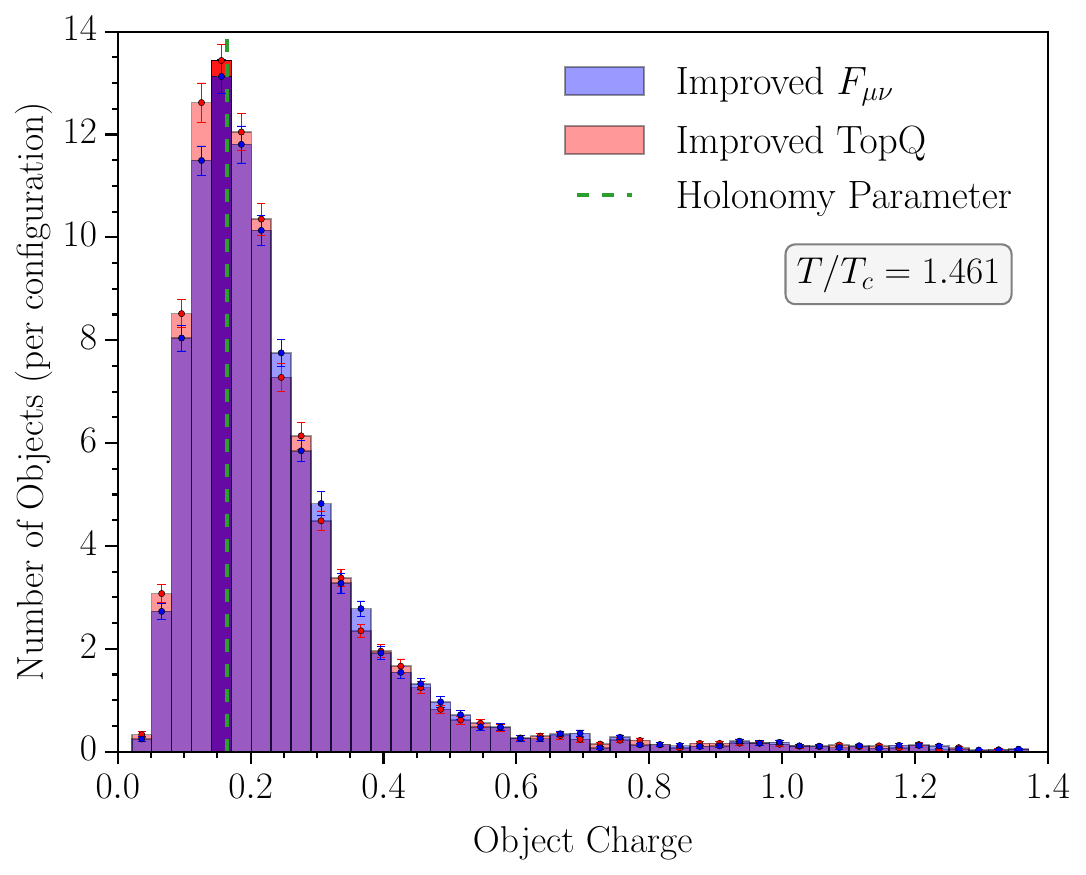}
    \includegraphics[width=0.478\linewidth]{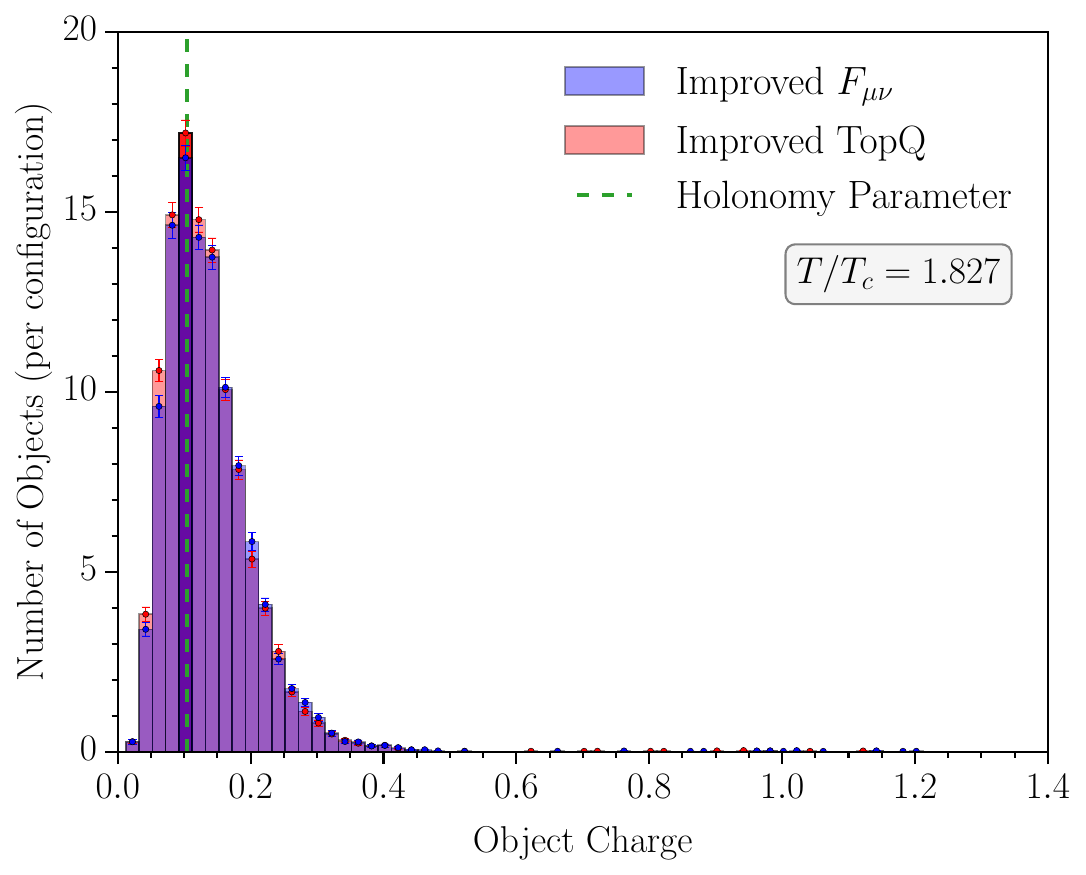}
    \caption{\label{fig:finiteTresults} Histograms showing the results of our algorithm with the hypercube dislocation filter applied to each of our finite-temperature ensembles: $N_t = 64$ (\textbf{top left}), $12$ (\textbf{top right}), $8$ (\textbf{middle left}), $6$ (\textbf{middle right}), $5$ (\textbf{bottom left}) and $4$ (\textbf{bottom right}). Below the critical temperature, the mode is roughly constant at just above $0.3$, but this shifts towards smaller values as $T$ increases above $T_c$. This behaviour is consistent with the free holonomy parameter in SU(3) (Sec.~\ref{subsec:instantondyons}), shown with the dashed vertical line. The results are calculated with a hypercube dislocation filter after a flow time $\tau=1.45$, the amount required to ensure consistency between the two improvement schemes considered.}
\end{figure*}

We present the charge histograms for each finite-temperature ensemble in Fig.~\ref{fig:finiteTresults}. The quantitative value of the mode for each temperature is provided in Table \ref{tab:ensembledetails}.
We find that below the critical temperature, the topological charge tends to be comprised of objects with net charges near $|Q|\approx 1/3$, with the modes for each of these ensembles situated very near each other. Being the ensemble closest to the critical temperature, the marginally smaller value for $N_t=8$ could arise from finite-volume effects resulting in a smooth crossover around $T_c$ instead of a discontinuous phase transition. That being said, given that the $N_t=8$ mode sits directly adjacent to the other modes below $T_c$, it could also be attributed to simple statistical fluctuations in the calculated charge values; this is more likely near $T_c$ due to challenges in the Markov chain around the phase transition.

Nevertheless, as the temperature increases into the deconfined phase, there is an undeniable shift in the calculated charges towards smaller values, with the modes visibly separated from each other. The largest decrease occurs in our first ensemble above $T_c$, where we find $|Q|\approx 0.2$, and this steadily continues to decline down to $|Q|\approx 0.1$ for the largest temperature considered here.

\section{Discussion} \label{sec:discussion}
Having presented our key results, we now proceed to compare with the instanton-dyon model for the topological structure of the gluon fields at finite temperature. We also investigate the temperature dependence of additional quantities such as object density and radial size, which are in general distinct from the charge contained within each such object. Throughout this section, all statistics are obtained using 100 bootstrap ensembles, with errors calculated through the standard deviation of the bootstrap estimates.

\subsection{Polyakov loop and holonomy}
The Polyakov loop is an order parameter for confinement in SU($N$) Yang-Mills theory defined for each spatial position $\mathbf{x}$ as
\begin{equation}
	P(\mathbf{x}) = \mathcal{P} \exp\left(ig \int_0^{1/T} dx_4 \, A_4(x) \right) \in \mathrm{SU}(N) \,,
\end{equation}
where $\mathcal{P}$ is the path-ordering operator. It exhibits a simple relation with the free energy of a single quark \cite{PolyakovFreeEnergy},
\begin{equation} \label{eq:PolyakovFq}
	\langle \tr P \rangle \sim \exp\left(-\frac{F_q}{T}\right) \,.
\end{equation}
From this one concludes $\langle\tr P\rangle=0$ below $T_c$ where confinement implies $F_q\to\infty$, whilst it jumps to a nonzero value above $T_c$.

The Polyakov loop at spatial infinity, also known as the holonomy, is a topological invariant and (up to gauge symmetry) can be written as \cite{InstantonDyonsIII}
\begin{equation} \label{eq:Pinfinity}
	\begin{gathered}
		P_\infty = \lim_{|\mathbf{x}|\to\infty} P(\mathbf{x}) = \exp\!\big[2\pi i\,\mathrm{diag}(\mu_1,\hdots,\mu_N)\big] \,, \\
		\mu_1 < \hdots < \mu_N < \mu_{N+1} \equiv \mu_1 + 1\,, \;\;\;\;\; \sum_{i=1}^N \mu_i = 0 \,.
	\end{gathered}
\end{equation}
Simply put, Eq.~(\ref{eq:Pinfinity}) says that the eigenvalues of $P_\infty$ lie on the unit circle within one rotation of $2\pi$, with the summation condition enforcing $\det P_\infty = 1$. Whilst at extremely high temperatures the holonomy is trivial, near and below $T_c$ it develops a nontrivial value, $P_\infty \neq \mathbb{I}$. In particular, the maximally nontrivial holonomy corresponds to $\tr P_\infty = 0$, which occurs in the confined phase; it is also known as the ``confining holonomy".

For convenience, one defines $N$ so-called ``holonomy parameters" as $\nu_i = \mu_{i+1}-\mu_i$. Following the definition of $\mu_{N+1}=\mu_1+1$, the holonomy parameters are constrained to sum to unity, leaving naively $(N-1)$ free parameters in SU($N$) gauge theory. However, as per Eq.~(\ref{eq:PolyakovFq}), $\langle \tr P\rangle$ is a physical quantity and must accordingly be real. This imposes an additional constraint on the set of holonomy parameters, which in the case of SU(3) leaves just a single parameter $\nu$ to uniquely specify each $\nu_i$. The relationship is summarised as \cite{SU3DyonConfinement}
\begin{equation} \label{eq:holonomy}
	\begin{gathered}
		\nu_1=\nu_2=\nu \,, \;\;\;\;\;\;\;\;\; \nu_3=1-2\nu \,, \\
		\langle P\rangle \equiv \frac{1}{3}\langle\tr P\rangle = \frac{1}{3} + \frac{2}{3}\cos(2\pi\nu) \,.
	\end{gathered}
\end{equation}
It is thus immediately clear that one finds $\nu = 1/3$ in the confined phase, whilst above $T_c$ the holonomy parameter decreases towards zero as the Polyakov loop becomes trivial. We note immediately that this quantitatively agrees with the mode of our charge histograms below $T_c$, and also displays initial qualitative agreement above $T_c$. This hints at a connection between the holonomy and charge of topological objects, which motivates calculating $\nu$ on each finite-temperature ensemble.

The Polyakov loop on the lattice is calculated as a product of temporal link variables over all sites in the temporal dimension,
\begin{equation}
	P(\mathbf{x}) = \prod_{x_4=1}^{N_t} U_4(\mathbf{x},x_4) \,.
\end{equation}
We exploit translational symmetry to calculate $\langle \tr P_\infty\rangle$ as the expectation of the spatially averaged Polyakov loop,
\begin{equation}
	\langle\tr P\rangle = \bigg\langle\frac{1}{V} \sum_\mathbf{x} \tr P(\mathbf{x}) \bigg\rangle \,.
\end{equation}
The pure gauge theory carries an additional complication in the form of a centre symmetry which does not exist in full QCD. The pure gauge action is invariant under centre transformations,
\begin{equation} \label{eq:centretrans}
	\begin{gathered}
		U_4(\mathbf{x},x_4) \longrightarrow z\,U_4(\mathbf{x},x_4) \text{ for fixed } x_4\,, \\
		z\in\mathbb{Z}_3 = \left\{ \exp\left(\frac{2\pi i}{3} \,m \right) \mathbb{I} \;\middle|\; m = -1,0,1 \right\} \,,
	\end{gathered}
\end{equation}
though the Polyakov loop transforms nontrivially under such transformations as
\begin{equation}
	P(\mathbf{x}) \longrightarrow z\,P(\mathbf{x}) \,.
\end{equation}
Thus if centre symmetry is preserved we must have $\langle \tr P(\mathbf{x})\rangle = 0$, and deconfinement corresponds to the spontaneous breaking of the centre symmetry.

Consequently, the Polyakov loop below $T_c$ is observed to exhibit a symmetry between the three centre phases of SU(3) \cite{CentreClustersI, CentreClustersII}. Above $T_c$, this symmetry is spontaneously broken with one of the three phases becoming dominant. In full QCD, the fermion determinant singles out $m=0$ as the preferred phase \cite{CentreClustersII}, ensuring the Polyakov loop remains real. On the other hand, in the pure gauge theory the dominant phase can vary on a configuration-to-configuration basis, meaning one would find $\langle\tr P\rangle=0$ even above $T_c$.

This is often overcome by taking the modulus of the Polyakov loop as the order parameter, though we take an alternative approach to remove the remaining symmetry by performing centre transformations \cite{CentreTransformation}. This can be interpreted as rotating the phase of $\tr P$ by $\pm\, 2\pi i/3$ to bring the dominant phase of each configuration to a phase of zero. Finally, we take the real part to discard any remnant imaginary part that should vanish in the ensemble average. This is subsequently taken to estimate $\langle\tr P\rangle$. Substituting the result into Eq.~(\ref{eq:holonomy}) gives a corresponding estimate of the free holonomy parameter. The final values are given in Table \ref{tab:ensembledetails} and illustrated in Fig.~\ref{fig:finiteTresults}.

We find that for each temperature considered here, the calculated values of the holonomy coincide remarkably well with the histogram modes for the charge contained within each distinct topological object. For each ensemble in the confined phase, we find $\langle P\rangle \approx 0$ as expected, and the associated holonomy parameter is $\nu = 1/3$. This is consistent with our findings below $T_c$, with the modes located near $1/3$. In the deconfined phase, where the holonomy parameter decreases away from $1/3$, we continue to find a strong agreement between its value and the histogram mode for each ensemble above $T_c$. This reveals an intrinsic connection between the holonomy of the field configurations and the distinct topological charges comprising their structure. Such a relationship is built into the instanton-dyon model, discussed in the following section.

\subsection{Instanton-dyons} \label{subsec:instantondyons}
Given the link between confinement and holonomy, it is natural to seek analytic solutions of the Yang-Mills equations that possess nontrivial asymptotic holonomy. This is realised by the caloron \cite{InstantonDyonsI, InstantonDyonsII, InstantonDyonsIII, InstantonDyonsIV}, a finite-temperature generalisation of the instanton. In SU($N$) gauge theory, calorons can be viewed as composed of $N$ monopole constituents known as dyons, whose structure depends on the value of the holonomy.

To be precise, the caloron is divided up into its constituent dyons according to the holonomy parameters $\nu_i$, such that the action of the $i$th dyon type is $S_i = S_0\,\nu_i$. Like the instanton itself, the dyons are self-dual such that their topological charges satisfy $|Q_i|=\nu_i$. Since the $\nu_i$ sum to unity, it is clear that summing the dyons' individual actions and topological charges recovers the single-instanton properties.

Thus, one finds a one-to-one relationship between the temperature of the system (through the Polyakov loop) and the charges of the dyons (through the free holonomy parameter). Two of the dyons have charges $|Q_i| = \nu$, whilst the third dyon has $|Q_3| = 1-2\nu$. This provides a prediction for the topological structure at finite temperature we can compare to our numerical findings. The agreement between the holonomy parameter $\nu$ and the charge histogram modes suggests a consistency with the presence of the first two dyon types. This can be interpreted as evidence that dyons form a significant part of the gluon fields' topological structure. At each temperature, the mode provides an indicator for the dominant contribution, with the distribution around the mode representing quantum fluctuations about the semiclassical dyon solution and systematic uncertainties in assigning topological charge density to the objects.

However, the simple decomposition described above holds exclusively for a single-caloron configuration. The fact that we observe consistency between the holonomy parameter and object charges on our configurations, which in general comprise an ensemble of positive and negative topological excitations, is a substantially stronger constraint. This is nonetheless allowed within the framework of instanton-dyons. Analytic ``multi-caloron" configurations of higher topological charge $|Q| = k$ have previously been constructed \cite{MulticaloronI, MulticaloronII, MulticaloronIII, MulticaloronIV, MulticaloronV}. These follow the natural expectation of decomposing into $kN$ constituents in SU($N$) gauge theory, with $k$ dyons of each of the $N$ types. All dyons of the $i$th type have the same mass \cite{InstantonDyonsIII, InstantonDyonsIV}, as determined by the corresponding holonomy parameter $\nu_i$ of the system.

In addition, the typical procedure for superposing calorons in modelling vacuum structure requires each distinct caloron to have the same holonomy \cite{CaloronSuperpositionI, CaloronSuperpositionII, CaloronSuperpositionIII}, and therefore their constituent dyons of the same type have matching charges. The system resulting from the superposition then features that same asymptotic holonomy.

These are both in a similar vein to the scenario reflected by our findings, where we observe a sharply peaked topological charge distribution located at $\nu_1 = \nu_2$. From this discussion, we can conclude our results admit the presence of the two dyons possessing charge  $|Q_1| = |Q_2| = \nu$, with fluctuations about this value as reflected by the charge histograms.

One would be prescient to point out the stark lack of charge values consistent with the third dyon in Fig.~\ref{fig:finiteTresults}, which would be $|Q_3|\approx 0.6$, $0.7$ and $0.8$ respectively for each temperature above $T_c$. However, this can be understood by considering the number densities of each dyon. In the SU(3) dyonic partition function each dyon is individually weighted by \cite{SU3DyonConfinement}
\begin{equation}
    d \sim e^{-S_0\nu_i}\, \nu_i^{\,\frac{8\nu_i}{3}-1}
\end{equation}
for holonomy parameter $\nu_i$. The outcome is a compounding effect where both terms favour small holonomy parameters, and the third dyon with $\nu_3=1-2\nu$ will be exponentially suppressed. For this reason we would expect our results to be overwhelmingly skewed towards the two dyons with charges equal to $\nu$, as we observe

One might query how this imbalance manifests in the topological structure, given the topological index is restricted to integer values on the periodic torus by the Atiyah-Singer index theorem \cite{IndexTheorem}. We find that the majority of our $N_t=4$ configurations converge to a net topological charge of zero under smoothing, which overcomes the problem entirely by representing an equal quantity of positive and negative topological charge density. The suppression of the third dyon is irrelevant in that context, needing only a balance of dyons with $Q=\pm\,\nu$. There are nevertheless a handful of configurations with small nonzero values for the topological index, which could be attributed to the presence of the larger third dyon. In fact, our highest-temperature histogram in Fig.~\ref{fig:finiteTresults} displays a small clustering of points near $|Q|=1$ which originate precisely from this small set of configurations, supporting this idea. Prior analysis of the topological content in the confined and deconfined phases substantiates this discussion \cite{TopologicalIndex}.

\subsection{Zero-temperature distribution} \label{subsec:zeroT}
All considered, we have found that the instanton-dyon model captures the essence of the observed finite-temperature physics. Even so, the plethora of other fractionally charged topological configurations $\sim \!\! 1/N$ \cite{TwistedSelfdual, TwistedSelfdualCoolingI, TwistedSelfdualCoolingII, FractionalReview, FractionalGeneralisationI, FractionalGeneralisationII, FractionalSUN, FractionalVortexI, FractionalVortexII, FractionalVortexIII, FractionalFibonacci} leave our zero-temperature results open to different interpretations. In fact, if the inverse temperature greatly exceeds the characteristic separation between their constituent dyons, then calorons simply resemble standard instantons \cite{InstantonDyonsIII,InstantonDyonsIV}.

In this regard, the peaks we observe near $\approx 1/3$ at low temperatures $T<T_c$ could reflect a significant contribution from ``ordinary" fractional instantons. Although the current methods do not provide a means to distinguish between these constructions, we can nonetheless conclude the confining vacuum is an ensemble of fractionally charged objects. Many of these configurations are analytically constructed on tori with different twisted boundary conditions, and it is natural to query whether they can arise on a torus without twist, as studied here. The resolution is that these are trivially also solutions on a torus comprising multiple periods of the original, smaller torus. It directly follows that fractional instantons can emerge on the standard periodic torus, with the requirement that they exist in groups of $N$ to conserve the integer topological index.

There is also the ``$\mathbb{Z}_N$ dyons" of Ref.~\cite{FractionalSUN}, which in contrast to instanton-dyons and fractional instantons are not predicted to be self-dual in general. This is allowed within the bounds of our present results.  At the level of smoothing which ensures consistency between our two topological charge definitions, $S/S_0$ is still approximately a factor of $3$ larger than $\sum\, |q(x)|$ for an average configuration. This demands the presence of non-self-dual topological objects comprising the gluon fields.

On that account, our low-temperature results could signal the presence of non-self-dual $\mathbb{Z}_N$ dyons. In addition, these configurations carry magnetic charge quantised by $\mathbb{Z}_N$, and consequently are relevant exclusively in the confined phase and a narrow region above $T_c$. In the deconfined phase, the increasing magnetic tension binds the $\mathbb{Z}_N$ dyons into dilute instantons at high temperatures \cite{FractionalSUN}. The small clustering of points near $|Q| = 1$ found at our highest temperature could also signal a contribution from these instantons, though the leading effect remains which aligns with instanton-dyons with a peak in the charge histogram at $|Q| = \nu$.

Further analysis could proceed by summing the action over each object individually using the partition of the lattice obtained herein, allowing an investigation into self-duality on a per-object basis. There is also the possibility that some topological objects look locally self-dual near their peak, though experience significant deviations from self-duality at the tails of their distributions due to interactions with surrounding objects. We leave such investigations open for future work.

\subsection{Number of objects} \label{subsec:numberofobjects}
\begin{figure}
    \centering
    \includegraphics[width=\linewidth]{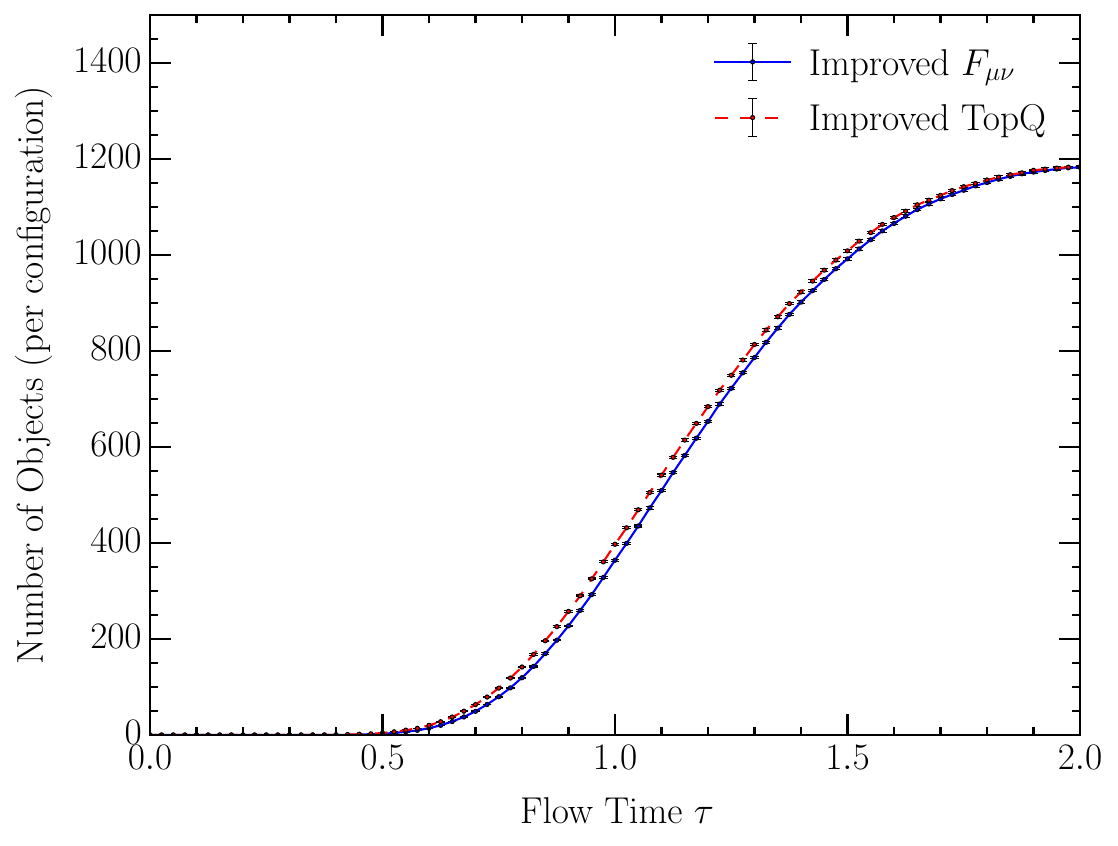}
    \caption{\label{fig:NobjSmoothing} The evolution of the observed number of topological objects under gradient flow after hypercubic dislocation filtering, for $N_t=64$. We show separate curves for each topological charge improvement scheme, highlighting a subtle difference between the two distributions which is reduced as we continue to smooth the fields.}
\end{figure}

In addition to the charge of the topological objects, another basic statistic available through our methods is the total number of objects per configuration. We start by investigating how the number of objects varies as we smooth the configurations. Performing this analysis is enlightening as it reflects the effect smoothing has on the gauge field. This evolution, as defined by the hypercube filter, is shown for the $N_t = 64$ ensemble in Fig.~\ref{fig:NobjSmoothing}; a very similar trend is observed for the other ensembles.

We find that initially there are zero topological objects that pass the hypercube filter. This quantitatively emphasises how ``raw" lattice configurations are comprised entirely of UV fluctuations on the scale of the lattice spacing which obscure the long-distance topological features of the gluon fields. Continuing to smooth gradually reveals larger-scale features at least the size of a hypercube, whilst the short-scale fluctuations are smoothed out. Eventually, this begins to plateau, with the number of objects stabilising as we approach a flow time of $\tau=2$. Although we have not explored beyond this point, one could imagine that the number of objects would eventually begin to decrease, which is necessary to coincide with the classical limit. Indeed, topological excitations are understood to undergo pair annihilation under extended smoothing \cite{CoolingII, CoolingIII, ImprovedFmunu, QuarkPropagationInstantons, InstantonsSmoothing}, and it has been revealed through visualisations how the excitations ``walk" across the lattice to annihilate with each other \cite{QCDVis}. This provides an intuitive picture for when one may wish to cease smoothing: after the majority of the UV fluctuations have been removed, but before the genuine features we are interested in begin to annihilate. Our method of matching the object charges between two topological charge definitions singles out the precise level required.

\begin{figure}
    \centering
    \includegraphics[width=\linewidth]{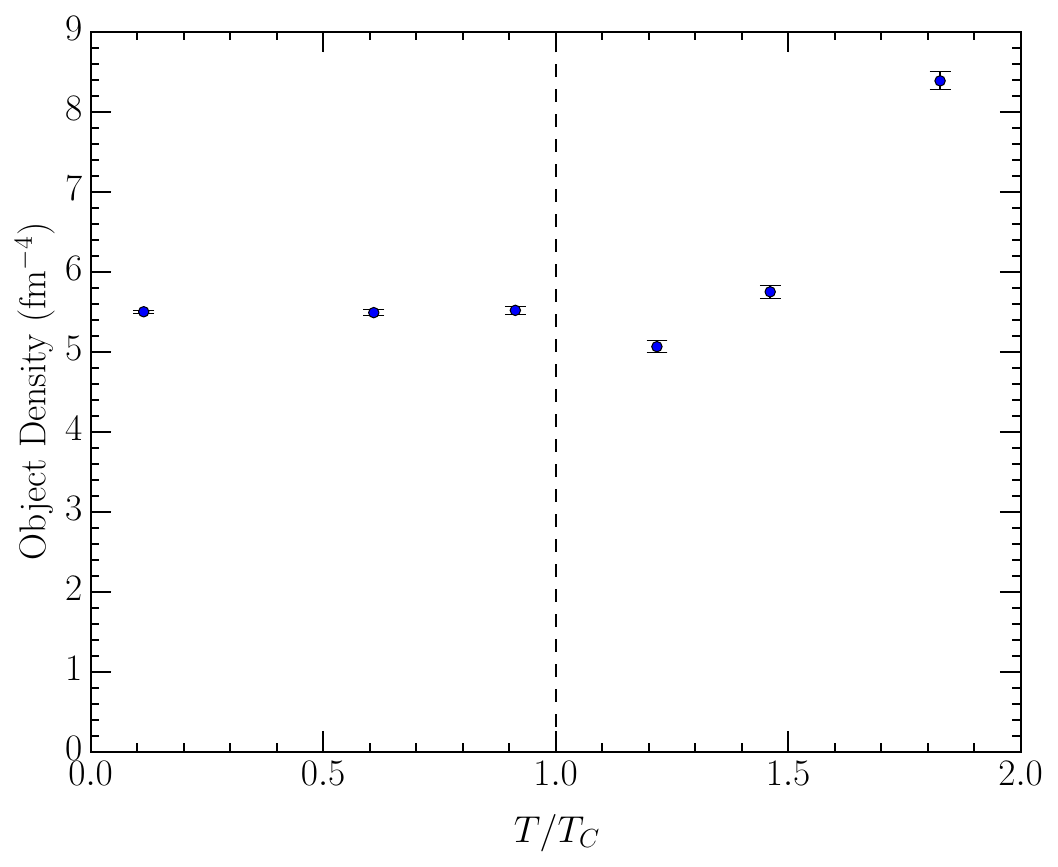}
    \caption{\label{fig:objectdensity} The object number density under hypercubic dislocation filtering for each $32^3\times N_t$ ensemble, calculated at the justified flow time $\tau=1.45$. There appears to be an initial drop in $n$ as we cross $T_c$, though it increases thereafter.}
\end{figure}

The number of objects also gives access to the object density, which we can compare across each temperature. For our purposes, we define the number density as
\begin{align}
    n = \frac{N}{V} \,, && V = (aN_s)^3 \times (aN_t) \,,
\end{align}
where $N$ is the total number of objects, and $V$ is the (physical) four-dimensional volume of the lattice. To resolve the slight ambiguity between the two topological charge definitions, as seen in Fig.~\ref{fig:NobjSmoothing}, we take the average of the two values as a unified estimate of the density, and combine half the difference with the statistical error in quadrature. The evolution of $n$ with temperature is shown in Fig.~\ref{fig:objectdensity}.

Whilst the density is found to be constant below the critical temperature, there is interesting behaviour above $T_c$. We observe a slight drop in the density as the phase transition is crossed, but it subsequently increases as the temperature climbs away from $T_c$. This points towards an increase in ``activity" in the gauge fields at very high temperatures.

However, we must note that the scaling of $n$ as $a\to 0$ depends on how one takes the continuum limit. The precise value of $n$ utilising the fixed lattice dislocation filter method is dependent on the lattice spacing. For instance, calculating $n$ with the hypercubic dislocation filter on our finer ($a=0.067\,$fm) ensemble gives $n\approx 17.46(37)$, over a factor of $3$ larger than the coarse $a=0.10\,$fm result of $5.498(19)$. Clearly, the topological objects being physically smaller allows them to be more densely packed. Alternatively, in the fixed scale continuum limit, the density remains insensitive to the lattice spacing. This motivates further investigation into the object density at a broader range of temperatures and lattice spacings.

\subsection{Root-mean-square radius} \label{subsec:rmsradius}
\begin{figure*}
    \centering
    \includegraphics[width=0.48\linewidth]{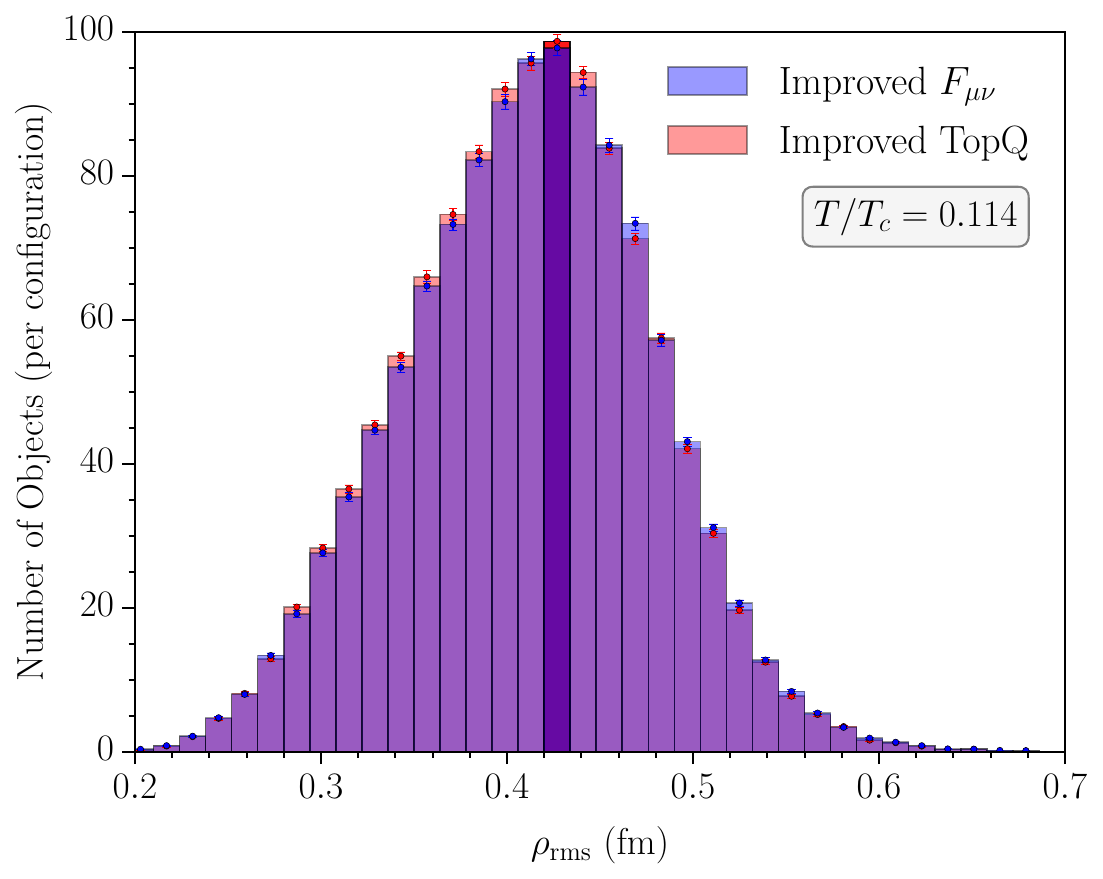}
    \includegraphics[width=0.48\linewidth]{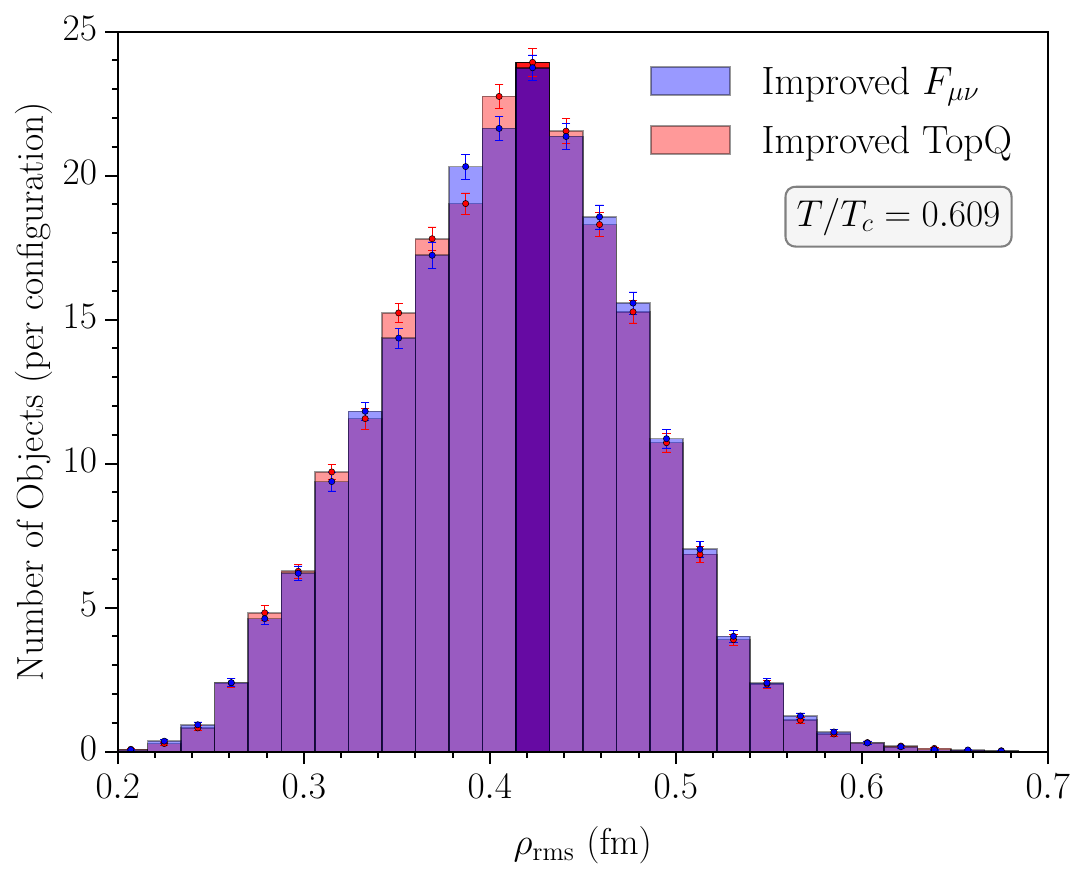}
    \includegraphics[width=0.48\linewidth]{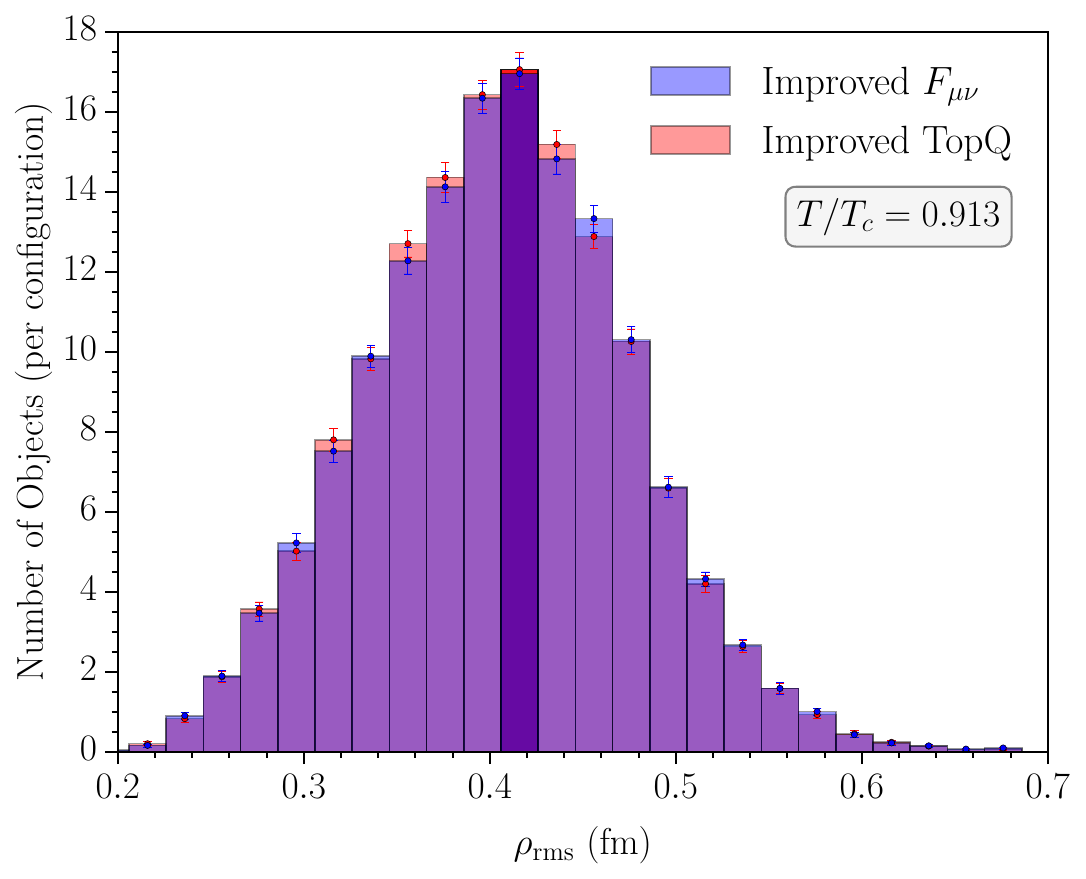}
    \includegraphics[width=0.48\linewidth]{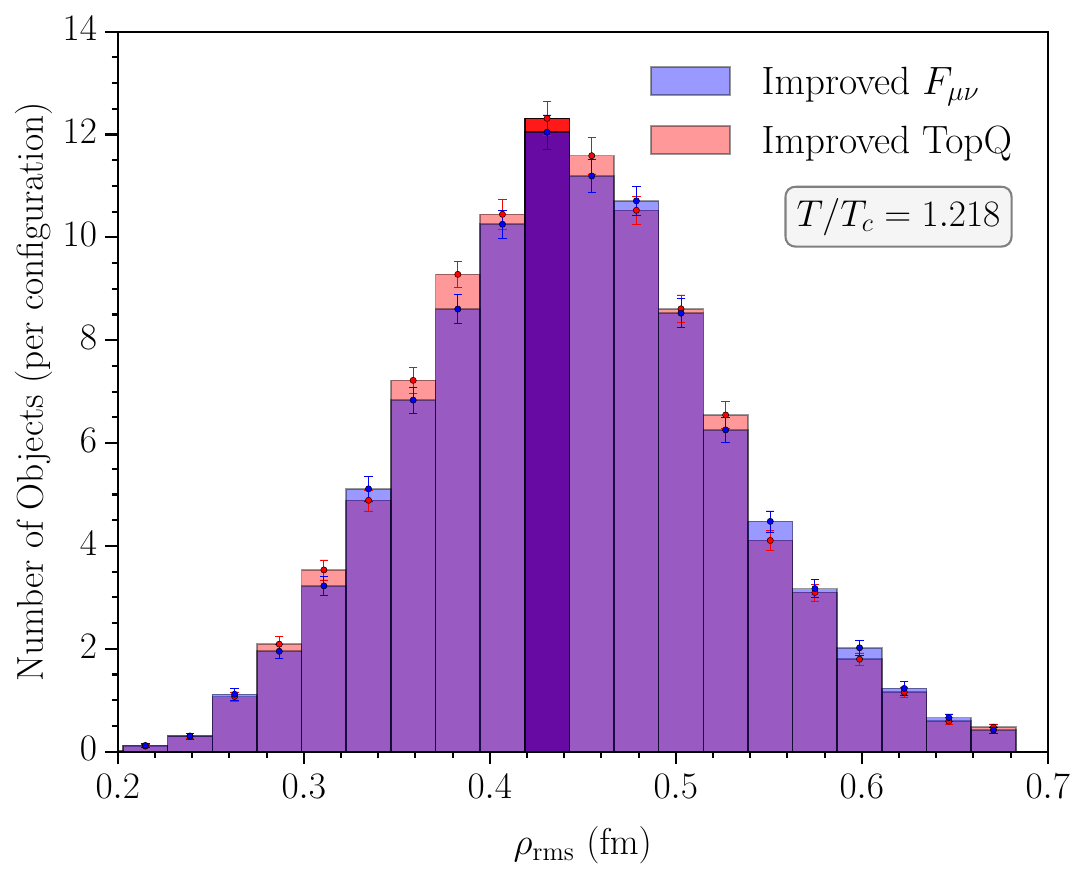}
    \includegraphics[width=0.48\linewidth]{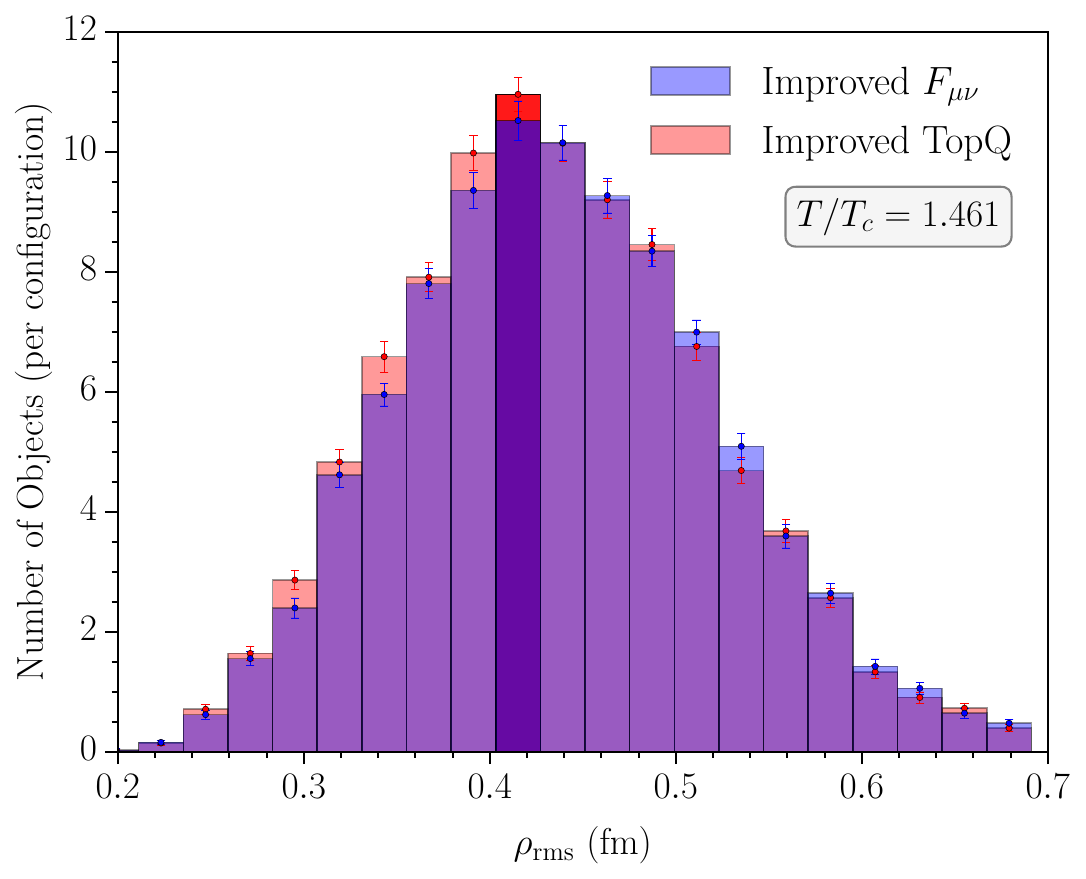}
    \includegraphics[width=0.48\linewidth]{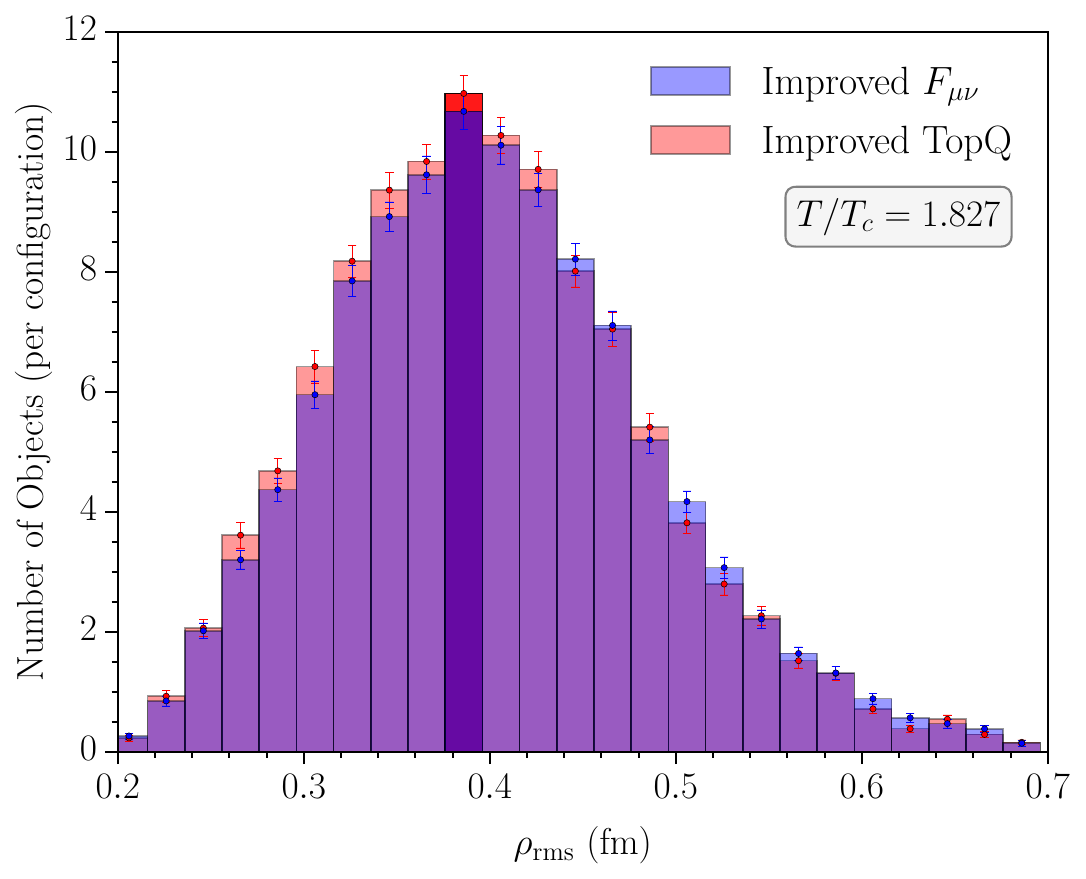}
    \caption{\label{fig:rmsradiusresults} Histograms showing the normalised RMS topological charge density radii for objects resolved on each of our finite-temperature ensembles: $N_t = 64$ (\textbf{top left}), $12$ (\textbf{top right}), $8$ (\textbf{middle left}), $6$ (\textbf{middle right}), $5$ (\textbf{bottom left}) and $4$ (\textbf{bottom right}). There are very minimal changes to the distribution as temperature is increased. The results are calculated with a hypercube dislocation filter after a flow time $\tau=1.45$, the amount required to ensure consistency between the topological charge modes of our two improvement schemes.}
\end{figure*}

Next, we investigate any variation in the radial size of the topological objects with temperature using the RMS radius defined in Eq.~(\ref{eq:rmsradius}). These histograms are displayed in Fig.~\ref{fig:rmsradiusresults}.
In direct contrast with the charges, we find there is minimal change in the distribution of radial sizes, besides a plausible shift at the highest temperature where the mode drops from $\approx 0.43$\,fm to $\approx 0.39$\,fm. Nevertheless, this could simply be a consequence of statistical variability, with the smaller lattice volumes above $T_c$ by default having less statistics. Either way, the drastic drop in charge values cannot be primarily accounted for by a corresponding decrease in radial size. For instance, taking the decrease at face value, the fractional reduction in volume taken up by a typical object would be $(0.39/0.43)^4 \approx 0.67$, which fails to account for the factor of $\approx 1/3$ cut in charge values by the highest temperature. Instead, the shift in charge values is likely due predominantly to a reduction in the height of the peaks in the topological charge. In other words, vacuum field fluctuations are suppressed at high temperature.

\subsection{Visualisations} \label{subsec:vis}
\begin{figure}
    \centering
    \includegraphics[width=\linewidth]{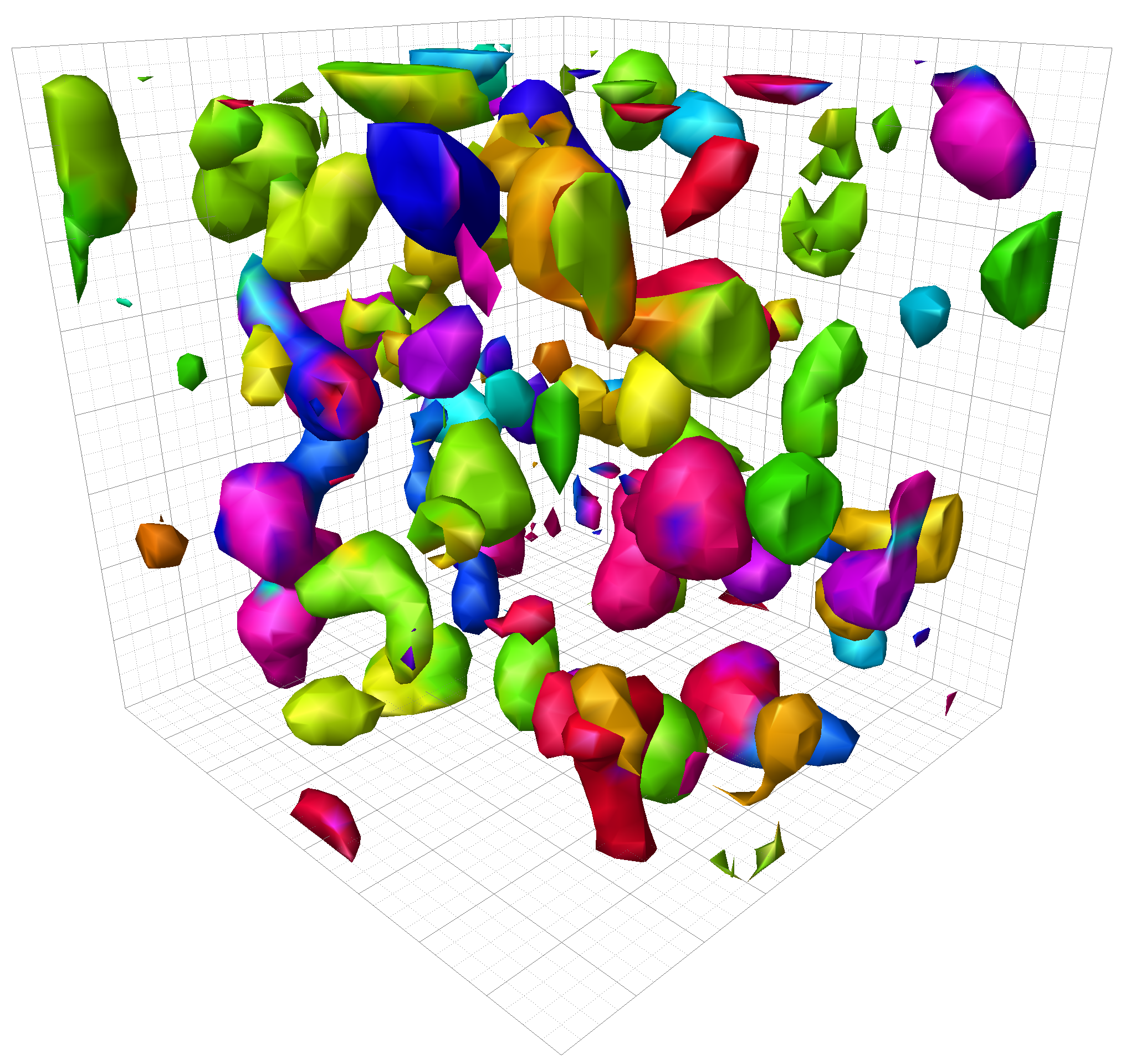}
    \includegraphics[width=\linewidth]{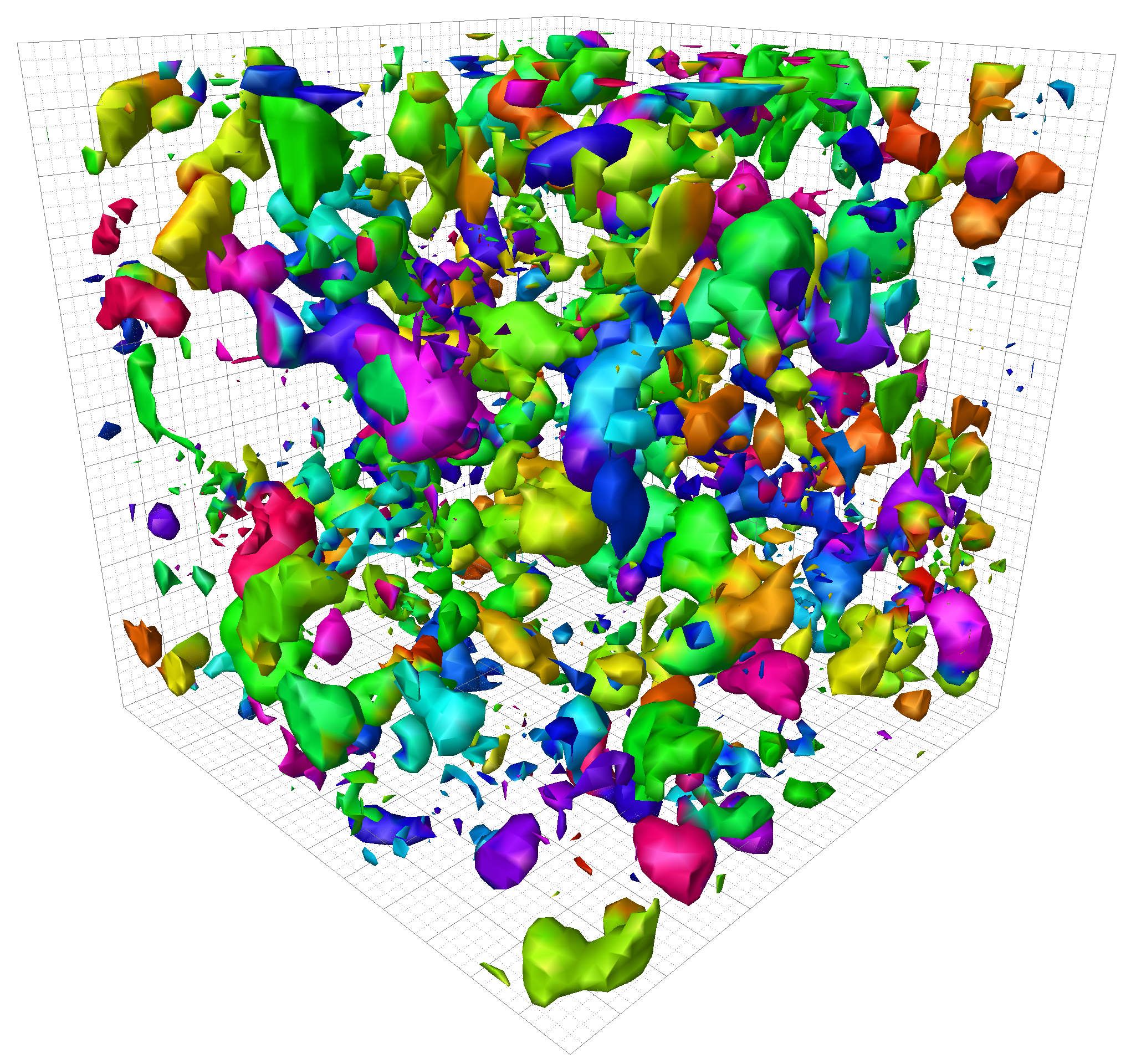}
    \caption{\label{fig:fullvis} Visualisations of the topological charge density on the $32^3 \times 64$ ensemble with $a = 0.10$\,fm (\textbf{top}) and the $48^3 \times 96$ ensemble with $a = 0.067$\,fm (\textbf{bottom}), produced at the flow times used to calculate our results with a hypercubic dislocation filter. The visualisations are constructed identically to Fig.~\ref{fig:classicalvis}, with the additional caveat that any points \textit{not} assigned an object number (due to dislocation filtering) are not rendered. Even though the topological charge density is coloured by the object number, due to the high number of objects in the four-dimensional volume the colours (shades of grey) are often very similar. The topological charge is visually more granular at the smaller lattice spacing, with a greater density of (radially) smaller objects despite the similar object charge distributions of Fig.~\ref{fig:hypercube}.}.
\end{figure}

As a final point of discussion, we visualise the topological charge density for the two different lattice spacings, allowing further insight into the nature of the algorithm. The visualisations are produced at the respective levels of smoothing for the fixed hypercube dislocation filter, under which there are interesting changes to the vacuum structure. This also qualitatively reveals the changes to the gauge field as the lattice spacing is decreased identified in Secs.~\ref{subsec:fixedcutoff} and \ref{subsec:numberofobjects}. A single temporal slice is presented for each lattice spacing in Fig.~\ref{fig:fullvis}.
In both cases, the effectiveness of the algorithm in capturing the behaviour around each peak is evident, with most ``lumps" consisting of a single colour. The instances of overlapping objects are also convincingly managed, with visible boundaries dividing the individual peaks. From these visualisations, it is clear we can be confident in the calculated numerical charge values as reliable estimates of the topological charge distribution.

Additionally, there is a substantial shift in the topological structure as the lattice spacing is decreased, consistent with the quantitative findings on object density and radial size. For one, the number of distinct objects has significantly increased, which matches the comment on the increase in object density in Sec.~\ref{subsec:numberofobjects} as the lattice spacing is decreased. Second, the lumps in the bottom visualisation are, on average, unarguably smaller in size than the top figure, coinciding with the decrease in radial size found in Sec.~\ref{subsec:fixedcutoff} by considering an RMS estimate of each object's radius. Clearly, shrinking the lattice spacing admits the existence of radially smaller topological features (but with the same net charges, as illustrated in Fig.~\ref{fig:hypercube}) which tend to comprise the volume, producing the increase in abundance of individual topological objects observed in Fig.~\ref{fig:fullvis}.

\section{Conclusion} \label{sec:conclusion}
In this work we have devised a novel method for identifying and calculating the net topological charge contained within topological objects. This was utilised to explore the evolution of the topological structure of SU(3) gauge theory with temperature. We obtained a distribution of charge values which was interpreted as quantum fluctuations around a semiclassical value, considered as the mode of the distribution. This was taken as an indicator of the underlying topological features.

The results exhibit a foundational consistency with the instanton-dyon model \cite{InstantonDyonsI, InstantonDyonsII, InstantonDyonsIII, InstantonDyonsIV} for the topological structure of SU($N$) gauge theory at finite temperature. They reveal distributions peaked near $\approx 1/3$ in the confined phase, and which decrease above the critical temperature in a trend matching the single free holonomy parameter in SU(3).

The lack of self-duality on the analysed configurations leaves our results open to a variety of fractional topological objects, such as $\mathbb{Z}_N$ dyons \cite{FractionalSUN} which are predicted to arise from quantum fluctuations of an effective action. Future work can explore the action at the individual-object level to provide a more detailed assessment of the types of objects present in the ground-state gluon fields.

We intend to extend the work presented here to SU($N$) gauge theory for $N\neq 3$, with a focus on $N=2$ and $4$, to determine whether there is any discernible change between the various numbers of colours and if it follows the instanton-dyon prediction for that particular value of $N$.

Insight into the large-scale topological structure of the Yang-Mills vacuum can also be obtained through eigenmodes of the Dirac operator, which can isolate semiclassical features of the gauge fields as an alternative to smoothing \cite{Eigenmodes}. Although previously studied in the fundamental representation, a recent analysis \cite{AFMAnalysis} has provided evidence in favour of eigenmodes in the adjoint representation, needing only a small number of modes to reconstruct the (semiclassical) topological charge density. The ``adjoint filtering method" (AFM) \cite{AFMI,AFMII} is used to filter out the UV fluctuations in the configurations, which has benefits and drawbacks compared to smoothing. For instance, the AFM has been shown to capture instanton and anti-instanton pairs which would otherwise annihilate under smoothing, though on occasion it also misses objects that are revealed under smoothing \cite{AFMAnalysis}. Looking forward, an especially quantitative approach could involve applying the algorithm presented here to the eigenmodes identified through the AFM. It will be interesting to learn if the reduced structure seen on individual eigenmodes offers any quantitative advantage in the process of identifying objects in the QCD ground-state fields.

\begin{acknowledgments}
This work was supported with supercomputing resources provided by the Phoenix High Performance Computing (HPC) service at the University of Adelaide. This research was undertaken with the assistance of resources and services from the National Computational Infrastructure (NCI), which is supported by the Australian Government. This research was supported by the Australian Research Council through Grant No. DP210103706. W.~K. was supported by the Pawsey Supercomputing Centre through the Pawsey Centre for Extreme Scale Readiness (PaCER) program.
\end{acknowledgments}

\appendix

\section{Cooling procedure} \label{app:cooling}
In this appendix, we summarise the cooling used to test the algorithm. Standard cooling starts with the local Wilson action \cite{WilsonAction} corresponding to the link $U_\mu(x)$,
\begin{align}
S(x) &= \beta \sum_{\nu\neq\mu} \left[1 - \frac{1}{3}\Re\tr\big(U_\mu(x) \Sigma_{\mu\nu}(x) \big)\right] \,, \label{eq:localaction} \\
\begin{split}
    \Sigma_{\mu\nu}(x) &= U_\nu(x+\hat{\mu}) U_\mu^\dagger(x+\hat{\nu}) U_\nu^\dagger(x) \\
    &\hphantom{=} + U_\nu^\dagger(x+\hat{\mu}-\hat{\nu}) U_\mu^\dagger(x-\hat{\nu}) U_\nu(x-\hat{\nu}) \,,
\end{split}
\end{align}
where $\Sigma_{\mu\nu}$ is the sum of the two $1\times 1$ staples in the $\mu$-$\nu$ plane touching $U_\mu(x)$. It is clear from Eq.~(\ref{eq:localaction}) that the local action will be minimised when
\begin{equation} \label{eq:coolingcriteria}
    \Re\tr \left[U_\mu(x) \sum_{\nu\neq\mu} \Sigma_{\mu\nu}(x) \right]
\end{equation}
is maximised, and hence cooling aims to update $U_\mu(x)$ to a new link such that Eq.~(\ref{eq:coolingcriteria}) is maximised. For our purposes, we implement improved cooling by including the contributions of the $1\times 1$, $2\times 2$ and $3\times 3$ loops touching $U_\mu(x)$ in the calculation of the staple sum,
\begin{equation}
    \begin{split} \label{eq:Oa4ImprovedStaples}
        \Sigma_{\mu\nu} &= \frac{3}{2} \sum \big(1\times 1 \text{ paths} \big) - \frac{3}{80} \sum \big(2\times 2 \text{ paths} \big) \\
        &+\, \frac{1}{810} \sum \big(3\times 3 \text{ paths} \big) \,.
    \end{split}
\end{equation}
This combination corresponds to minimising an $\mathcal{O}(a^4)$-improved discretised action operator \cite{ImprovedActionTopQ}.

In SU(2), this update is a straightforward projection of the staple sum onto SU(2),
\begin{equation}
\begin{split} \label{eq:su2cooling}
    U_\mu(x) \longrightarrow U'_\mu(x) &= \Bigg(\frac{1}{k} \sum_{\nu\neq\mu} \Sigma_{\mu\nu}(x) \Bigg)^{\!\dagger} \,, \\
    k^2 &= \det \sum_{\nu\neq\mu} \Sigma_{\mu\nu}(x) \,.
\end{split}
\end{equation}
The strategy in SU(3) is more involved, but is achieved based on the Cabibbo-Marinari pseudo-heat bath algorithm \cite{CabibboMarinari} by iterating over SU(2) subgroups and performing the above projection step for each one \cite{SU3SmoothingCalibration}. First, the staples are closed through multiplication by the link $U_\mu(x)$,
\begin{equation}
    \Omega = U_\mu(x) \sum_{\nu\neq\mu} \Sigma_{\mu\nu}(x) \,.
\end{equation}
One then defines the below $2\times 2$ submatrix from the upper-left block of the $3\times 3$ matrix $\Omega$,
\begin{equation}
    \Omega_{2\times 2} = \frac{1}{2} \begin{bmatrix}
        \Omega_{11} + \Omega_{22}^* & \Omega_{12} - \Omega_{21}^* \\
        \Omega_{21} - \Omega_{12}^* & \Omega_{11}^* + \Omega_{22}
    \end{bmatrix} \,.
\end{equation}
This is cooled $\Omega_{2\times 2} \longrightarrow \Omega_{2\times 2}'$ according to Eq.~(\ref{eq:su2cooling}) and embedded in the upper-left $2\times 2$ block of an SU(3) matrix $\Omega'$. Finally, the link is updated via
\begin{equation}
    U_\mu(x) \longrightarrow U_\mu'(x) = \Omega' U_\mu(x) \,.
\end{equation}
This process is subsequently repeated for the other two diagonal SU(2) subgroups, thereby covering SU(3). Additionally, it is possible to perform multiple loops over the three SU(2) subgroups for a single update of $U_\mu(x)$ to provide better convergence to the true link which maximises Eq.~(\ref{eq:coolingcriteria}) \cite{ImprovedSmoothing}. We perform eight such cycles per update, which has previously been justified as a suitable choice to optimise the projected link $U_\mu'(x)$ \cite{CoolingLoops}.

\bibliography{main}

\providecommand{\noopsort}[1]{}\providecommand{\singleletter}[1]{#1}%
\begin{thebibliography}{97}%
\makeatletter
\providecommand \@ifxundefined [1]{%
 \@ifx{#1\undefined}
}%
\providecommand \@ifnum [1]{%
 \ifnum #1\expandafter \@firstoftwo
 \else \expandafter \@secondoftwo
 \fi
}%
\providecommand \@ifx [1]{%
 \ifx #1\expandafter \@firstoftwo
 \else \expandafter \@secondoftwo
 \fi
}%
\providecommand \natexlab [1]{#1}%
\providecommand \enquote  [1]{``#1''}%
\providecommand \bibnamefont  [1]{#1}%
\providecommand \bibfnamefont [1]{#1}%
\providecommand \citenamefont [1]{#1}%
\providecommand \href@noop [0]{\@secondoftwo}%
\providecommand \href [0]{\begingroup \@sanitize@url \@href}%
\providecommand \@href[1]{\@@startlink{#1}\@@href}%
\providecommand \@@href[1]{\endgroup#1\@@endlink}%
\providecommand \@sanitize@url [0]{\catcode `\\12\catcode `\$12\catcode
  `\&12\catcode `\#12\catcode `\^12\catcode `\_12\catcode `\%12\relax}%
\providecommand \@@startlink[1]{}%
\providecommand \@@endlink[0]{}%
\providecommand \url  [0]{\begingroup\@sanitize@url \@url }%
\providecommand \@url [1]{\endgroup\@href {#1}{\urlprefix }}%
\providecommand \urlprefix  [0]{URL }%
\providecommand \Eprint [0]{\href }%
\providecommand \doibase [0]{https://doi.org/}%
\providecommand \selectlanguage [0]{\@gobble}%
\providecommand \bibinfo  [0]{\@secondoftwo}%
\providecommand \bibfield  [0]{\@secondoftwo}%
\providecommand \translation [1]{[#1]}%
\providecommand \BibitemOpen [0]{}%
\providecommand \bibitemStop [0]{}%
\providecommand \bibitemNoStop [0]{.\EOS\space}%
\providecommand \EOS [0]{\spacefactor3000\relax}%
\providecommand \BibitemShut  [1]{\csname bibitem#1\endcsname}%
\let\auto@bib@innerbib\@empty
\bibitem [{\citenamefont {Dosch}\ and\ \citenamefont
  {Simonov}(1988)}]{WilsonLoopAreaLaw}%
  \BibitemOpen
  \bibfield  {author} {\bibinfo {author} {\bibfnamefont {H.~G.}\ \bibnamefont
  {Dosch}}\ and\ \bibinfo {author} {\bibfnamefont {Y.~A.}\ \bibnamefont
  {Simonov}},\ }\bibfield  {title} {\bibinfo {title} {{The area law of the
  Wilson loop and vacuum field correlators}},\ }\href
  {https://doi.org/10.1016/0370-2693(88)91675-9} {\bibfield  {journal}
  {\bibinfo  {journal} {Phys. Lett. B}\ }\textbf {\bibinfo {volume} {205}},\
  \bibinfo {pages} {339} (\bibinfo {year} {1988})}\BibitemShut {NoStop}%
\bibitem [{\citenamefont {Bali}\ \emph {et~al.}(2005)\citenamefont {Bali},
  \citenamefont {Neff}, \citenamefont {Duessel}, \citenamefont {Lippert},\ and\
  \citenamefont {Schilling}}]{StringBreaking}%
  \BibitemOpen
  \bibfield  {author} {\bibinfo {author} {\bibfnamefont {G.~S.}\ \bibnamefont
  {Bali}}, \bibinfo {author} {\bibfnamefont {H.}~\bibnamefont {Neff}}, \bibinfo
  {author} {\bibfnamefont {T.}~\bibnamefont {Duessel}}, \bibinfo {author}
  {\bibfnamefont {T.}~\bibnamefont {Lippert}},\ and\ \bibinfo {author}
  {\bibfnamefont {K.}~\bibnamefont {Schilling}} (\bibinfo {collaboration}
  {SESAM Collaboration}),\ }\bibfield  {title} {\bibinfo {title} {{Observation
  of string breaking in QCD}},\ }\href
  {https://doi.org/10.1103/PhysRevD.71.114513} {\bibfield  {journal} {\bibinfo
  {journal} {Phys. Rev. D}\ }\textbf {\bibinfo {volume} {71}},\ \bibinfo
  {pages} {114513} (\bibinfo {year} {2005})},\ \Eprint
  {https://arxiv.org/abs/hep-lat/0505012} {arXiv:hep-lat/0505012} \BibitemShut
  {NoStop}%
\bibitem [{\citenamefont {Aubin}\ and\ \citenamefont
  {Ogilvie}(2003)}]{GluonPropConfinementI}%
  \BibitemOpen
  \bibfield  {author} {\bibinfo {author} {\bibfnamefont {C.~A.}\ \bibnamefont
  {Aubin}}\ and\ \bibinfo {author} {\bibfnamefont {M.~C.}\ \bibnamefont
  {Ogilvie}},\ }\bibfield  {title} {\bibinfo {title} {{Lattice gauge fixing and
  the violation of spectral positivity}},\ }\href
  {https://doi.org/10.1016/j.physletb.2003.07.034} {\bibfield  {journal}
  {\bibinfo  {journal} {Phys. Lett. B}\ }\textbf {\bibinfo {volume} {570}},\
  \bibinfo {pages} {59} (\bibinfo {year} {2003})},\ \Eprint
  {https://arxiv.org/abs/hep-lat/0306012} {arXiv:hep-lat/0306012} \BibitemShut
  {NoStop}%
\bibitem [{\citenamefont {Bowman}\ \emph {et~al.}(2007)\citenamefont {Bowman},
  \citenamefont {Heller}, \citenamefont {Leinweber}, \citenamefont
  {Parappilly}, \citenamefont {Sternbeck}, \citenamefont {von Smekal},
  \citenamefont {Williams},\ and\ \citenamefont
  {Zhang}}]{GluonPropConfinementII}%
  \BibitemOpen
  \bibfield  {author} {\bibinfo {author} {\bibfnamefont {P.~O.}\ \bibnamefont
  {Bowman}}, \bibinfo {author} {\bibfnamefont {U.~M.}\ \bibnamefont {Heller}},
  \bibinfo {author} {\bibfnamefont {D.~B.}\ \bibnamefont {Leinweber}}, \bibinfo
  {author} {\bibfnamefont {M.~B.}\ \bibnamefont {Parappilly}}, \bibinfo
  {author} {\bibfnamefont {A.}~\bibnamefont {Sternbeck}}, \bibinfo {author}
  {\bibfnamefont {L.}~\bibnamefont {von Smekal}}, \bibinfo {author}
  {\bibfnamefont {A.~G.}\ \bibnamefont {Williams}},\ and\ \bibinfo {author}
  {\bibfnamefont {J.-b.}\ \bibnamefont {Zhang}},\ }\bibfield  {title} {\bibinfo
  {title} {{Scaling behavior and positivity violation of the gluon propagator
  in full QCD}},\ }\href {https://doi.org/10.1103/PhysRevD.76.094505}
  {\bibfield  {journal} {\bibinfo  {journal} {Phys. Rev. D}\ }\textbf {\bibinfo
  {volume} {76}},\ \bibinfo {pages} {094505} (\bibinfo {year} {2007})},\
  \Eprint {https://arxiv.org/abs/hep-lat/0703022} {arXiv:hep-lat/0703022}
  \BibitemShut {NoStop}%
\bibitem [{\citenamefont {Biddle}\ \emph {et~al.}(2022)\citenamefont {Biddle},
  \citenamefont {Kamleh},\ and\ \citenamefont
  {Leinweber}}]{GluonPropConfinementIII}%
  \BibitemOpen
  \bibfield  {author} {\bibinfo {author} {\bibfnamefont {J.~C.}\ \bibnamefont
  {Biddle}}, \bibinfo {author} {\bibfnamefont {W.}~\bibnamefont {Kamleh}},\
  and\ \bibinfo {author} {\bibfnamefont {D.~B.}\ \bibnamefont {Leinweber}},\
  }\bibfield  {title} {\bibinfo {title} {{Impact of dynamical fermions on the
  center vortex gluon propagator}},\ }\href
  {https://doi.org/10.1103/PhysRevD.106.014506} {\bibfield  {journal} {\bibinfo
   {journal} {Phys. Rev. D}\ }\textbf {\bibinfo {volume} {106}},\ \bibinfo
  {pages} {014506} (\bibinfo {year} {2022})},\ \Eprint
  {https://arxiv.org/abs/2206.02320} {arXiv:2206.02320 [hep-lat]} \BibitemShut
  {NoStop}%
\bibitem [{\citenamefont {Belavin}\ \emph {et~al.}(1975)\citenamefont
  {Belavin}, \citenamefont {Polyakov}, \citenamefont {Schwartz},\ and\
  \citenamefont {Tyupkin}}]{YangMillsSolution}%
  \BibitemOpen
  \bibfield  {author} {\bibinfo {author} {\bibfnamefont {A.~A.}\ \bibnamefont
  {Belavin}}, \bibinfo {author} {\bibfnamefont {A.~M.}\ \bibnamefont
  {Polyakov}}, \bibinfo {author} {\bibfnamefont {A.~S.}\ \bibnamefont
  {Schwartz}},\ and\ \bibinfo {author} {\bibfnamefont {Y.~S.}\ \bibnamefont
  {Tyupkin}},\ }\bibfield  {title} {\bibinfo {title} {{Pseudoparticle solutions
  of the Yang-Mills equations}},\ }\href
  {https://doi.org/10.1016/0370-2693(75)90163-X} {\bibfield  {journal}
  {\bibinfo  {journal} {Phys. Lett. B}\ }\textbf {\bibinfo {volume} {59}},\
  \bibinfo {pages} {85} (\bibinfo {year} {1975})}\BibitemShut {NoStop}%
\bibitem [{\citenamefont
  {Shuryak}(1982{\natexlab{a}})}]{InstantonLiquidModelI}%
  \BibitemOpen
  \bibfield  {author} {\bibinfo {author} {\bibfnamefont {E.~V.}\ \bibnamefont
  {Shuryak}},\ }\bibfield  {title} {\bibinfo {title} {{The role of instantons
  in quantum chromodynamics: (I). Physical vacuum}},\ }\href
  {https://doi.org/10.1016/0550-3213(82)90478-3} {\bibfield  {journal}
  {\bibinfo  {journal} {Nucl. Phys. B}\ }\textbf {\bibinfo {volume} {203}},\
  \bibinfo {pages} {93} (\bibinfo {year} {1982}{\natexlab{a}})}\BibitemShut
  {NoStop}%
\bibitem [{\citenamefont
  {Shuryak}(1982{\natexlab{b}})}]{InstantonLiquidModelII}%
  \BibitemOpen
  \bibfield  {author} {\bibinfo {author} {\bibfnamefont {E.~V.}\ \bibnamefont
  {Shuryak}},\ }\bibfield  {title} {\bibinfo {title} {{The role of instantons
  in quantum chromodynamics: (II). Hadronic structure}},\ }\href
  {https://doi.org/10.1016/0550-3213(82)90479-5} {\bibfield  {journal}
  {\bibinfo  {journal} {Nucl. Phys. B}\ }\textbf {\bibinfo {volume} {203}},\
  \bibinfo {pages} {116} (\bibinfo {year} {1982}{\natexlab{b}})}\BibitemShut
  {NoStop}%
\bibitem [{\citenamefont
  {Shuryak}(1982{\natexlab{c}})}]{InstantonLiquidModelIII}%
  \BibitemOpen
  \bibfield  {author} {\bibinfo {author} {\bibfnamefont {E.~V.}\ \bibnamefont
  {Shuryak}},\ }\bibfield  {title} {\bibinfo {title} {{The role of instantons
  in quantum chromodynamics: (III). Quark-gluon plasma}},\ }\href
  {https://doi.org/10.1016/0550-3213(82)90480-1} {\bibfield  {journal}
  {\bibinfo  {journal} {Nucl. Phys. B}\ }\textbf {\bibinfo {volume} {203}},\
  \bibinfo {pages} {140} (\bibinfo {year} {1982}{\natexlab{c}})}\BibitemShut
  {NoStop}%
\bibitem [{\citenamefont {Shuryak}(1989)}]{ILMConfinementI}%
  \BibitemOpen
  \bibfield  {author} {\bibinfo {author} {\bibfnamefont {E.~V.}\ \bibnamefont
  {Shuryak}},\ }\bibfield  {title} {\bibinfo {title} {{Instantons in {QCD}:
  (III). Quark propagators and mesons containing heavy quarks}},\ }\href
  {https://doi.org/10.1016/0550-3213(89)90093-X} {\bibfield  {journal}
  {\bibinfo  {journal} {Nucl. Phys. B}\ }\textbf {\bibinfo {volume} {328}},\
  \bibinfo {pages} {85} (\bibinfo {year} {1989})}\BibitemShut {NoStop}%
\bibitem [{\citenamefont {Diakonov}\ \emph {et~al.}(1989)\citenamefont
  {Diakonov}, \citenamefont {Petrov},\ and\ \citenamefont
  {Pobylitsa}}]{ILMConfinementII}%
  \BibitemOpen
  \bibfield  {author} {\bibinfo {author} {\bibfnamefont {D.}~\bibnamefont
  {Diakonov}}, \bibinfo {author} {\bibfnamefont {V.~Y.}\ \bibnamefont
  {Petrov}},\ and\ \bibinfo {author} {\bibfnamefont {P.~V.}\ \bibnamefont
  {Pobylitsa}},\ }\bibfield  {title} {\bibinfo {title} {{The Wilson loop and
  heavy-quark potential in the instanton vacuum}},\ }\href
  {https://doi.org/10.1016/0370-2693(89)91213-6} {\bibfield  {journal}
  {\bibinfo  {journal} {Phys. Lett. B}\ }\textbf {\bibinfo {volume} {226}},\
  \bibinfo {pages} {372} (\bibinfo {year} {1989})}\BibitemShut {NoStop}%
\bibitem [{\citenamefont {Sch\"afer}\ and\ \citenamefont
  {Shuryak}(1998)}]{ILMConfinementIII}%
  \BibitemOpen
  \bibfield  {author} {\bibinfo {author} {\bibfnamefont {T.}~\bibnamefont
  {Sch\"afer}}\ and\ \bibinfo {author} {\bibfnamefont {E.~V.}\ \bibnamefont
  {Shuryak}},\ }\bibfield  {title} {\bibinfo {title} {{Instantons in QCD}},\
  }\href {https://doi.org/10.1103/RevModPhys.70.323} {\bibfield  {journal}
  {\bibinfo  {journal} {Rev. Mod. Phys.}\ }\textbf {\bibinfo {volume} {70}},\
  \bibinfo {pages} {323} (\bibinfo {year} {1998})},\ \Eprint
  {https://arxiv.org/abs/hep-ph/9610451} {arXiv:hep-ph/9610451} \BibitemShut
  {NoStop}%
\bibitem [{\citenamefont {'t~Hooft}(1981)}]{TwistedSelfdual}%
  \BibitemOpen
  \bibfield  {author} {\bibinfo {author} {\bibfnamefont {G.}~\bibnamefont
  {'t~Hooft}},\ }\bibfield  {title} {\bibinfo {title} {{Some twisted self-dual
  solutions for the Yang-Mills equations on a hypertorus}},\ }\href
  {https://doi.org/10.1007/BF01208900} {\bibfield  {journal} {\bibinfo
  {journal} {Commun. Math. Phys.}\ }\textbf {\bibinfo {volume} {81}},\ \bibinfo
  {pages} {267} (\bibinfo {year} {1981})}\BibitemShut {NoStop}%
\bibitem [{\citenamefont {Gonz\'alez-Arroyo}\ \emph {et~al.}(1995)\citenamefont
  {Gonz\'alez-Arroyo}, \citenamefont {Mart\'inez},\ and\ \citenamefont
  {Montero}}]{TwistedSelfdualCoolingI}%
  \BibitemOpen
  \bibfield  {author} {\bibinfo {author} {\bibfnamefont {A.}~\bibnamefont
  {Gonz\'alez-Arroyo}}, \bibinfo {author} {\bibfnamefont {P.}~\bibnamefont
  {Mart\'inez}},\ and\ \bibinfo {author} {\bibfnamefont {A.}~\bibnamefont
  {Montero}},\ }\bibfield  {title} {\bibinfo {title} {{Gauge invariant
  structures and confinement}},\ }\href
  {https://doi.org/10.1016/0370-2693(95)01056-V} {\bibfield  {journal}
  {\bibinfo  {journal} {Phys. Lett. B}\ }\textbf {\bibinfo {volume} {359}},\
  \bibinfo {pages} {159} (\bibinfo {year} {1995})},\ \Eprint
  {https://arxiv.org/abs/hep-lat/9507006} {arXiv:hep-lat/9507006} \BibitemShut
  {NoStop}%
\bibitem [{\citenamefont {Gonz\'alez-Arroyo}\ and\ \citenamefont
  {Mart\'inez}(1996)}]{TwistedSelfdualCoolingII}%
  \BibitemOpen
  \bibfield  {author} {\bibinfo {author} {\bibfnamefont {A.}~\bibnamefont
  {Gonz\'alez-Arroyo}}\ and\ \bibinfo {author} {\bibfnamefont {P.}~\bibnamefont
  {Mart\'inez}},\ }\bibfield  {title} {\bibinfo {title} {{Investigating
  Yang-Mills theory and confinement as a function of the spatial volume}},\
  }\href {https://doi.org/10.1016/0550-3213(95)00601-X} {\bibfield  {journal}
  {\bibinfo  {journal} {Nucl. Phys. B}\ }\textbf {\bibinfo {volume} {459}},\
  \bibinfo {pages} {337} (\bibinfo {year} {1996})},\ \Eprint
  {https://arxiv.org/abs/hep-lat/9507001} {arXiv:hep-lat/9507001} \BibitemShut
  {NoStop}%
\bibitem [{\citenamefont {Garc\'ia~P\'erez}\ \emph {et~al.}(1990)\citenamefont
  {Garc\'ia~P\'erez}, \citenamefont {Gonz\'alez-Arroyo},\ and\ \citenamefont
  {S{\"o}derberg}}]{TwistedSelfdualStudyI}%
  \BibitemOpen
  \bibfield  {author} {\bibinfo {author} {\bibfnamefont {M.}~\bibnamefont
  {Garc\'ia~P\'erez}}, \bibinfo {author} {\bibfnamefont {A.}~\bibnamefont
  {Gonz\'alez-Arroyo}},\ and\ \bibinfo {author} {\bibfnamefont
  {B.}~\bibnamefont {S{\"o}derberg}},\ }\bibfield  {title} {\bibinfo {title}
  {{Minimum action solutions for SU(2) gauge theory on the torus with
  non-orthogonal twist}},\ }\href
  {https://doi.org/10.1016/0370-2693(90)90106-G} {\bibfield  {journal}
  {\bibinfo  {journal} {Phys. Lett. B}\ }\textbf {\bibinfo {volume} {235}},\
  \bibinfo {pages} {117} (\bibinfo {year} {1990})}\BibitemShut {NoStop}%
\bibitem [{\citenamefont {Garc\'ia~P\'erez}\ and\ \citenamefont
  {Gonz\'alez-Arroyo}(1993)}]{TwistedSelfdualStudyII}%
  \BibitemOpen
  \bibfield  {author} {\bibinfo {author} {\bibfnamefont {M.}~\bibnamefont
  {Garc\'ia~P\'erez}}\ and\ \bibinfo {author} {\bibfnamefont {A.}~\bibnamefont
  {Gonz\'alez-Arroyo}},\ }\bibfield  {title} {\bibinfo {title} {{Numerical
  study of Yang-Mills classical solutions on the twisted torus}},\ }\href
  {https://doi.org/10.1088/0305-4470/26/11/015} {\bibfield  {journal} {\bibinfo
   {journal} {J. Phys. A}\ }\textbf {\bibinfo {volume} {26}},\ \bibinfo {pages}
  {2667} (\bibinfo {year} {1993})},\ \Eprint
  {https://arxiv.org/abs/hep-lat/9206016} {arXiv:hep-lat/9206016} \BibitemShut
  {NoStop}%
\bibitem [{\citenamefont {Garc\'ia~P\'erez}\ \emph {et~al.}(1998)\citenamefont
  {Garc\'ia~P\'erez}, \citenamefont {Gonz\'alez-Arroyo}, \citenamefont
  {Montero},\ and\ \citenamefont {Pena}}]{TwistedSelfdualStudyIII}%
  \BibitemOpen
  \bibfield  {author} {\bibinfo {author} {\bibfnamefont {M.}~\bibnamefont
  {Garc\'ia~P\'erez}}, \bibinfo {author} {\bibfnamefont {A.}~\bibnamefont
  {Gonz\'alez-Arroyo}}, \bibinfo {author} {\bibfnamefont {A.}~\bibnamefont
  {Montero}},\ and\ \bibinfo {author} {\bibfnamefont {C.}~\bibnamefont
  {Pena}},\ }\bibfield  {title} {\bibinfo {title} {{Yang-Mills classical
  solutions and fermionic zero modes from lattice calculations}},\ }\href
  {https://doi.org/10.1016/S0920-5632(97)00814-1} {\bibfield  {journal}
  {\bibinfo  {journal} {Nucl. Phys. B Proc. Suppl.}\ }\textbf {\bibinfo
  {volume} {63}},\ \bibinfo {pages} {501} (\bibinfo {year} {1998})},\ \Eprint
  {https://arxiv.org/abs/hep-lat/9709107} {arXiv:hep-lat/9709107} \BibitemShut
  {NoStop}%
\bibitem [{\citenamefont
  {Montero}(2000{\natexlab{a}})}]{TwistedSelfdualStudyIV}%
  \BibitemOpen
  \bibfield  {author} {\bibinfo {author} {\bibfnamefont {A.}~\bibnamefont
  {Montero}},\ }\bibfield  {title} {\bibinfo {title} {{Numerical analysis of
  fractional charge solutions on the torus}},\ }\href
  {https://doi.org/10.1088/1126-6708/2000/05/022} {\bibfield  {journal}
  {\bibinfo  {journal} {J. High Energy Phys.}\ }\textbf {\bibinfo {volume}
  {05}},\ \bibinfo {pages} {022} (\bibinfo {year} {2000})},\ \Eprint
  {https://arxiv.org/abs/hep-lat/0004009} {arXiv:hep-lat/0004009} \BibitemShut
  {NoStop}%
\bibitem [{\citenamefont {Gonz\'alez-Arroyo}(2023)}]{FractionalReview}%
  \BibitemOpen
  \bibfield  {author} {\bibinfo {author} {\bibfnamefont {A.}~\bibnamefont
  {Gonz\'alez-Arroyo}},\ }\bibfield  {title} {\bibinfo {title} {{On the
  fractional instanton liquid picture of the Yang-Mills vacuum and
  Confinement}},\ }\href@noop {} {\  (\bibinfo {year} {2023})},\ \Eprint
  {https://arxiv.org/abs/2302.12356} {arXiv:2302.12356 [hep-th]} \BibitemShut
  {NoStop}%
\bibitem [{\citenamefont
  {Gonz\'alez-Arroyo}(2020)}]{FractionalGeneralisationI}%
  \BibitemOpen
  \bibfield  {author} {\bibinfo {author} {\bibfnamefont {A.}~\bibnamefont
  {Gonz\'alez-Arroyo}},\ }\bibfield  {title} {\bibinfo {title} {{Constructing
  SU({$N$}) fractional instantons}},\ }\href
  {https://doi.org/10.1007/JHEP02(2020)137} {\bibfield  {journal} {\bibinfo
  {journal} {J. High Energy Phys.}\ }\textbf {\bibinfo {volume} {02}},\
  \bibinfo {pages} {137} (\bibinfo {year} {2020})},\ \Eprint {https://arxiv.org/abs/1910.12565}
  {arXiv:1910.12565 [hep-th]} \BibitemShut {NoStop}%
\bibitem [{\citenamefont {Anber}\ and\ \citenamefont
  {Poppitz}(2023)}]{FractionalGeneralisationII}%
  \BibitemOpen
  \bibfield  {author} {\bibinfo {author} {\bibfnamefont {M.~M.}\ \bibnamefont
  {Anber}}\ and\ \bibinfo {author} {\bibfnamefont {E.}~\bibnamefont
  {Poppitz}},\ }\bibfield  {title} {\bibinfo {title} {{Multi-fractional
  instantons in SU({$N$}) Yang-Mills theory on the twisted {$\mathbb{T}^4$}}},\
  }\href {https://doi.org/10.1007/JHEP09(2023)095} {\bibfield  {journal}
  {\bibinfo  {journal} {J. High Energy Phys.}\ }\textbf {\bibinfo {volume}
  {09}},\ \bibinfo {pages} {095} (\bibinfo {year} {2023})},\ \Eprint {https://arxiv.org/abs/2307.04795}
  {arXiv:2307.04795 [hep-th]} \BibitemShut {NoStop}%
\bibitem [{\citenamefont {Lee}(1998)}]{InstantonDyonsI}%
  \BibitemOpen
  \bibfield  {author} {\bibinfo {author} {\bibfnamefont {K.-M.}\ \bibnamefont
  {Lee}},\ }\bibfield  {title} {\bibinfo {title} {{Instantons and magnetic
  monopoles on {$R^3 \times S^1$} with arbitrary simple gauge groups}},\ }\href
  {https://doi.org/10.1016/S0370-2693(98)00283-4} {\bibfield  {journal}
  {\bibinfo  {journal} {Phys. Lett. B}\ }\textbf {\bibinfo {volume} {426}},\
  \bibinfo {pages} {323} (\bibinfo {year} {1998})},\ \Eprint
  {https://arxiv.org/abs/hep-th/9802012} {arXiv:hep-th/9802012} \BibitemShut
  {NoStop}%
\bibitem [{\citenamefont {Lee}\ and\ \citenamefont
  {Lu}(1998)}]{InstantonDyonsII}%
  \BibitemOpen
  \bibfield  {author} {\bibinfo {author} {\bibfnamefont {K.-M.}\ \bibnamefont
  {Lee}}\ and\ \bibinfo {author} {\bibfnamefont {C.-h.}\ \bibnamefont {Lu}},\
  }\bibfield  {title} {\bibinfo {title} {{SU(2) calorons and magnetic
  monopoles}},\ }\href {https://doi.org/10.1103/PhysRevD.58.025011} {\bibfield
  {journal} {\bibinfo  {journal} {Phys. Rev. D}\ }\textbf {\bibinfo {volume}
  {58}},\ \bibinfo {pages} {025011} (\bibinfo {year} {1998})},\ \Eprint
  {https://arxiv.org/abs/hep-th/9802108} {arXiv:hep-th/9802108} \BibitemShut
  {NoStop}%
\bibitem [{\citenamefont {Kraan}\ and\ \citenamefont {van
  Baal}(1998{\natexlab{a}})}]{InstantonDyonsIII}%
  \BibitemOpen
  \bibfield  {author} {\bibinfo {author} {\bibfnamefont {T.~C.}\ \bibnamefont
  {Kraan}}\ and\ \bibinfo {author} {\bibfnamefont {P.}~\bibnamefont {van
  Baal}},\ }\bibfield  {title} {\bibinfo {title} {{Monopole constituents inside
  {$SU(n)$} calorons}},\ }\href {https://doi.org/10.1016/S0370-2693(98)00799-0}
  {\bibfield  {journal} {\bibinfo  {journal} {Phys. Lett. B}\ }\textbf
  {\bibinfo {volume} {435}},\ \bibinfo {pages} {389} (\bibinfo {year}
  {1998}{\natexlab{a}})},\ \Eprint {https://arxiv.org/abs/hep-th/9806034}
  {arXiv:hep-th/9806034} \BibitemShut {NoStop}%
\bibitem [{\citenamefont {Kraan}\ and\ \citenamefont {van
  Baal}(1998{\natexlab{b}})}]{InstantonDyonsIV}%
  \BibitemOpen
  \bibfield  {author} {\bibinfo {author} {\bibfnamefont {T.~C.}\ \bibnamefont
  {Kraan}}\ and\ \bibinfo {author} {\bibfnamefont {P.}~\bibnamefont {van
  Baal}},\ }\bibfield  {title} {\bibinfo {title} {{Periodic instantons with
  non-trivial holonomy}},\ }\href
  {https://doi.org/10.1016/S0550-3213(98)00590-2} {\bibfield  {journal}
  {\bibinfo  {journal} {Nucl. Phys. B}\ }\textbf {\bibinfo {volume} {533}},\
  \bibinfo {pages} {627} (\bibinfo {year} {1998}{\natexlab{b}})},\ \Eprint
  {https://arxiv.org/abs/hep-th/9805168} {arXiv:hep-th/9805168} \BibitemShut
  {NoStop}%
\bibitem [{\citenamefont {Diakonov}\ and\ \citenamefont
  {Petrov}(2007)}]{ConfiningDyons}%
  \BibitemOpen
  \bibfield  {author} {\bibinfo {author} {\bibfnamefont {D.}~\bibnamefont
  {Diakonov}}\ and\ \bibinfo {author} {\bibfnamefont {V.}~\bibnamefont
  {Petrov}},\ }\bibfield  {title} {\bibinfo {title} {{Confining ensemble of
  dyons}},\ }\href {https://doi.org/10.1103/PhysRevD.76.056001} {\bibfield
  {journal} {\bibinfo  {journal} {Phys. Rev. D}\ }\textbf {\bibinfo {volume}
  {76}},\ \bibinfo {pages} {056001} (\bibinfo {year} {2007})},\ \Eprint
  {https://arxiv.org/abs/0704.3181} {arXiv:0704.3181 [hep-th]} \BibitemShut
  {NoStop}%
\bibitem [{\citenamefont {Diakonov}\ and\ \citenamefont
  {Petrov}(2011)}]{DyonConfinementDeconfinement}%
  \BibitemOpen
  \bibfield  {author} {\bibinfo {author} {\bibfnamefont {D.}~\bibnamefont
  {Diakonov}}\ and\ \bibinfo {author} {\bibfnamefont {V.}~\bibnamefont
  {Petrov}},\ }\bibfield  {title} {\bibinfo {title} {{Confinement and
  deconfinement for any gauge group from dyons viewpoint}},\ }\href
  {https://doi.org/10.1063/1.3574944} {\bibfield  {journal} {\bibinfo
  {journal} {AIP Conf. Proc.}\ }\textbf {\bibinfo {volume} {1343}},\ \bibinfo
  {pages} {69} (\bibinfo {year} {2011})},\ \Eprint
  {https://arxiv.org/abs/1011.5636} {arXiv:1011.5636 [hep-th]} \BibitemShut
  {NoStop}%
\bibitem [{\citenamefont {Liu}\ \emph {et~al.}(2015)\citenamefont {Liu},
  \citenamefont {Shuryak},\ and\ \citenamefont {Zahed}}]{DyonLiquidModel}%
  \BibitemOpen
  \bibfield  {author} {\bibinfo {author} {\bibfnamefont {Y.}~\bibnamefont
  {Liu}}, \bibinfo {author} {\bibfnamefont {E.}~\bibnamefont {Shuryak}},\ and\
  \bibinfo {author} {\bibfnamefont {I.}~\bibnamefont {Zahed}},\ }\bibfield
  {title} {\bibinfo {title} {{Confining dyon-antidyon Coulomb liquid model.
  I.}},\ }\href {https://doi.org/10.1103/PhysRevD.92.085006} {\bibfield
  {journal} {\bibinfo  {journal} {Phys. Rev. D}\ }\textbf {\bibinfo {volume}
  {92}},\ \bibinfo {pages} {085006} (\bibinfo {year} {2015})},\ \Eprint
  {https://arxiv.org/abs/1503.03058} {arXiv:1503.03058 [hep-ph]} \BibitemShut
  {NoStop}%
\bibitem [{\citenamefont {Shuryak}\ and\ \citenamefont
  {Sulejmanpasic}(2013)}]{SU2DyonEnsemble}%
  \BibitemOpen
  \bibfield  {author} {\bibinfo {author} {\bibfnamefont {E.}~\bibnamefont
  {Shuryak}}\ and\ \bibinfo {author} {\bibfnamefont {T.}~\bibnamefont
  {Sulejmanpasic}},\ }\bibfield  {title} {\bibinfo {title} {{Holonomy potential
  and confinement from a simple model of the gauge topology}},\ }\href
  {https://doi.org/10.1016/j.physletb.2013.08.014} {\bibfield  {journal}
  {\bibinfo  {journal} {Phys. Lett. B}\ }\textbf {\bibinfo {volume} {726}},\
  \bibinfo {pages} {257} (\bibinfo {year} {2013})},\ \Eprint
  {https://arxiv.org/abs/1305.0796} {arXiv:1305.0796 [hep-ph]} \BibitemShut
  {NoStop}%
\bibitem [{\citenamefont {Larsen}\ and\ \citenamefont
  {Shuryak}(2015)}]{SU2DyonConfinementI}%
  \BibitemOpen
  \bibfield  {author} {\bibinfo {author} {\bibfnamefont {R.}~\bibnamefont
  {Larsen}}\ and\ \bibinfo {author} {\bibfnamefont {E.}~\bibnamefont
  {Shuryak}},\ }\bibfield  {title} {\bibinfo {title} {{Interacting ensemble of
  the instanton-dyons and the deconfinement phase transition in the SU(2) gauge
  theory}},\ }\href {https://doi.org/10.1103/PhysRevD.92.094022} {\bibfield
  {journal} {\bibinfo  {journal} {Phys. Rev. D}\ }\textbf {\bibinfo {volume}
  {92}},\ \bibinfo {pages} {094022} (\bibinfo {year} {2015})},\ \Eprint
  {https://arxiv.org/abs/1504.03341} {arXiv:1504.03341 [hep-ph]} \BibitemShut
  {NoStop}%
\bibitem [{\citenamefont {Lopez-Ruiz}\ \emph {et~al.}(2018)\citenamefont
  {Lopez-Ruiz}, \citenamefont {Jiang},\ and\ \citenamefont
  {Liao}}]{SU2DyonConfinementII}%
  \BibitemOpen
  \bibfield  {author} {\bibinfo {author} {\bibfnamefont {M.~A.}\ \bibnamefont
  {Lopez-Ruiz}}, \bibinfo {author} {\bibfnamefont {Y.}~\bibnamefont {Jiang}},\
  and\ \bibinfo {author} {\bibfnamefont {J.}~\bibnamefont {Liao}},\ }\bibfield
  {title} {\bibinfo {title} {{Confinement, holonomy and correlated
  instanton-dyon ensemble: SU(2) Yang-Mills theory}},\ }\href
  {https://doi.org/10.1103/PhysRevD.97.054026} {\bibfield  {journal} {\bibinfo
  {journal} {Phys. Rev. D}\ }\textbf {\bibinfo {volume} {97}},\ \bibinfo
  {pages} {054026} (\bibinfo {year} {2018})},\ \Eprint
  {https://arxiv.org/abs/1611.02539} {arXiv:1611.02539 [hep-ph]} \BibitemShut
  {NoStop}%
\bibitem [{\citenamefont {Lopez-Ruiz}\ \emph {et~al.}(2019)\citenamefont
  {Lopez-Ruiz}, \citenamefont {Jiang},\ and\ \citenamefont
  {Liao}}]{SU2DyonConfinementIII}%
  \BibitemOpen
  \bibfield  {author} {\bibinfo {author} {\bibfnamefont {M.~A.}\ \bibnamefont
  {Lopez-Ruiz}}, \bibinfo {author} {\bibfnamefont {Y.}~\bibnamefont {Jiang}},\
  and\ \bibinfo {author} {\bibfnamefont {J.}~\bibnamefont {Liao}},\ }\bibfield
  {title} {\bibinfo {title} {{Confinement from correlated instanton-dyon
  ensemble in SU(2) Yang-Mills theory}},\ }\href
  {https://doi.org/10.1103/PhysRevD.99.114013} {\bibfield  {journal} {\bibinfo
  {journal} {Phys. Rev. D}\ }\textbf {\bibinfo {volume} {99}},\ \bibinfo
  {pages} {114013} (\bibinfo {year} {2019})},\ \Eprint
  {https://arxiv.org/abs/1903.02684} {arXiv:1903.02684 [hep-ph]} \BibitemShut
  {NoStop}%
\bibitem [{\citenamefont {DeMartini}\ and\ \citenamefont
  {Shuryak}(2021)}]{SU3DyonConfinement}%
  \BibitemOpen
  \bibfield  {author} {\bibinfo {author} {\bibfnamefont {D.}~\bibnamefont
  {DeMartini}}\ and\ \bibinfo {author} {\bibfnamefont {E.}~\bibnamefont
  {Shuryak}},\ }\bibfield  {title} {\bibinfo {title} {{Deconfinement phase
  transition in the {$SU(3)$} instanton-dyon ensemble}},\ }\href
  {https://doi.org/10.1103/PhysRevD.104.054010} {\bibfield  {journal} {\bibinfo
   {journal} {Phys. Rev. D}\ }\textbf {\bibinfo {volume} {104}},\ \bibinfo
  {pages} {054010} (\bibinfo {year} {2021})},\ \Eprint
  {https://arxiv.org/abs/2102.11321} {arXiv:2102.11321 [hep-ph]} \BibitemShut
  {NoStop}%
\bibitem [{\citenamefont {Gonz\'alez-Arroyo}\ and\ \citenamefont
  {Montero}(1998)}]{FractionalVortexI}%
  \BibitemOpen
  \bibfield  {author} {\bibinfo {author} {\bibfnamefont {A.}~\bibnamefont
  {Gonz\'alez-Arroyo}}\ and\ \bibinfo {author} {\bibfnamefont {A.}~\bibnamefont
  {Montero}},\ }\bibfield  {title} {\bibinfo {title} {{Self-dual vortex-like
  configurations in SU(2) Yang-Mills theory}},\ }\href
  {https://doi.org/10.1016/S0370-2693(98)01229-5} {\bibfield  {journal}
  {\bibinfo  {journal} {Phys. Lett. B}\ }\textbf {\bibinfo {volume} {442}},\
  \bibinfo {pages} {273} (\bibinfo {year} {1998})},\ \Eprint
  {https://arxiv.org/abs/hep-th/9809037} {arXiv:hep-th/9809037} \BibitemShut
  {NoStop}%
\bibitem [{\citenamefont {Montero}(1999)}]{FractionalVortexII}%
  \BibitemOpen
  \bibfield  {author} {\bibinfo {author} {\bibfnamefont {A.}~\bibnamefont
  {Montero}},\ }\bibfield  {title} {\bibinfo {title} {{Study of SU(3)
  vortex-like configurations with a new maximal center gauge fixing method}},\
  }\href {https://doi.org/10.1016/S0370-2693(99)01113-2} {\bibfield  {journal}
  {\bibinfo  {journal} {Phys. Lett. B}\ }\textbf {\bibinfo {volume} {467}},\
  \bibinfo {pages} {106} (\bibinfo {year} {1999})},\ \Eprint
  {https://arxiv.org/abs/hep-lat/9906010} {arXiv:hep-lat/9906010} \BibitemShut
  {NoStop}%
\bibitem [{\citenamefont {Montero}(2000{\natexlab{b}})}]{FractionalVortexIII}%
  \BibitemOpen
  \bibfield  {author} {\bibinfo {author} {\bibfnamefont {A.}~\bibnamefont
  {Montero}},\ }\bibfield  {title} {\bibinfo {title} {{Vortex configurations in
  the large {$N$} limit}},\ }\href
  {https://doi.org/10.1016/S0370-2693(00)00572-4} {\bibfield  {journal}
  {\bibinfo  {journal} {Phys. Lett. B}\ }\textbf {\bibinfo {volume} {483}},\
  \bibinfo {pages} {309} (\bibinfo {year} {2000}{\natexlab{b}})},\ \Eprint
  {https://arxiv.org/abs/hep-lat/0004002} {arXiv:hep-lat/0004002} \BibitemShut
  {NoStop}%
\bibitem [{\citenamefont {Dasilva~Gol\'an}\ and\ \citenamefont
  {Garc\'\i{}a~P\'erez}(2022)}]{FractionalFibonacci}%
  \BibitemOpen
  \bibfield  {author} {\bibinfo {author} {\bibfnamefont {J.}~\bibnamefont
  {Dasilva~Gol\'an}}\ and\ \bibinfo {author} {\bibfnamefont {M.}~\bibnamefont
  {Garc\'\i{}a~P\'erez}},\ }\bibfield  {title} {\bibinfo {title} {{SU({$N$})
  fractional instantons and the Fibonacci sequence}},\ }\href
  {https://doi.org/10.1007/JHEP12(2022)109} {\bibfield  {journal} {\bibinfo
  {journal} {J. High Energy Phys.}\ }\textbf {\bibinfo {volume} {12}},\
  \bibinfo {pages} {109} (\bibinfo {year} {2022})},\ \Eprint {https://arxiv.org/abs/2208.07133}
  {arXiv:2208.07133 [hep-th]} \BibitemShut {NoStop}%
\bibitem [{\citenamefont {Nair}\ and\ \citenamefont
  {Pisarski}(2023)}]{FractionalSUN}%
  \BibitemOpen
  \bibfield  {author} {\bibinfo {author} {\bibfnamefont {V.~P.}\ \bibnamefont
  {Nair}}\ and\ \bibinfo {author} {\bibfnamefont {R.~D.}\ \bibnamefont
  {Pisarski}},\ }\bibfield  {title} {\bibinfo {title} {{Fractional topological
  charge in {$SU(N)$} gauge theories without dynamical quarks}},\ }\href
  {https://doi.org/10.1103/PhysRevD.108.074007} {\bibfield  {journal} {\bibinfo
   {journal} {Phys. Rev. D}\ }\textbf {\bibinfo {volume} {108}},\ \bibinfo
  {pages} {074007} (\bibinfo {year} {2023})},\ \Eprint
  {https://arxiv.org/abs/2206.11284} {arXiv:2206.11284 [hep-th]} \BibitemShut
  {NoStop}%
\bibitem [{\citenamefont {Ilgenfritz}\ \emph {et~al.}(2005)\citenamefont
  {Ilgenfritz}, \citenamefont {Martemyanov}, \citenamefont
  {M{\"u}ller-Preussker},\ and\ \citenamefont {Veselov}}]{ThresholdAnalysisI}%
  \BibitemOpen
  \bibfield  {author} {\bibinfo {author} {\bibfnamefont {E.~M.}\ \bibnamefont
  {Ilgenfritz}}, \bibinfo {author} {\bibfnamefont {B.~V.}\ \bibnamefont
  {Martemyanov}}, \bibinfo {author} {\bibfnamefont {M.}~\bibnamefont
  {M{\"u}ller-Preussker}},\ and\ \bibinfo {author} {\bibfnamefont {A.~I.}\
  \bibnamefont {Veselov}},\ }\bibfield  {title} {\bibinfo {title} {{Monopole
  content of topological clusters: Have Kraan-van Baal calorons been found?}},\
  }\href {https://doi.org/10.1103/PhysRevD.71.034505} {\bibfield  {journal}
  {\bibinfo  {journal} {Phys. Rev. D}\ }\textbf {\bibinfo {volume} {71}},\
  \bibinfo {pages} {034505} (\bibinfo {year} {2005})},\ \Eprint
  {https://arxiv.org/abs/hep-lat/0412028} {arXiv:hep-lat/0412028} \BibitemShut
  {NoStop}%
\bibitem [{\citenamefont {Bornyakov}\ \emph {et~al.}(2007)\citenamefont
  {Bornyakov}, \citenamefont {Ilgenfritz}, \citenamefont {Martemyanov},
  \citenamefont {Morozov}, \citenamefont {M{\"u}ller-Preussker},\ and\
  \citenamefont {Veselov}}]{ThresholdAnalysisII}%
  \BibitemOpen
  \bibfield  {author} {\bibinfo {author} {\bibfnamefont {V.~G.}\ \bibnamefont
  {Bornyakov}}, \bibinfo {author} {\bibfnamefont {E.~M.}\ \bibnamefont
  {Ilgenfritz}}, \bibinfo {author} {\bibfnamefont {B.~V.}\ \bibnamefont
  {Martemyanov}}, \bibinfo {author} {\bibfnamefont {S.~M.}\ \bibnamefont
  {Morozov}}, \bibinfo {author} {\bibfnamefont {M.}~\bibnamefont
  {M{\"u}ller-Preussker}},\ and\ \bibinfo {author} {\bibfnamefont {A.~I.}\
  \bibnamefont {Veselov}},\ }\bibfield  {title} {\bibinfo {title} {{Calorons
  and dyons at the thermal phase transition analyzed by overlap fermions}},\
  }\href {https://doi.org/10.1103/PhysRevD.76.054505} {\bibfield  {journal}
  {\bibinfo  {journal} {Phys. Rev. D}\ }\textbf {\bibinfo {volume} {76}},\
  \bibinfo {pages} {054505} (\bibinfo {year} {2007})},\ \Eprint
  {https://arxiv.org/abs/0706.4206} {arXiv:0706.4206 [hep-lat]} \BibitemShut
  {NoStop}%
\bibitem [{\citenamefont {Bornyakov}\ \emph {et~al.}(2009)\citenamefont
  {Bornyakov}, \citenamefont {Ilgenfritz}, \citenamefont {Martemyanov},\ and\
  \citenamefont {M{\"u}ller-Preussker}}]{ThresholdAnalysisIII}%
  \BibitemOpen
  \bibfield  {author} {\bibinfo {author} {\bibfnamefont {V.~G.}\ \bibnamefont
  {Bornyakov}}, \bibinfo {author} {\bibfnamefont {E.~M.}\ \bibnamefont
  {Ilgenfritz}}, \bibinfo {author} {\bibfnamefont {B.~V.}\ \bibnamefont
  {Martemyanov}},\ and\ \bibinfo {author} {\bibfnamefont {M.}~\bibnamefont
  {M{\"u}ller-Preussker}},\ }\bibfield  {title} {\bibinfo {title} {{Dyonic
  picture of topological objects in the deconfined phase}},\ }\href
  {https://doi.org/10.1103/PhysRevD.79.034506} {\bibfield  {journal} {\bibinfo
  {journal} {Phys. Rev. D}\ }\textbf {\bibinfo {volume} {79}},\ \bibinfo
  {pages} {034506} (\bibinfo {year} {2009})},\ \Eprint
  {https://arxiv.org/abs/0809.2142} {arXiv:0809.2142 [hep-lat]} \BibitemShut
  {NoStop}%
\bibitem [{\citenamefont {Bornyakov}\ \emph {et~al.}(2015)\citenamefont
  {Bornyakov}, \citenamefont {Ilgenfritz}, \citenamefont {Martemyanov},\ and\
  \citenamefont {M{\"u}ller-Preussker}}]{ThresholdAnalysisIV}%
  \BibitemOpen
  \bibfield  {author} {\bibinfo {author} {\bibfnamefont {V.~G.}\ \bibnamefont
  {Bornyakov}}, \bibinfo {author} {\bibfnamefont {E.~M.}\ \bibnamefont
  {Ilgenfritz}}, \bibinfo {author} {\bibfnamefont {B.~V.}\ \bibnamefont
  {Martemyanov}},\ and\ \bibinfo {author} {\bibfnamefont {M.}~\bibnamefont
  {M{\"u}ller-Preussker}},\ }\bibfield  {title} {\bibinfo {title} {{Dyon
  structures in the deconfinement phase of lattice gluodynamics: Topological
  clusters, holonomies, and Abelian monopoles}},\ }\href
  {https://doi.org/10.1103/PhysRevD.91.074505} {\bibfield  {journal} {\bibinfo
  {journal} {Phys. Rev. D}\ }\textbf {\bibinfo {volume} {91}},\ \bibinfo
  {pages} {074505} (\bibinfo {year} {2015})},\ \Eprint
  {https://arxiv.org/abs/1410.4632} {arXiv:1410.4632 [hep-lat]} \BibitemShut
  {NoStop}%
\bibitem [{\citenamefont {Ilgenfritz}\ \emph {et~al.}(2014)\citenamefont
  {Ilgenfritz}, \citenamefont {Martemyanov},\ and\ \citenamefont
  {M\"uller-Preussker}}]{ThresholdAnalysisV}%
  \BibitemOpen
  \bibfield  {author} {\bibinfo {author} {\bibfnamefont {E.~M.}\ \bibnamefont
  {Ilgenfritz}}, \bibinfo {author} {\bibfnamefont {B.~V.}\ \bibnamefont
  {Martemyanov}},\ and\ \bibinfo {author} {\bibfnamefont {M.}~\bibnamefont
  {M\"uller-Preussker}},\ }\bibfield  {title} {\bibinfo {title} {{Topology near
  the transition temperature in lattice gluodynamics analyzed by low lying
  modes of the overlap Dirac operator}},\ }\href
  {https://doi.org/10.1103/PhysRevD.89.054503} {\bibfield  {journal} {\bibinfo
  {journal} {Phys. Rev. D}\ }\textbf {\bibinfo {volume} {89}},\ \bibinfo
  {pages} {054503} (\bibinfo {year} {2014})},\ \Eprint
  {https://arxiv.org/abs/1309.7850} {arXiv:1309.7850 [hep-lat]} \BibitemShut
  {NoStop}%
\bibitem [{\citenamefont {Bornyakov}\ \emph {et~al.}(2016)\citenamefont
  {Bornyakov}, \citenamefont {Ilgenfritz}, \citenamefont {Martemyanov},\ and\
  \citenamefont {M\"uller-Preussker}}]{ThresholdAnalysisVI}%
  \BibitemOpen
  \bibfield  {author} {\bibinfo {author} {\bibfnamefont {V.~G.}\ \bibnamefont
  {Bornyakov}}, \bibinfo {author} {\bibfnamefont {E.~M.}\ \bibnamefont
  {Ilgenfritz}}, \bibinfo {author} {\bibfnamefont {B.~V.}\ \bibnamefont
  {Martemyanov}},\ and\ \bibinfo {author} {\bibfnamefont {M.}~\bibnamefont
  {M\"uller-Preussker}},\ }\bibfield  {title} {\bibinfo {title} {{Dyons near
  the transition temperature in lattice QCD}},\ }\href
  {https://doi.org/10.1103/PhysRevD.93.074508} {\bibfield  {journal} {\bibinfo
  {journal} {Phys. Rev. D}\ }\textbf {\bibinfo {volume} {93}},\ \bibinfo
  {pages} {074508} (\bibinfo {year} {2016})},\ \Eprint
  {https://arxiv.org/abs/1512.03217} {arXiv:1512.03217 [hep-lat]} \BibitemShut
  {NoStop}%
\bibitem [{\citenamefont {Bornyakov}\ \emph {et~al.}(2018)\citenamefont
  {Bornyakov}, \citenamefont {Ilgenfritz},\ and\ \citenamefont
  {Martemyanov}}]{ThresholdAnalysisVII}%
  \BibitemOpen
  \bibfield  {author} {\bibinfo {author} {\bibfnamefont {V.~G.}\ \bibnamefont
  {Bornyakov}}, \bibinfo {author} {\bibfnamefont {E.~M.}\ \bibnamefont
  {Ilgenfritz}},\ and\ \bibinfo {author} {\bibfnamefont {B.~V.}\ \bibnamefont
  {Martemyanov}},\ }\bibfield  {title} {\bibinfo {title} {{Dyons near the
  transition temperature in $SU(3)$ lattice gluodynamics}},\ }\href
  {https://doi.org/10.1088/1361-6471/aab796} {\bibfield  {journal} {\bibinfo
  {journal} {J. Phys. G}\ }\textbf {\bibinfo {volume} {45}},\ \bibinfo {pages}
  {055006} (\bibinfo {year} {2018})},\ \Eprint
  {https://arxiv.org/abs/1711.04476} {arXiv:1711.04476 [hep-lat]} \BibitemShut
  {NoStop}%
\bibitem [{\citenamefont {Bonnet}\ \emph {et~al.}(2000)\citenamefont {Bonnet},
  \citenamefont {Fitzhenry}, \citenamefont {Leinweber}, \citenamefont
  {Stanford},\ and\ \citenamefont {Williams}}]{SU3SmoothingCalibration}%
  \BibitemOpen
  \bibfield  {author} {\bibinfo {author} {\bibfnamefont {F.~D.~R.}\
  \bibnamefont {Bonnet}}, \bibinfo {author} {\bibfnamefont {P.}~\bibnamefont
  {Fitzhenry}}, \bibinfo {author} {\bibfnamefont {D.~B.}\ \bibnamefont
  {Leinweber}}, \bibinfo {author} {\bibfnamefont {M.~R.}\ \bibnamefont
  {Stanford}},\ and\ \bibinfo {author} {\bibfnamefont {A.~G.}\ \bibnamefont
  {Williams}},\ }\bibfield  {title} {\bibinfo {title} {{Calibration of smearing
  and cooling algorithms in SU(3)-color gauge theory}},\ }\href
  {https://doi.org/10.1103/PhysRevD.62.094509} {\bibfield  {journal} {\bibinfo
  {journal} {Phys. Rev. D}\ }\textbf {\bibinfo {volume} {62}},\ \bibinfo
  {pages} {094509} (\bibinfo {year} {2000})},\ \Eprint
  {https://arxiv.org/abs/hep-lat/0001018} {arXiv:hep-lat/0001018} \BibitemShut
  {NoStop}%
\bibitem [{\citenamefont {Leinweber}(2001)}]{QCDVis}%
  \BibitemOpen
  \bibfield  {author} {\bibinfo {author} {\bibfnamefont {D.~B.}\ \bibnamefont
  {Leinweber}},\ }\bibfield  {title} {\bibinfo {title} {{Visualizations of the
  QCD vacuum}},\ }in\ \href {https://doi.org/10.1142/4462} {\emph {\bibinfo
  {booktitle} {{Proceedings of the Workshop on Lightcone QCD and
  Nonperturbative Hadron Physics}}}}\ (\bibinfo  {publisher} {World Scientific
  Publishing},\ \bibinfo {address} {Singapore},\ \bibinfo {year} {2001})\ pp.\
  \bibinfo {pages} {138--143},\ \Eprint {https://arxiv.org/abs/hep-lat/0004025}
  {arXiv:hep-lat/0004025} \BibitemShut {NoStop}%
\bibitem [{\citenamefont {Berg}(1981)}]{CoolingI}%
  \BibitemOpen
  \bibfield  {author} {\bibinfo {author} {\bibfnamefont {B.}~\bibnamefont
  {Berg}},\ }\bibfield  {title} {\bibinfo {title} {{Dislocations and
  topological background in the lattice O(3) $\sigma$-model}},\ }\href
  {https://doi.org/10.1016/0370-2693(81)90518-9} {\bibfield  {journal}
  {\bibinfo  {journal} {Phys. Lett. B}\ }\textbf {\bibinfo {volume} {104}},\
  \bibinfo {pages} {475} (\bibinfo {year} {1981})}\BibitemShut {NoStop}%
\bibitem [{\citenamefont {Teper}(1985)}]{CoolingII}%
  \BibitemOpen
  \bibfield  {author} {\bibinfo {author} {\bibfnamefont {M.}~\bibnamefont
  {Teper}},\ }\bibfield  {title} {\bibinfo {title} {{Instantons in the
  quantized SU(2) vacuum: A lattice Monte Carlo investigation}},\ }\href
  {https://doi.org/10.1016/0370-2693(85)90939-6} {\bibfield  {journal}
  {\bibinfo  {journal} {Phys. Lett. B}\ }\textbf {\bibinfo {volume} {162}},\
  \bibinfo {pages} {357} (\bibinfo {year} {1985})}\BibitemShut {NoStop}%
\bibitem [{\citenamefont {Ilgenfritz}\ \emph {et~al.}(1986)\citenamefont
  {Ilgenfritz}, \citenamefont {Laursen}, \citenamefont {Schierholz},
  \citenamefont {M{\"u}ller-Preu{\ss}ker},\ and\ \citenamefont
  {Schiller}}]{CoolingIII}%
  \BibitemOpen
  \bibfield  {author} {\bibinfo {author} {\bibfnamefont {E.-M.}\ \bibnamefont
  {Ilgenfritz}}, \bibinfo {author} {\bibfnamefont {M.~L.}\ \bibnamefont
  {Laursen}}, \bibinfo {author} {\bibfnamefont {G.}~\bibnamefont {Schierholz}},
  \bibinfo {author} {\bibfnamefont {M.}~\bibnamefont
  {M{\"u}ller-Preu{\ss}ker}},\ and\ \bibinfo {author} {\bibfnamefont
  {H.}~\bibnamefont {Schiller}},\ }\bibfield  {title} {\bibinfo {title} {{First
  evidence for the existence of instantons in the quantized SU(2) lattice
  vacuum}},\ }\href {https://doi.org/10.1016/0550-3213(86)90265-8} {\bibfield
  {journal} {\bibinfo  {journal} {Nucl. Phys. B}\ }\textbf {\bibinfo {volume}
  {268}},\ \bibinfo {pages} {693} (\bibinfo {year} {1986})}\BibitemShut
  {NoStop}%
\bibitem [{\citenamefont {Bilson-Thompson}\ \emph {et~al.}(2003)\citenamefont
  {Bilson-Thompson}, \citenamefont {Leinweber},\ and\ \citenamefont
  {Williams}}]{ImprovedFmunu}%
  \BibitemOpen
  \bibfield  {author} {\bibinfo {author} {\bibfnamefont {S.~O.}\ \bibnamefont
  {Bilson-Thompson}}, \bibinfo {author} {\bibfnamefont {D.~B.}\ \bibnamefont
  {Leinweber}},\ and\ \bibinfo {author} {\bibfnamefont {A.~G.}\ \bibnamefont
  {Williams}},\ }\bibfield  {title} {\bibinfo {title} {{Highly improved lattice
  field-strength tensor}},\ }\href
  {https://doi.org/10.1016/S0003-4916(03)00009-5} {\bibfield  {journal}
  {\bibinfo  {journal} {Ann. Phys.}\ }\textbf {\bibinfo {volume} {304}},\
  \bibinfo {pages} {1} (\bibinfo {year} {2003})},\ \Eprint
  {https://arxiv.org/abs/hep-lat/0203008} {arXiv:hep-lat/0203008} \BibitemShut
  {NoStop}%
\bibitem [{\citenamefont {Witten}(1977)}]{Multi-instantonSolutionI}%
  \BibitemOpen
  \bibfield  {author} {\bibinfo {author} {\bibfnamefont {E.}~\bibnamefont
  {Witten}},\ }\bibfield  {title} {\bibinfo {title} {{Some Exact
  Multipseudoparticle Solutions of Classical Yang-Mills Theory}},\ }\href
  {https://doi.org/10.1103/PhysRevLett.38.121} {\bibfield  {journal} {\bibinfo
  {journal} {Phys. Rev. Lett.}\ }\textbf {\bibinfo {volume} {38}},\ \bibinfo
  {pages} {121} (\bibinfo {year} {1977})}\BibitemShut {NoStop}%
\bibitem [{\citenamefont {Atiyah}\ \emph {et~al.}(1978)\citenamefont {Atiyah},
  \citenamefont {Hitchin}, \citenamefont {Drinfeld},\ and\ \citenamefont
  {Manin}}]{Multi-instantonSolutionII}%
  \BibitemOpen
  \bibfield  {author} {\bibinfo {author} {\bibfnamefont {M.~F.}\ \bibnamefont
  {Atiyah}}, \bibinfo {author} {\bibfnamefont {N.~J.}\ \bibnamefont {Hitchin}},
  \bibinfo {author} {\bibfnamefont {V.~G.}\ \bibnamefont {Drinfeld}},\ and\
  \bibinfo {author} {\bibfnamefont {Y.~I.}\ \bibnamefont {Manin}},\ }\bibfield
  {title} {\bibinfo {title} {{Construction of instantons}},\ }\href
  {https://doi.org/10.1016/0375-9601(78)90141-X} {\bibfield  {journal}
  {\bibinfo  {journal} {Phys. Lett. A}\ }\textbf {\bibinfo {volume} {65}},\
  \bibinfo {pages} {185} (\bibinfo {year} {1978})}\BibitemShut {NoStop}%
\bibitem [{\citenamefont {Garc{\'i}a~P{\'e}rez}\ \emph
  {et~al.}(2000{\natexlab{a}})\citenamefont {Garc{\'i}a~P{\'e}rez},
  \citenamefont {Kov{\'a}cs},\ and\ \citenamefont {van
  Baal}}]{OverlappingInstantonsI}%
  \BibitemOpen
  \bibfield  {author} {\bibinfo {author} {\bibfnamefont {M.}~\bibnamefont
  {Garc{\'i}a~P{\'e}rez}}, \bibinfo {author} {\bibfnamefont {T.~G.}\
  \bibnamefont {Kov{\'a}cs}},\ and\ \bibinfo {author} {\bibfnamefont
  {P.}~\bibnamefont {van Baal}},\ }\bibfield  {title} {\bibinfo {title}
  {{Instanton size distribution}},\ }\href
  {https://doi.org/10.1016/S0370-2693(99)01451-3} {\bibfield  {journal}
  {\bibinfo  {journal} {Phys. Lett. B}\ }\textbf {\bibinfo {volume} {472}},\
  \bibinfo {pages} {295} (\bibinfo {year} {2000}{\natexlab{a}})},\ \Eprint
  {https://arxiv.org/abs/hep-ph/9911485} {arXiv:hep-ph/9911485} \BibitemShut
  {NoStop}%
\bibitem [{\citenamefont {Garc{\'i}a~P{\'e}rez}\ \emph
  {et~al.}(2000{\natexlab{b}})\citenamefont {Garc{\'i}a~P{\'e}rez},
  \citenamefont {Kov{\'a}cs},\ and\ \citenamefont {van
  Baal}}]{OverlappingInstantonsII}%
  \BibitemOpen
  \bibfield  {author} {\bibinfo {author} {\bibfnamefont {M.}~\bibnamefont
  {Garc{\'i}a~P{\'e}rez}}, \bibinfo {author} {\bibfnamefont {T.~G.}\
  \bibnamefont {Kov{\'a}cs}},\ and\ \bibinfo {author} {\bibfnamefont
  {P.}~\bibnamefont {van Baal}},\ }\bibfield  {title} {\bibinfo {title}
  {{Overlapping instantons}},\ }in\ \href {https://doi.org/10.1142/4563} {\emph
  {\bibinfo {booktitle} {{Proceedings of the 4th Workshop on Continuous
  Advances in QCD}}}}\ (\bibinfo  {publisher} {World Scientific Publishing},\
  \bibinfo {address} {Singapore},\ \bibinfo {year} {2000})\ pp.\ \bibinfo
  {pages} {79--89},\ \Eprint {https://arxiv.org/abs/hep-ph/0006155}
  {arXiv:hep-ph/0006155} \BibitemShut {NoStop}%
\bibitem [{\citenamefont {Falcioni}\ \emph {et~al.}(1985)\citenamefont
  {Falcioni}, \citenamefont {Paciello}, \citenamefont {Parisi},\ and\
  \citenamefont {Taglienti}}]{APEI}%
  \BibitemOpen
  \bibfield  {author} {\bibinfo {author} {\bibfnamefont {M.}~\bibnamefont
  {Falcioni}}, \bibinfo {author} {\bibfnamefont {M.~L.}\ \bibnamefont
  {Paciello}}, \bibinfo {author} {\bibfnamefont {G.}~\bibnamefont {Parisi}},\
  and\ \bibinfo {author} {\bibfnamefont {B.}~\bibnamefont {Taglienti}},\
  }\bibfield  {title} {\bibinfo {title} {{Again on SU(3) glueball mass}},\
  }\href {https://doi.org/10.1016/0550-3213(85)90280-9} {\bibfield  {journal}
  {\bibinfo  {journal} {Nucl. Phys. B}\ }\textbf {\bibinfo {volume} {251}},\
  \bibinfo {pages} {624} (\bibinfo {year} {1985})}\BibitemShut {NoStop}%
\bibitem [{\citenamefont {Albanese}\ \emph {et~al.}(1987)\citenamefont
  {Albanese} \emph {et~al.}}]{APEII}%
  \BibitemOpen
  \bibfield  {author} {\bibinfo {author} {\bibfnamefont {M.}~\bibnamefont
  {Albanese}} \emph {et~al.} (\bibinfo {collaboration} {APE Collaboration}),\
  }\bibfield  {title} {\bibinfo {title} {{Glueball masses and string tension in
  lattice QCD}},\ }\href {https://doi.org/10.1016/0370-2693(87)91160-9}
  {\bibfield  {journal} {\bibinfo  {journal} {Phys. Lett. B}\ }\textbf
  {\bibinfo {volume} {192}},\ \bibinfo {pages} {163} (\bibinfo {year}
  {1987})}\BibitemShut {NoStop}%
\bibitem [{\citenamefont {Morningstar}\ and\ \citenamefont
  {Peardon}(2004)}]{Stout-linkSmearing}%
  \BibitemOpen
  \bibfield  {author} {\bibinfo {author} {\bibfnamefont {C.}~\bibnamefont
  {Morningstar}}\ and\ \bibinfo {author} {\bibfnamefont {M.~J.}\ \bibnamefont
  {Peardon}},\ }\bibfield  {title} {\bibinfo {title} {{Analytic smearing of
  SU(3) link variables in lattice QCD}},\ }\href
  {https://doi.org/10.1103/PhysRevD.69.054501} {\bibfield  {journal} {\bibinfo
  {journal} {Phys. Rev. D}\ }\textbf {\bibinfo {volume} {69}},\ \bibinfo
  {pages} {054501} (\bibinfo {year} {2004})},\ \Eprint
  {https://arxiv.org/abs/hep-lat/0311018} {arXiv:hep-lat/0311018} \BibitemShut
  {NoStop}%
\bibitem [{\citenamefont {L\"uscher}(2010{\natexlab{a}})}]{GradientFlowI}%
  \BibitemOpen
  \bibfield  {author} {\bibinfo {author} {\bibfnamefont {M.}~\bibnamefont
  {L\"uscher}},\ }\bibfield  {title} {\bibinfo {title} {{Trivializing Maps, the
  Wilson Flow and the HMC Algorithm}},\ }\href
  {https://doi.org/10.1007/s00220-009-0953-7} {\bibfield  {journal} {\bibinfo
  {journal} {Commun. Math. Phys.}\ }\textbf {\bibinfo {volume} {293}},\
  \bibinfo {pages} {899} (\bibinfo {year} {2010}{\natexlab{a}})},\ \Eprint
  {https://arxiv.org/abs/0907.5491} {arXiv:0907.5491 [hep-lat]} \BibitemShut
  {NoStop}%
\bibitem [{\citenamefont {L\"uscher}(2010{\natexlab{b}})}]{GradientFlowII}%
  \BibitemOpen
  \bibfield  {author} {\bibinfo {author} {\bibfnamefont {M.}~\bibnamefont
  {L\"uscher}},\ }\bibfield  {title} {\bibinfo {title} {{Properties and uses of
  the Wilson flow in lattice QCD}},\ }\href
  {https://doi.org/10.1007/JHEP08(2010)071} {\bibfield  {journal} {\bibinfo
  {journal} {J. High Energy Phys.}\ }\textbf {\bibinfo {volume} {08}},\
  \bibinfo {pages} {071} (\bibinfo {year} {2010})},\ \bibinfo {note} {[Erratum: J. High Energy Phys. 03,
  092 (2014)]},\ \Eprint {https://arxiv.org/abs/1006.4518} {arXiv:1006.4518
  [hep-lat]} \BibitemShut {NoStop}%
\bibitem [{\citenamefont {Garc{\"i}a~P{\"e}rez}\ \emph
  {et~al.}(1994)\citenamefont {Garc{\"i}a~P{\"e}rez}, \citenamefont
  {Gonz\'alez-Arroyo}, \citenamefont {Snippe},\ and\ \citenamefont {van
  Baal}}]{Over-improvedCooling}%
  \BibitemOpen
  \bibfield  {author} {\bibinfo {author} {\bibfnamefont {M.}~\bibnamefont
  {Garc{\"i}a~P{\"e}rez}}, \bibinfo {author} {\bibfnamefont {A.}~\bibnamefont
  {Gonz\'alez-Arroyo}}, \bibinfo {author} {\bibfnamefont {J.~R.}\ \bibnamefont
  {Snippe}},\ and\ \bibinfo {author} {\bibfnamefont {P.}~\bibnamefont {van
  Baal}},\ }\bibfield  {title} {\bibinfo {title} {{Instantons from
  over-improved cooling}},\ }\href
  {https://doi.org/10.1016/0550-3213(94)90631-9} {\bibfield  {journal}
  {\bibinfo  {journal} {Nucl. Phys. B}\ }\textbf {\bibinfo {volume} {413}},\
  \bibinfo {pages} {535} (\bibinfo {year} {1994})},\ \Eprint
  {https://arxiv.org/abs/hep-lat/9309009} {arXiv:hep-lat/9309009} \BibitemShut
  {NoStop}%
\bibitem [{\citenamefont {Moran}\ and\ \citenamefont
  {Leinweber}(2008)}]{Over-improvedStout-linkSmearing}%
  \BibitemOpen
  \bibfield  {author} {\bibinfo {author} {\bibfnamefont {P.~J.}\ \bibnamefont
  {Moran}}\ and\ \bibinfo {author} {\bibfnamefont {D.~B.}\ \bibnamefont
  {Leinweber}},\ }\bibfield  {title} {\bibinfo {title} {{Over-improved
  stout-link smearing}},\ }\href {https://doi.org/10.1103/PhysRevD.77.094501}
  {\bibfield  {journal} {\bibinfo  {journal} {Phys. Rev. D}\ }\textbf {\bibinfo
  {volume} {77}},\ \bibinfo {pages} {094501} (\bibinfo {year} {2008})},\
  \Eprint {https://arxiv.org/abs/0801.1165} {arXiv:0801.1165 [hep-lat]}
  \BibitemShut {NoStop}%
\bibitem [{\citenamefont {Bonati}\ and\ \citenamefont
  {D'Elia}(2014)}]{GradientFlowCoolingComparison}%
  \BibitemOpen
  \bibfield  {author} {\bibinfo {author} {\bibfnamefont {C.}~\bibnamefont
  {Bonati}}\ and\ \bibinfo {author} {\bibfnamefont {M.}~\bibnamefont
  {D'Elia}},\ }\bibfield  {title} {\bibinfo {title} {{Comparison of the
  gradient flow with cooling in $SU(3)$ pure gauge theory}},\ }\href
  {https://doi.org/10.1103/PhysRevD.89.105005} {\bibfield  {journal} {\bibinfo
  {journal} {Phys. Rev. D}\ }\textbf {\bibinfo {volume} {89}},\ \bibinfo
  {pages} {105005} (\bibinfo {year} {2014})},\ \Eprint
  {https://arxiv.org/abs/1401.2441} {arXiv:1401.2441 [hep-lat]} \BibitemShut
  {NoStop}%
\bibitem [{\citenamefont {Thomas}\ \emph {et~al.}(2015)\citenamefont {Thomas},
  \citenamefont {Kamleh},\ and\ \citenamefont
  {Leinweber}}]{GradientFlowSmoothingConnectionI}%
  \BibitemOpen
  \bibfield  {author} {\bibinfo {author} {\bibfnamefont {S.~D.}\ \bibnamefont
  {Thomas}}, \bibinfo {author} {\bibfnamefont {W.}~\bibnamefont {Kamleh}},\
  and\ \bibinfo {author} {\bibfnamefont {D.~B.}\ \bibnamefont {Leinweber}},\
  }\bibfield  {title} {\bibinfo {title} {{Instanton contributions to the
  low-lying hadron mass spectrum}},\ }\href
  {https://doi.org/10.1103/PhysRevD.92.094515} {\bibfield  {journal} {\bibinfo
  {journal} {Phys. Rev. D}\ }\textbf {\bibinfo {volume} {92}},\ \bibinfo
  {pages} {094515} (\bibinfo {year} {2015})},\ \Eprint
  {https://arxiv.org/abs/1410.7105} {arXiv:1410.7105 [hep-lat]} \BibitemShut
  {NoStop}%
\bibitem [{\citenamefont {Nagatsuka}\ \emph {et~al.}(2023)\citenamefont
  {Nagatsuka}, \citenamefont {Sakai},\ and\ \citenamefont
  {Sasaki}}]{GradientFlowSmoothingConnectionII}%
  \BibitemOpen
  \bibfield  {author} {\bibinfo {author} {\bibfnamefont {M.}~\bibnamefont
  {Nagatsuka}}, \bibinfo {author} {\bibfnamefont {K.}~\bibnamefont {Sakai}},\
  and\ \bibinfo {author} {\bibfnamefont {S.}~\bibnamefont {Sasaki}},\
  }\bibfield  {title} {\bibinfo {title} {{Equivalence between the Wilson flow
  and stout-link smearing}},\ }\href
  {https://doi.org/10.1103/PhysRevD.108.094506} {\bibfield  {journal} {\bibinfo
   {journal} {Phys. Rev. D}\ }\textbf {\bibinfo {volume} {108}},\ \bibinfo
  {pages} {094506} (\bibinfo {year} {2023})},\ \Eprint
  {https://arxiv.org/abs/2303.09938} {arXiv:2303.09938 [hep-lat]} \BibitemShut
  {NoStop}%
\bibitem [{\citenamefont {Ilgenfritz}\ \emph {et~al.}(2008)\citenamefont
  {Ilgenfritz}, \citenamefont {Leinweber}, \citenamefont {Moran}, \citenamefont
  {Koller}, \citenamefont {Schierholz},\ and\ \citenamefont
  {Weinberg}}]{Eigenmodes}%
  \BibitemOpen
  \bibfield  {author} {\bibinfo {author} {\bibfnamefont {E.~M.}\ \bibnamefont
  {Ilgenfritz}}, \bibinfo {author} {\bibfnamefont {D.}~\bibnamefont
  {Leinweber}}, \bibinfo {author} {\bibfnamefont {P.}~\bibnamefont {Moran}},
  \bibinfo {author} {\bibfnamefont {K.}~\bibnamefont {Koller}}, \bibinfo
  {author} {\bibfnamefont {G.}~\bibnamefont {Schierholz}},\ and\ \bibinfo
  {author} {\bibfnamefont {V.}~\bibnamefont {Weinberg}},\ }\bibfield  {title}
  {\bibinfo {title} {{Vacuum structure revealed by over-improved stout-link
  smearing compared with the overlap analysis for quenched QCD}},\ }\href
  {https://doi.org/10.1103/PhysRevD.77.074502} {\bibfield  {journal} {\bibinfo
  {journal} {Phys. Rev. D}\ }\textbf {\bibinfo {volume} {77}},\ \bibinfo
  {pages} {074502} (\bibinfo {year} {2008})},\ \bibinfo {note} {[Erratum: Phys.
  Rev. D 77, 099902 (2008)]},\ \Eprint {https://arxiv.org/abs/0801.1725}
  {arXiv:0801.1725 [hep-lat]} \BibitemShut {NoStop}%
\bibitem [{\citenamefont {de~Forcrand}\ \emph {et~al.}(1997)\citenamefont
  {de~Forcrand}, \citenamefont {Garc{\"i}a~P{\"e}rez},\ and\ \citenamefont
  {Stamatescu}}]{ImprovedActionTopQ}%
  \BibitemOpen
  \bibfield  {author} {\bibinfo {author} {\bibfnamefont {P.}~\bibnamefont
  {de~Forcrand}}, \bibinfo {author} {\bibfnamefont {M.}~\bibnamefont
  {Garc{\"i}a~P{\"e}rez}},\ and\ \bibinfo {author} {\bibfnamefont {I.-O.}\
  \bibnamefont {Stamatescu}},\ }\bibfield  {title} {\bibinfo {title} {{Topology
  of the SU(2) vacuum: a lattice study using improved cooling}},\ }\href
  {https://doi.org/10.1016/S0550-3213(97)00275-7} {\bibfield  {journal}
  {\bibinfo  {journal} {Nucl. Phys. B}\ }\textbf {\bibinfo {volume} {499}},\
  \bibinfo {pages} {409} (\bibinfo {year} {1997})},\ \Eprint
  {https://arxiv.org/abs/hep-lat/9701012} {arXiv:hep-lat/9701012} \BibitemShut
  {NoStop}%
\bibitem [{\citenamefont {de~Forcrand}\ \emph {et~al.}(1996)\citenamefont
  {de~Forcrand}, \citenamefont {Garc{\"i}a~P{\"e}rez},\ and\ \citenamefont
  {Stamatescu}}]{5-loopImprovedCooling}%
  \BibitemOpen
  \bibfield  {author} {\bibinfo {author} {\bibfnamefont {P.}~\bibnamefont
  {de~Forcrand}}, \bibinfo {author} {\bibfnamefont {M.}~\bibnamefont
  {Garc{\"i}a~P{\"e}rez}},\ and\ \bibinfo {author} {\bibfnamefont {I.-O.}\
  \bibnamefont {Stamatescu}},\ }\bibfield  {title} {\bibinfo {title} {{Improved
  cooling algorithm for gauge theories}},\ }\href
  {https://doi.org/10.1016/0920-5632(96)00172-7} {\bibfield  {journal}
  {\bibinfo  {journal} {Nucl. Phys. B Proc. Suppl.}\ }\textbf {\bibinfo
  {volume} {47}},\ \bibinfo {pages} {777} (\bibinfo {year} {1996})},\ \Eprint
  {https://arxiv.org/abs/hep-lat/9509064} {arXiv:hep-lat/9509064} \BibitemShut
  {NoStop}%
\bibitem [{\citenamefont {Karsch}(2000)}]{CriticalTemp}%
  \BibitemOpen
  \bibfield  {author} {\bibinfo {author} {\bibfnamefont {F.}~\bibnamefont
  {Karsch}},\ }\bibfield  {title} {\bibinfo {title} {{Lattice QCD at finite
  temperature and density}},\ }\href
  {https://doi.org/10.1016/S0920-5632(00)91591-3} {\bibfield  {journal}
  {\bibinfo  {journal} {Nucl. Phys. B Proc. Suppl.}\ }\textbf {\bibinfo
  {volume} {83}},\ \bibinfo {pages} {14} (\bibinfo {year} {2000})},\ \Eprint
  {https://arxiv.org/abs/hep-lat/9909006} {arXiv:hep-lat/9909006} \BibitemShut
  {NoStop}%
\bibitem [{\citenamefont {Duane}\ \emph {et~al.}(1987)\citenamefont {Duane},
  \citenamefont {Kennedy}, \citenamefont {Pendleton},\ and\ \citenamefont
  {Roweth}}]{HMCI}%
  \BibitemOpen
  \bibfield  {author} {\bibinfo {author} {\bibfnamefont {S.}~\bibnamefont
  {Duane}}, \bibinfo {author} {\bibfnamefont {A.~D.}\ \bibnamefont {Kennedy}},
  \bibinfo {author} {\bibfnamefont {B.~J.}\ \bibnamefont {Pendleton}},\ and\
  \bibinfo {author} {\bibfnamefont {D.}~\bibnamefont {Roweth}},\ }\bibfield
  {title} {\bibinfo {title} {{Hybrid Monte Carlo}},\ }\href
  {https://doi.org/10.1016/0370-2693(87)91197-X} {\bibfield  {journal}
  {\bibinfo  {journal} {Phys. Lett. B}\ }\textbf {\bibinfo {volume} {195}},\
  \bibinfo {pages} {216} (\bibinfo {year} {1987})}\BibitemShut {NoStop}%
\bibitem [{\citenamefont {Kamleh}\ \emph {et~al.}(2004)\citenamefont {Kamleh},
  \citenamefont {Leinweber},\ and\ \citenamefont {Williams}}]{HMCII}%
  \BibitemOpen
  \bibfield  {author} {\bibinfo {author} {\bibfnamefont {W.}~\bibnamefont
  {Kamleh}}, \bibinfo {author} {\bibfnamefont {D.~B.}\ \bibnamefont
  {Leinweber}},\ and\ \bibinfo {author} {\bibfnamefont {A.~G.}\ \bibnamefont
  {Williams}},\ }\bibfield  {title} {\bibinfo {title} {{Hybrid Monte Carlo
  algorithm with fat link fermion actions}},\ }\href
  {https://doi.org/10.1103/PhysRevD.70.014502} {\bibfield  {journal} {\bibinfo
  {journal} {Phys. Rev. D}\ }\textbf {\bibinfo {volume} {70}},\ \bibinfo
  {pages} {014502} (\bibinfo {year} {2004})},\ \Eprint
  {https://arxiv.org/abs/hep-lat/0403019} {arXiv:hep-lat/0403019} \BibitemShut
  {NoStop}%
\bibitem [{\citenamefont {Iwasaki}(1983)}]{IwasakiActionI}%
  \BibitemOpen
  \bibfield  {author} {\bibinfo {author} {\bibfnamefont {Y.}~\bibnamefont
  {Iwasaki}},\ }\bibfield  {title} {\bibinfo {title} {{Renormalization Group
  Analysis of Lattice Theories and Improved Lattice Action. II.
  Four-dimensional non-Abelian SU(N) gauge model}},\ }\href@noop {} {\
  (\bibinfo {year} {1983})},\ \Eprint {https://arxiv.org/abs/1111.7054}
  {arXiv:1111.7054 [hep-lat]} \BibitemShut {NoStop}%
\bibitem [{\citenamefont {Iwasaki}\ \emph {et~al.}(1997)\citenamefont
  {Iwasaki}, \citenamefont {Kanaya}, \citenamefont {Kaneko},\ and\
  \citenamefont {Yoshi{\'e}}}]{IwasakiActionII}%
  \BibitemOpen
  \bibfield  {author} {\bibinfo {author} {\bibfnamefont {Y.}~\bibnamefont
  {Iwasaki}}, \bibinfo {author} {\bibfnamefont {K.}~\bibnamefont {Kanaya}},
  \bibinfo {author} {\bibfnamefont {T.}~\bibnamefont {Kaneko}},\ and\ \bibinfo
  {author} {\bibfnamefont {T.}~\bibnamefont {Yoshi{\'e}}},\ }\bibfield  {title}
  {\bibinfo {title} {{Scaling in SU(3) pure gauge theory with a
  renormalization-group-improved action}},\ }\href
  {https://doi.org/10.1103/PhysRevD.56.151} {\bibfield  {journal} {\bibinfo
  {journal} {Phys. Rev. D}\ }\textbf {\bibinfo {volume} {56}},\ \bibinfo
  {pages} {151} (\bibinfo {year} {1997})},\ \Eprint
  {https://arxiv.org/abs/hep-lat/9610023} {arXiv:hep-lat/9610023} \BibitemShut
  {NoStop}%
\bibitem [{\citenamefont {McLerran}\ and\ \citenamefont
  {Svetitsky}(1981)}]{PolyakovFreeEnergy}%
  \BibitemOpen
  \bibfield  {author} {\bibinfo {author} {\bibfnamefont {L.~D.}\ \bibnamefont
  {McLerran}}\ and\ \bibinfo {author} {\bibfnamefont {B.}~\bibnamefont
  {Svetitsky}},\ }\bibfield  {title} {\bibinfo {title} {{Quark liberation at
  high temperature: A Monte Carlo study of SU(2) gauge theory}},\ }\href
  {https://doi.org/10.1103/PhysRevD.24.450} {\bibfield  {journal} {\bibinfo
  {journal} {Phys. Rev. D}\ }\textbf {\bibinfo {volume} {24}},\ \bibinfo
  {pages} {450} (\bibinfo {year} {1981})}\BibitemShut {NoStop}%
\bibitem [{\citenamefont {Gattringer}\ and\ \citenamefont
  {Schmidt}(2011)}]{CentreClustersI}%
  \BibitemOpen
  \bibfield  {author} {\bibinfo {author} {\bibfnamefont {C.}~\bibnamefont
  {Gattringer}}\ and\ \bibinfo {author} {\bibfnamefont {A.}~\bibnamefont
  {Schmidt}},\ }\bibfield  {title} {\bibinfo {title} {{Center clusters in the
  Yang-Mills vacuum}},\ }\href {https://doi.org/10.1007/JHEP01(2011)051}
  {\bibfield  {journal} {\bibinfo  {journal} {J. High Energy Phys.}\ }\textbf
  {\bibinfo {volume} {01}},\ \bibinfo {pages} {051} (\bibinfo {year} {2011})},\ \Eprint
  {https://arxiv.org/abs/1011.2329} {arXiv:1011.2329 [hep-lat]} \BibitemShut
  {NoStop}%
\bibitem [{\citenamefont {Danzer}\ \emph {et~al.}(2010)\citenamefont {Danzer},
  \citenamefont {Gattringer}, \citenamefont {Borsanyi},\ and\ \citenamefont
  {Fodor}}]{CentreClustersII}%
  \BibitemOpen
  \bibfield  {author} {\bibinfo {author} {\bibfnamefont {J.}~\bibnamefont
  {Danzer}}, \bibinfo {author} {\bibfnamefont {C.}~\bibnamefont {Gattringer}},
  \bibinfo {author} {\bibfnamefont {S.}~\bibnamefont {Borsanyi}},\ and\
  \bibinfo {author} {\bibfnamefont {Z.}~\bibnamefont {Fodor}},\ }\bibfield
  {title} {\bibinfo {title} {{Center clusters and their percolation properties
  in lattice QCD}},\ }\href {https://doi.org/10.22323/1.105.0176} {\bibfield
  {journal} {\bibinfo  {journal} {Proc. Sci.}\ }\textbf {\bibinfo {volume}
  {LATTICE2010}},\ \bibinfo {pages} {176} (\bibinfo {year} {2010})},\ \Eprint
  {https://arxiv.org/abs/1010.5073} {arXiv:1010.5073 [hep-lat]} \BibitemShut
  {NoStop}%
\bibitem [{\citenamefont {Stokes}\ \emph {et~al.}(2014)\citenamefont {Stokes},
  \citenamefont {Kamleh},\ and\ \citenamefont
  {Leinweber}}]{CentreTransformation}%
  \BibitemOpen
  \bibfield  {author} {\bibinfo {author} {\bibfnamefont {F.~M.}\ \bibnamefont
  {Stokes}}, \bibinfo {author} {\bibfnamefont {W.}~\bibnamefont {Kamleh}},\
  and\ \bibinfo {author} {\bibfnamefont {D.~B.}\ \bibnamefont {Leinweber}},\
  }\bibfield  {title} {\bibinfo {title} {{Visualizations of coherent center
  domains in local Polyakov loops}},\ }\href
  {https://doi.org/10.1016/j.aop.2014.05.002} {\bibfield  {journal} {\bibinfo
  {journal} {Ann. Phys.}\ }\textbf {\bibinfo {volume} {348}},\ \bibinfo {pages}
  {341} (\bibinfo {year} {2014})},\ \Eprint {https://arxiv.org/abs/1312.0991}
  {arXiv:1312.0991 [hep-lat]} \BibitemShut {NoStop}%
\bibitem [{\citenamefont {Bruckmann}\ and\ \citenamefont {van
  Baal}(2002)}]{MulticaloronI}%
  \BibitemOpen
  \bibfield  {author} {\bibinfo {author} {\bibfnamefont {F.}~\bibnamefont
  {Bruckmann}}\ and\ \bibinfo {author} {\bibfnamefont {P.}~\bibnamefont {van
  Baal}},\ }\bibfield  {title} {\bibinfo {title} {{Multi-caloron solutions}},\
  }\href {https://doi.org/10.1016/S0550-3213(02)00834-9} {\bibfield  {journal}
  {\bibinfo  {journal} {Nucl. Phys. B}\ }\textbf {\bibinfo {volume} {645}},\
  \bibinfo {pages} {105} (\bibinfo {year} {2002})},\ \Eprint
  {https://arxiv.org/abs/hep-th/0209010} {arXiv:hep-th/0209010} \BibitemShut
  {NoStop}%
\bibitem [{\citenamefont {Bruckmann}\ \emph {et~al.}(2003)\citenamefont
  {Bruckmann}, \citenamefont {N{\'o}gr{\'a}di},\ and\ \citenamefont {van
  Baal}}]{MulticaloronII}%
  \BibitemOpen
  \bibfield  {author} {\bibinfo {author} {\bibfnamefont {F.}~\bibnamefont
  {Bruckmann}}, \bibinfo {author} {\bibfnamefont {D.}~\bibnamefont
  {N{\'o}gr{\'a}di}},\ and\ \bibinfo {author} {\bibfnamefont {P.}~\bibnamefont
  {van Baal}},\ }\bibfield  {title} {\bibinfo {title} {{Constituent monopoles
  through the eyes of fermion zero-modes}},\ }\href
  {https://doi.org/10.1016/S0550-3213(03)00531-5} {\bibfield  {journal}
  {\bibinfo  {journal} {Nucl. Phys. B}\ }\textbf {\bibinfo {volume} {666}},\
  \bibinfo {pages} {197} (\bibinfo {year} {2003})},\ \Eprint
  {https://arxiv.org/abs/hep-th/0305063} {arXiv:hep-th/0305063} \BibitemShut
  {NoStop}%
\bibitem [{\citenamefont {Bruckmann}\ \emph {et~al.}(2004)\citenamefont
  {Bruckmann}, \citenamefont {N{\'o}gr{\'a}di},\ and\ \citenamefont {van
  Baal}}]{MulticaloronIII}%
  \BibitemOpen
  \bibfield  {author} {\bibinfo {author} {\bibfnamefont {F.}~\bibnamefont
  {Bruckmann}}, \bibinfo {author} {\bibfnamefont {D.}~\bibnamefont
  {N{\'o}gr{\'a}di}},\ and\ \bibinfo {author} {\bibfnamefont {P.}~\bibnamefont
  {van Baal}},\ }\bibfield  {title} {\bibinfo {title} {{Higher charge calorons
  with non-trivial holonomy}},\ }\href
  {https://doi.org/10.1016/j.nuclphysb.2004.07.038} {\bibfield  {journal}
  {\bibinfo  {journal} {Nucl. Phys. B}\ }\textbf {\bibinfo {volume} {698}},\
  \bibinfo {pages} {233} (\bibinfo {year} {2004})},\ \Eprint
  {https://arxiv.org/abs/hep-th/0404210} {arXiv:hep-th/0404210} \BibitemShut
  {NoStop}%
\bibitem [{\citenamefont {Bruckmann}\ \emph {et~al.}(2005)\citenamefont
  {Bruckmann}, \citenamefont {Ilgenfritz}, \citenamefont {Martemyanov},
  \citenamefont {M{\"u}ller-Preussker}, \citenamefont {N{\'o}gr{\'a}di},
  \citenamefont {Peschka},\ and\ \citenamefont {van Baal}}]{MulticaloronIV}%
  \BibitemOpen
  \bibfield  {author} {\bibinfo {author} {\bibfnamefont {F.}~\bibnamefont
  {Bruckmann}}, \bibinfo {author} {\bibfnamefont {E.~M.}\ \bibnamefont
  {Ilgenfritz}}, \bibinfo {author} {\bibfnamefont {B.~V.}\ \bibnamefont
  {Martemyanov}}, \bibinfo {author} {\bibfnamefont {M.}~\bibnamefont
  {M{\"u}ller-Preussker}}, \bibinfo {author} {\bibfnamefont {D.}~\bibnamefont
  {N{\'o}gr{\'a}di}}, \bibinfo {author} {\bibfnamefont {D.}~\bibnamefont
  {Peschka}},\ and\ \bibinfo {author} {\bibfnamefont {P.}~\bibnamefont {van
  Baal}},\ }\bibfield  {title} {\bibinfo {title} {{Calorons with non-trivial
  holonomy on and off the lattice}},\ }\href
  {https://doi.org/10.1016/j.nuclphysbps.2004.11.268} {\bibfield  {journal}
  {\bibinfo  {journal} {Nucl. Phys. B Proc. Suppl.}\ }\textbf {\bibinfo
  {volume} {140}},\ \bibinfo {pages} {635} (\bibinfo {year} {2005})},\ \Eprint
  {https://arxiv.org/abs/hep-lat/0408036} {arXiv:hep-lat/0408036} \BibitemShut
  {NoStop}%
\bibitem [{\citenamefont {Nogradi}(2005)}]{MulticaloronV}%
  \BibitemOpen
  \bibfield  {author} {\bibinfo {author} {\bibfnamefont {D.}~\bibnamefont
  {Nogradi}},\ }\emph {\bibinfo {title} {{Multi-calorons and their moduli}}},\
  \href@noop {} {Ph.D. thesis} (\bibinfo {year} {2005}),\ \Eprint
  {https://arxiv.org/abs/hep-th/0511125} {arXiv:hep-th/0511125} \BibitemShut
  {NoStop}%
\bibitem [{\citenamefont {Gerhold}\ \emph
  {et~al.}(2007{\natexlab{a}})\citenamefont {Gerhold}, \citenamefont
  {Ilgenfritz},\ and\ \citenamefont
  {M{\"u}ller-Preussker}}]{CaloronSuperpositionI}%
  \BibitemOpen
  \bibfield  {author} {\bibinfo {author} {\bibfnamefont {P.}~\bibnamefont
  {Gerhold}}, \bibinfo {author} {\bibfnamefont {E.~M.}\ \bibnamefont
  {Ilgenfritz}},\ and\ \bibinfo {author} {\bibfnamefont {M.}~\bibnamefont
  {M{\"u}ller-Preussker}},\ }\bibfield  {title} {\bibinfo {title} {{An
  {$SU(2)$} KvBLL caloron gas model and confinement}},\ }\href
  {https://doi.org/10.1016/j.nuclphysb.2006.10.003} {\bibfield  {journal}
  {\bibinfo  {journal} {Nucl. Phys. B}\ }\textbf {\bibinfo {volume} {760}},\
  \bibinfo {pages} {1} (\bibinfo {year} {2007}{\natexlab{a}})},\ \Eprint
  {https://arxiv.org/abs/hep-ph/0607315} {arXiv:hep-ph/0607315} \BibitemShut
  {NoStop}%
\bibitem [{\citenamefont {Gerhold}\ \emph
  {et~al.}(2007{\natexlab{b}})\citenamefont {Gerhold}, \citenamefont
  {Ilgenfritz},\ and\ \citenamefont
  {M{\"u}ller-Preussker}}]{CaloronSuperpositionII}%
  \BibitemOpen
  \bibfield  {author} {\bibinfo {author} {\bibfnamefont {P.}~\bibnamefont
  {Gerhold}}, \bibinfo {author} {\bibfnamefont {E.-M.}\ \bibnamefont
  {Ilgenfritz}},\ and\ \bibinfo {author} {\bibfnamefont {M.}~\bibnamefont
  {M{\"u}ller-Preussker}},\ }\bibfield  {title} {\bibinfo {title} {{Improved
  superposition schemes for approximate multi-caloron configurations}},\ }\href
  {https://doi.org/10.1016/j.nuclphysb.2007.04.003} {\bibfield  {journal}
  {\bibinfo  {journal} {Nucl. Phys. B}\ }\textbf {\bibinfo {volume} {774}},\
  \bibinfo {pages} {268} (\bibinfo {year} {2007}{\natexlab{b}})},\ \Eprint
  {https://arxiv.org/abs/hep-ph/0610426} {arXiv:hep-ph/0610426} \BibitemShut
  {NoStop}%
\bibitem [{\citenamefont {Bruckmann}\ \emph {et~al.}(2010)\citenamefont
  {Bruckmann}, \citenamefont {Ilgenfritz}, \citenamefont {Martemyanov},\ and\
  \citenamefont {Zhang}}]{CaloronSuperpositionIII}%
  \BibitemOpen
  \bibfield  {author} {\bibinfo {author} {\bibfnamefont {F.}~\bibnamefont
  {Bruckmann}}, \bibinfo {author} {\bibfnamefont {E.-M.}\ \bibnamefont
  {Ilgenfritz}}, \bibinfo {author} {\bibfnamefont {B.}~\bibnamefont
  {Martemyanov}},\ and\ \bibinfo {author} {\bibfnamefont {B.}~\bibnamefont
  {Zhang}},\ }\bibfield  {title} {\bibinfo {title} {{Vortex structure of
  {$SU(2)$} calorons}},\ }\href {https://doi.org/10.1103/PhysRevD.81.074501}
  {\bibfield  {journal} {\bibinfo  {journal} {Phys. Rev. D}\ }\textbf {\bibinfo
  {volume} {81}},\ \bibinfo {pages} {074501} (\bibinfo {year} {2010})},\
  \Eprint {https://arxiv.org/abs/0912.4186} {arXiv:0912.4186 [hep-th]}
  \BibitemShut {NoStop}%
\bibitem [{\citenamefont {Atiyah}\ and\ \citenamefont
  {Singer}(1969)}]{IndexTheorem}%
  \BibitemOpen
  \bibfield  {author} {\bibinfo {author} {\bibfnamefont {M.~F.}\ \bibnamefont
  {Atiyah}}\ and\ \bibinfo {author} {\bibfnamefont {I.~M.}\ \bibnamefont
  {Singer}},\ }\bibfield  {title} {\bibinfo {title} {{The index of elliptic
  operators on compact manifolds}},\ }\href
  {https://doi.org/10.1090/S0002-9904-1963-10957-X} {\bibfield  {journal}
  {\bibinfo  {journal} {Bull. Am. Math. Soc.}\ }\textbf {\bibinfo {volume}
  {69}},\ \bibinfo {pages} {422} (\bibinfo {year} {1969})}\BibitemShut
  {NoStop}%
\bibitem [{\citenamefont {Bornyakov}\ \emph {et~al.}(2013)\citenamefont
  {Bornyakov}, \citenamefont {Ilgenfritz}, \citenamefont {Martemyanov},
  \citenamefont {Mitrjushkin},\ and\ \citenamefont
  {M\"uller-Preussker}}]{TopologicalIndex}%
  \BibitemOpen
  \bibfield  {author} {\bibinfo {author} {\bibfnamefont {V.~G.}\ \bibnamefont
  {Bornyakov}}, \bibinfo {author} {\bibfnamefont {E.~M.}\ \bibnamefont
  {Ilgenfritz}}, \bibinfo {author} {\bibfnamefont {B.~V.}\ \bibnamefont
  {Martemyanov}}, \bibinfo {author} {\bibfnamefont {V.~K.}\ \bibnamefont
  {Mitrjushkin}},\ and\ \bibinfo {author} {\bibfnamefont {M.}~\bibnamefont
  {M\"uller-Preussker}},\ }\bibfield  {title} {\bibinfo {title} {{Topology
  across the finite temperature transition studied by overimproved cooling in
  gluodynamics and QCD}},\ }\href {https://doi.org/10.1103/PhysRevD.87.114508}
  {\bibfield  {journal} {\bibinfo  {journal} {Phys. Rev. D}\ }\textbf {\bibinfo
  {volume} {87}},\ \bibinfo {pages} {114508} (\bibinfo {year} {2013})},\
  \Eprint {https://arxiv.org/abs/1304.0935} {arXiv:1304.0935 [hep-lat]}
  \BibitemShut {NoStop}%
\bibitem [{\citenamefont {Trewartha}\ \emph {et~al.}(2013)\citenamefont
  {Trewartha}, \citenamefont {Kamleh}, \citenamefont {Leinweber},\ and\
  \citenamefont {Roberts}}]{QuarkPropagationInstantons}%
  \BibitemOpen
  \bibfield  {author} {\bibinfo {author} {\bibfnamefont {A.}~\bibnamefont
  {Trewartha}}, \bibinfo {author} {\bibfnamefont {W.}~\bibnamefont {Kamleh}},
  \bibinfo {author} {\bibfnamefont {D.}~\bibnamefont {Leinweber}},\ and\
  \bibinfo {author} {\bibfnamefont {D.~S.}\ \bibnamefont {Roberts}},\
  }\bibfield  {title} {\bibinfo {title} {{Quark propagation in the instantons
  of lattice QCD}},\ }\href {https://doi.org/10.1103/PhysRevD.88.034501}
  {\bibfield  {journal} {\bibinfo  {journal} {Phys. Rev. D}\ }\textbf {\bibinfo
  {volume} {88}},\ \bibinfo {pages} {034501} (\bibinfo {year} {2013})},\
  \Eprint {https://arxiv.org/abs/1306.3283} {arXiv:1306.3283 [hep-lat]}
  \BibitemShut {NoStop}%
\bibitem [{\citenamefont {Trewartha}\ \emph {et~al.}(2015)\citenamefont
  {Trewartha}, \citenamefont {Kamleh},\ and\ \citenamefont
  {Leinweber}}]{InstantonsSmoothing}%
  \BibitemOpen
  \bibfield  {author} {\bibinfo {author} {\bibfnamefont {A.}~\bibnamefont
  {Trewartha}}, \bibinfo {author} {\bibfnamefont {W.}~\bibnamefont {Kamleh}},\
  and\ \bibinfo {author} {\bibfnamefont {D.}~\bibnamefont {Leinweber}},\
  }\bibfield  {title} {\bibinfo {title} {{Connection between center vortices
  and instantons through gauge-field smoothing}},\ }\href
  {https://doi.org/10.1103/PhysRevD.92.074507} {\bibfield  {journal} {\bibinfo
  {journal} {Phys. Rev. D}\ }\textbf {\bibinfo {volume} {92}},\ \bibinfo
  {pages} {074507} (\bibinfo {year} {2015})},\ \Eprint
  {https://arxiv.org/abs/1509.05518} {arXiv:1509.05518 [hep-lat]} \BibitemShut
  {NoStop}%
\bibitem [{\citenamefont {Soler}\ \emph {et~al.}(2024)\citenamefont {Soler},
  \citenamefont {Bergner},\ and\ \citenamefont
  {Gonz\'alez-Arroyo}}]{AFMAnalysis}%
  \BibitemOpen
  \bibfield  {author} {\bibinfo {author} {\bibfnamefont {I.}~\bibnamefont
  {Soler}}, \bibinfo {author} {\bibfnamefont {G.}~\bibnamefont {Bergner}},\
  and\ \bibinfo {author} {\bibfnamefont {A.}~\bibnamefont
  {Gonz\'alez-Arroyo}},\ }\bibfield  {title} {\bibinfo {title} {{Extracting
  Yang-Mills topological structures with adjoint modes}},\ }\href
  {https://doi.org/10.22323/1.453.0378} {\bibfield  {journal} {\bibinfo
  {journal} {Proc. Sci.}\ }\textbf {\bibinfo {volume} {LATTICE2023}},\ \bibinfo
  {pages} {378} (\bibinfo {year} {2024})},\ \Eprint
  {https://arxiv.org/abs/2312.06308} {arXiv:2312.06308 [hep-lat]} \BibitemShut
  {NoStop}%
\bibitem [{\citenamefont {Gonz\'alez-Arroyo}\ and\ \citenamefont
  {Kirchner}(2006)}]{AFMI}%
  \BibitemOpen
  \bibfield  {author} {\bibinfo {author} {\bibfnamefont {A.}~\bibnamefont
  {Gonz\'alez-Arroyo}}\ and\ \bibinfo {author} {\bibfnamefont {R.}~\bibnamefont
  {Kirchner}},\ }\bibfield  {title} {\bibinfo {title} {{Adjoint modes as probes
  of gauge field structure}},\ }\href
  {https://doi.org/10.1088/1126-6708/2006/01/029} {\bibfield  {journal}
  {\bibinfo  {journal} {J. High Energy Phys.}\ }\textbf {\bibinfo {volume}
  {01}},\ \bibinfo {pages} {029} (\bibinfo {year} {2006})},\ \Eprint
  {https://arxiv.org/abs/hep-lat/0507036} {arXiv:hep-lat/0507036} \BibitemShut
  {NoStop}%
\bibitem [{\citenamefont {Garc{\'i}a~P{\'e}rez}\ \emph
  {et~al.}(2011)\citenamefont {Garc{\'i}a~P{\'e}rez}, \citenamefont
  {Gonz\'alez-Arroyo},\ and\ \citenamefont {Sastre}}]{AFMII}%
  \BibitemOpen
  \bibfield  {author} {\bibinfo {author} {\bibfnamefont {M.}~\bibnamefont
  {Garc{\'i}a~P{\'e}rez}}, \bibinfo {author} {\bibfnamefont {A.}~\bibnamefont
  {Gonz\'alez-Arroyo}},\ and\ \bibinfo {author} {\bibfnamefont
  {A.}~\bibnamefont {Sastre}},\ }\bibfield  {title} {\bibinfo {title}
  {{Ultraviolet filtering of lattice configurations and applications to Monte
  Carlo dynamics}},\ }\href {https://doi.org/10.1007/JHEP07(2011)034}
  {\bibfield  {journal} {\bibinfo  {journal} {J. High Energy Phys.}\ }\textbf
  {\bibinfo {volume} {07}},\ \bibinfo {pages} {034} (\bibinfo {year} {2011})},\ \Eprint
  {https://arxiv.org/abs/1103.5999} {arXiv:1103.5999 [hep-lat]} \BibitemShut
  {NoStop}%
\bibitem [{\citenamefont {Wilson}(1974)}]{WilsonAction}%
  \BibitemOpen
  \bibfield  {author} {\bibinfo {author} {\bibfnamefont {K.~G.}\ \bibnamefont
  {Wilson}},\ }\bibfield  {title} {\bibinfo {title} {{Confinement of quarks}},\
  }\href {https://doi.org/10.1103/PhysRevD.10.2445} {\bibfield  {journal}
  {\bibinfo  {journal} {Phys. Rev. D}\ }\textbf {\bibinfo {volume} {10}},\
  \bibinfo {pages} {2445} (\bibinfo {year} {1974})}\BibitemShut {NoStop}%
\bibitem [{\citenamefont {Cabibbo}\ and\ \citenamefont
  {Marinari}(1982)}]{CabibboMarinari}%
  \BibitemOpen
  \bibfield  {author} {\bibinfo {author} {\bibfnamefont {N.}~\bibnamefont
  {Cabibbo}}\ and\ \bibinfo {author} {\bibfnamefont {E.}~\bibnamefont
  {Marinari}},\ }\bibfield  {title} {\bibinfo {title} {{A new method for
  updating SU({$N$}) matrices in computer simulations of gauge theories}},\
  }\href {https://doi.org/10.1016/0370-2693(82)90696-7} {\bibfield  {journal}
  {\bibinfo  {journal} {Phys. Lett. B}\ }\textbf {\bibinfo {volume} {119}},\
  \bibinfo {pages} {387} (\bibinfo {year} {1982})}\BibitemShut {NoStop}%
\bibitem [{\citenamefont {Bonnet}\ \emph {et~al.}(2002)\citenamefont {Bonnet},
  \citenamefont {Leinweber}, \citenamefont {Williams},\ and\ \citenamefont
  {Zanotti}}]{ImprovedSmoothing}%
  \BibitemOpen
  \bibfield  {author} {\bibinfo {author} {\bibfnamefont {F.~D.~R.}\
  \bibnamefont {Bonnet}}, \bibinfo {author} {\bibfnamefont {D.~B.}\
  \bibnamefont {Leinweber}}, \bibinfo {author} {\bibfnamefont {A.~G.}\
  \bibnamefont {Williams}},\ and\ \bibinfo {author} {\bibfnamefont {J.~M.}\
  \bibnamefont {Zanotti}},\ }\bibfield  {title} {\bibinfo {title} {{Improved
  smoothing algorithms for lattice gauge theory}},\ }\href
  {https://doi.org/10.1103/PhysRevD.65.114510} {\bibfield  {journal} {\bibinfo
  {journal} {Phys. Rev. D}\ }\textbf {\bibinfo {volume} {65}},\ \bibinfo
  {pages} {114510} (\bibinfo {year} {2002})},\ \Eprint
  {https://arxiv.org/abs/hep-lat/0106023} {arXiv:hep-lat/0106023} \BibitemShut
  {NoStop}%
\bibitem [{\citenamefont {Zanotti}\ \emph {et~al.}(2005)\citenamefont
  {Zanotti}, \citenamefont {Lasscock}, \citenamefont {Leinweber},\ and\
  \citenamefont {Williams}}]{CoolingLoops}%
  \BibitemOpen
  \bibfield  {author} {\bibinfo {author} {\bibfnamefont {J.~M.}\ \bibnamefont
  {Zanotti}}, \bibinfo {author} {\bibfnamefont {B.}~\bibnamefont {Lasscock}},
  \bibinfo {author} {\bibfnamefont {D.~B.}\ \bibnamefont {Leinweber}},\ and\
  \bibinfo {author} {\bibfnamefont {A.~G.}\ \bibnamefont {Williams}},\
  }\bibfield  {title} {\bibinfo {title} {{Scaling of fat-link irrelevant-clover
  fermions}},\ }\href {https://doi.org/10.1103/PhysRevD.71.034510} {\bibfield
  {journal} {\bibinfo  {journal} {Phys. Rev. D}\ }\textbf {\bibinfo {volume}
  {71}},\ \bibinfo {pages} {034510} (\bibinfo {year} {2005})},\ \Eprint
  {https://arxiv.org/abs/hep-lat/0405015} {arXiv:hep-lat/0405015} \BibitemShut
  {NoStop}%
\end{thebibliography}%

\end{document}